\documentclass[journal=dce]{CUP-JNL-DTM}%

\usepackage{graphicx}
\usepackage{multicol,multirow}
\usepackage{amsmath,amssymb,amsfonts}
\usepackage{mathrsfs}
\usepackage{amsthm}
\usepackage{rotating}
\usepackage{appendix}
\usepackage{ifpdf}
\usepackage[T1]{fontenc}
\usepackage{newtxtext}
\usepackage{newtxmath}
\usepackage{textcomp}
\usepackage{xcolor}
\usepackage{lipsum}
\usepackage{lineno}
\usepackage[colorlinks,allcolors=blue]{hyperref}
\usepackage[framemethod=tikz]{mdframed}
\usepackage[ruled,vlined,english]{algorithm2e}
\usepackage{ragged2e} % in your preamble

\newcommand{\circlewhite}{\raisebox{-1pt}{\tikz{\draw[color=black,line width = 1pt] (0,0) circle [radius=0.1];}}}

\definecolor{myyellow}{RGB}{237, 177, 32}
\newcommand{\yellowline}{\raisebox{2pt}{\tikz{\draw[-,myyellow, solid, line width = 1pt] (0,0) -- (5mm,0);}}}

\newcommand{\redline}{\raisebox{2pt}{\tikz{\draw[-,red,solid,line width = 1pt](0,0) -- (5mm,0);}}}

\definecolor{greenlime}{RGB}{50,205,50}
\newcommand{\greenlimelinedotted}{\raisebox{2pt}{\tikz{\draw[-,greenlime,dotted,line width = 1.5pt](0,0) -- (5mm,0);}}}

\newcommand{\greenlimeline}{\raisebox{2pt}{\tikz{\draw[-,greenlime,solid,line width = 1pt](0,0) -- (5mm,0);}}}

\definecolor{orange}{RGB}{255,165,0}
\newcommand{\orangeline}{\raisebox{2pt}{\tikz{\draw[-,orange,solid,line width = 1pt](0,0) -- (5mm,0);}}}

\newcommand{\blueline}{\raisebox{2pt}{\tikz{\draw[-.,blue,solid,line width = 1pt](0,0) -- (5mm,0);}}}

\definecolor{greySTRONG}{RGB}{127.5,127.5,127.5}
\newcommand{\greylinedottedstrong}{\raisebox{2pt}{\tikz{\draw[-,greySTRONG,dotted,line width = 1.5pt](0,0) -- (5mm,0)}}}

\newcommand{\greylinesolidstrong}{\raisebox{2pt}{\tikz{\draw[-,greySTRONG,solid,line width = 1.5pt](0,0) -- (5mm,0)}}}

\definecolor{magenta}{RGB}{255.0, 0, 255.0}
\newcommand{\magentaline}{\raisebox{2pt}{\tikz{\draw[-,magenta,solid,line width = 1pt](0,0) -- (5mm,0);}}}

\theoremstyle{definition}

\numberwithin{equation}{section}

\articletype{RESEARCH ARTICLE}
\jyear{2025}
\begin{document}

\begin{Frontmatter}

%\title[Article Title]{A Data-informed Immersed Boundary Method Using Online Supervised Machine Learning with Streaming Data}
\title[Article Title]{Towards Streaming Prediction of Oscillatory Flows: A Data Assimilation and Machine Learning Approach}

% There is no need to include ORCID IDs in your .pdf; this information is captured by the submission portal when a manuscript is submitted. 

% I will need to indicate that I am the corresponding author
\author[1]{Miguel M. Valero}
\author[1]{Marcello Meldi}
%\author[2]{Author Name3}

\authormark{M. M. Valero and M. Meldi}

\address[1]{\orgname{Univ. Lille, CNRS, ONERA, Arts et Métiers ParisTech, Centrale Lille, UMR 9014- LMFL- Laboratoire de Mécanique des fluides de Lille - Kampé de Feriet}, \orgaddress{ \postcode{F-59000}, \city{Lille}, \country{France}}} 
\keywords[Corresponding author]{Miguel M. Valero \email{miguel.martinez@centralelille.fr}} 

%\address[2]{\orgdiv{Division}, \orgname{Organization}, \orgaddress{\city{City}, \postcode{Pincode}, \state{State},  \country{Country}}. \email{name2@email.com}}comment

\authormark{M. M. Valero and M. Meldi.}


%\keywords[MSC Codes]{\codes[Primary]{CODE1}; \codes[Secondary]{CODE2, CODE3}}

\abstract{Data-driven methods have demonstrated strong predictive capabilities in fluid mechanics, yet most current applications still focus on simplified configurations, often characterised by statistical stationarity or limited temporal variability. This work proposes a methodology that combines Data Assimilation (DA) and Machine Learning (ML) to predict flow configurations that exhibit cyclic behaviour over time. Starting from limited, sparse high-fidelity measurements and a low-fidelity numerical model, the DA approach performs data fusion to obtain complete and accurate flow state estimations in time. This complete dataset is used to train multiple ML tools, which are applied across different phases of the flow cycle to augment the model's predictions when high-fidelity data might not be available for the DA application. The methodology is applied to the analysis of an oscillating cylinder in a laminar regime using a sliding-window approach, in which separate models are trained for specific flow conditions to ensure each model specialises in flow dynamics representative of a phase of the oscillation period. This phase-resolved learning enables the efficient capture of transient features that would be challenging for a single global model. The results highlight the potential of this method to study complex flow configurations with oscillatory features in which neither the flow nor the cycle is known a priori, in particular by exploiting real-time training and updates, as is commonly done in digital twins, which require continuous model correction and adaptation.}

\end{Frontmatter}

%% UNCOMMENT FOR DCE %%
%\section*{Impact Statement}
%Some Data journals (DAP, DCE) require an `Impact Statement' section. Comment out this section if it is not required.

% Some math journals (FLO) require a table of contents. Comment out this line if no ToC is needed.
%\localtableofcontents
%\linenumbers

\section{Introduction}
\label{sec:Introduction}

Over the past decade, methods based on data usage (data-driven, data-informed, data-aware...) provided advancement on a wide range of problems in computational and experimental fluid mechanics. Such approaches have been used to develop equation-free or partially learned flow representations through reduced-order modelling \citep{Kutz2023} and system identification techniques \citep{Brunton2016_pnas}, to design and optimise estimation and control strategies in unsteady flows \citep{Noack2011, Brunton2015_amr}, and to improve or complement classical closure models in both Reynolds-Averaged Navier-Stokes (RANS) and Large Eddy Simulations (LES) turbulence frameworks \citep{Pope2000_cambridge, Duraisamy2019_arfm}. These strategies include methods that extract information from limited measurements (sparse sensing), reconstruct flow fields, learn model relationships through supervised training, and adaptively correct predictions in real time, and share the common goal of exploiting high-fidelity or experimental data to enhance predictive capability, stability, or computational efficiency across diverse flow regimes.

%Despite decades of progress in turbulence modelling \citep{Pope2000_cambridge,Duraisamy2019_arfm}, the development of new classical approaches has notably slowed in recent years, with several authors suggesting that expectations for major breakthroughs in physical models in the near term remain low \citep{Spalart2016_taj}. Moreover, existing models, whether based on Reynolds-Averaged Navier-Stokes (RANS) closures \citep{Wilcox2006_DCW,Xiao2019_pas} or Large Eddy Simulation (LES) subgrid-scale formulations \citep{Sagaut2006}, are typically calibrated for specific flow conditions, such as canonical zero-pressure-gradient flat-plate boundary layers or fully developed channel flows. Consequently, their predictive accuracy often degrades when applied to realistic configurations featuring complex flow features, in which one can find strong adverse pressure gradients, massive flow separation, or unsteady interactions among multiple flow regions. These persistent model limitations have motivated the pursuit of innovative, ``data-informed'' or ``data-driven'' modelling strategies \citep{Mendez2023}, in which high-fidelity numerical or experimental data are incorporated to infer model parameters, correct closure terms, and replace empirical coefficients. Such methods aim to enhance or complement conventional turbulence modelling frameworks.

Within these data-driven approaches, measurements or flow reconstructions, commonly referred to as \emph{sensors}, play a central role. In Data Assimilation (DA) \citep{Asch2016_siam}, sensor information, also known as \emph{observations}, is continuously integrated into the numerical model to correct the evolving flow state \citep{Mons2016_jcp, Villanueva2024_ijhff,  Jeanney2025} and, in some formulations, to adjust unknown spatially distributed physical model parameters \citep{Franceschini2020_prf, BenAli2022_jweia, Moussie2024_ftc}. Machine Learning (ML) methods \citep{Brunton2020_arfm} typically use sensor data \emph{offline} to train models that identify patterns and extract flow features \citep{Strofer2018_cicp}, super-resolution reconstruction \citep{Fukami2019_jfm}, or emulate governing equations \citep{Raissi2019_jcp}. While both paradigms rely on the quality and representativeness of sensor information, their effectiveness in both cases strongly depends on how well the available data captures the relevant flow dynamics.

In this context, several key challenges emerge when extending these data-driven techniques beyond idealised conditions. Actually, most current studies evaluating DA and ML frameworks in Computational Fluid Dynamics (CFD) still focus on academic test cases characterised by statistically steady flow conditions \citep{Duraisamy2019_arfm}. Hence, further developments are required to enable the study of real-world applications, which often deviate from these assumptions and involve inherently unsteady, time-dependent phenomena. In such scenarios, additional complexities arise, including:

(i) Sensors might be moving, or the relative distance between the sensors and the regions of physical interest (such as surfaces for integral evaluations, wake structures, or detachment zones) can vary with time, as in the case of moving bodies. For example, in the study of oscillating or pitching airfoils, the instantaneous positions of pressure sensors mounted on the surface or in the wake change continuously with respect to the characteristic flow features \citep{PiccoloSerafim2025_aiaa}. This relative motion modifies the local sampling of flow quantities and complicates both state estimation (or field reconstruction) and model training, particularly when the flow exhibits large variations in separation and reattachment over a cycle.

(ii) Beyond sensor-related issues, unsteady configurations often involve complex interactions across spatial and temporal scales, such as vortex shedding influencing upstream boundary-layer development or unsteady separation affecting distant wake regions. These nonlocal and phase-dependent couplings evolve during the motion cycle and cannot be captured by purely local measurements or static mappings. For DA, this complicates the propagation of observational information across the domain, whereas for ML, it challenges the construction of models that remain valid as these long-range dependencies shift over time. Accounting for such dynamic multi-scale interactions requires assimilation and learning strategies capable of representing nonlocality and capturing correlations that change with the instantaneous flow state.

(iii) Local extreme events might occur, leading to external intermittency, time-varying statistics, and the necessity for streaming data-processing methodologies. These factors also highlight the limitations of offline-trained ML models, which may fail to predict high-intensity perturbations or transient phenomena absent from the original training dataset. As the underlying dynamics evolve, previously learned dependencies may become less relevant, and new correlations can emerge on shorter time scales. Capturing these instantaneous and evolving relationships requires online adaptation mechanisms that can integrate new information as it becomes available. 

To analyse the challenges previously listed, window-based strategies provide a natural framework for maintaining model relevance in nonstationary flow configurations. In the present work, we aim to employ the augmented information estimated by a sequential DA technique, namely an Ensemble Kalman Filter combining state estimation with parameter optimisation \citep{Evensen2009_ieee}, to train multiple black-box models using the Random Forest Regression algorithm \citep{Breiman2001_ml}, within a sliding-window framework. The test case considered involves a flow with periodic behaviour, allowing each window to focus on locally coherent dynamics within the oscillation cycle. In this sense, the windowing strategy can provide a foundation for a temporal \emph{mixture of experts} (MoE) framework, with different regressors specialising in distinct phases of the motion, conceptually similar to spatial MoE approaches previously explored in complex or multi-region flows \citep{Yuksel2012_tnnls}. This configuration shares conceptual similarities with previous variational DA studies of rotating cylinders in laminar regimes \citep{Mons2017_jfm}, which likewise aimed to couple physical models with assimilation strategies to improve flow reconstruction and control. However, to the best of our knowledge, no prior work has applied DA or ML to configurations involving translating cylinders, where the body motion directly modifies the near-wake dynamics and introduces greater temporal variability. This methodology represents an intermediate step toward real-world applications, as it leaves behind the common assumption of statistical stationarity and enables the analysis of inherently unsteady, time-dependent flows.

Within this scope, the paper is structured as follows. \S\ref{sec:numIngredients} describes the numerical ingredients and methodologies employed. The test case used for the analyses is introduced in \S\ref{sec:test_case}. Strategies aimed at enhancing predictive capabilities through DA and ML are then presented in \S\ref{sec:DA_experiment} and \S\ref{sec:ML_experiment}, respectively. Finally, conclusions and perspectives are discussed in  \S\ref{sec:Conclusions}.

\section{Numerical ingredients}
\label{sec:numIngredients}

\subsection{Numerical solver and Immersed Boundary Method}
\label{sec:numericalSolver}

The Navier--Stokes equations for incompressible flows applied for Newtonian fluids and negligible temperature gradients can be written as (\ref{eqn:mass_eq})--(\ref{eqn:momentum_eq}), with (\ref{eqn:momentum_eq}) expressed in vector notation:

\begin{eqnarray}
    \boldsymbol{\nabla} \cdot \boldsymbol{u} &=& 0 
    \label{eqn:mass_eq} \\
    \frac{\partial \boldsymbol{u}}{\partial t} + (\boldsymbol{u} \cdot \boldsymbol{\nabla}) \boldsymbol{u} &=& \nabla p + \nu \nabla^2 \boldsymbol{u} + \boldsymbol{f}_P
    \label{eqn:momentum_eq}
\end{eqnarray}
where $\boldsymbol{u}$ is the velocity, $t$ is the time, $\nu$ is the kinematic viscosity, and $p=P/\rho$ is the normalised pressure field $P$ with respect to density $\rho$. In low-Reynolds-number regimes, the flow can exhibit two-dimensional features, as observed in the present investigation. The flow is analysed in a plane where $z=constant$. Therefore, the position vector is $\boldsymbol{x} = (x,y)$ in Cartesian coordinates where $x$ and $y$ are the streamwise and crosswise positions, respectively. The velocity field is represented as $\boldsymbol{u} = (u_x, u_y)$. One can include an additional source term to account for body forces such as gravity, centrifugal, Coriolis, or electromagnetic contributions. In the present analysis, $\boldsymbol{f}_P$ represents a penalisation term for a classical Immersed Boundary Method (IBM). $\boldsymbol{f}_P$ accounts for the presence of immersed rigid solid obstacles, whose bodies are located in a region $\Omega_b$. \citet{Angot1999} showed by means of asymptotic analysis that the solution of the velocity field with this source term converges to the standard incompressible Navier--Stokes equations with a Dirichlet boundary condition at the immersed boundary for the velocity. The penalisation term is then defined as:

\begin{equation}
     \boldsymbol{f}_P = \left\{
        \begin{array}{ll}
        \boldsymbol{0} & \textrm{if } \boldsymbol{x} \in \Omega_f \\
        -\cfrac{1}{\alpha_p \Delta t} \left(\boldsymbol{u} - \boldsymbol{u_{ib}} \right) & \textrm{if } \boldsymbol{x} \in \Omega_b
        \end{array} \right.
    \label{eqn:forcePenDarcy}
\end{equation}
Here, $\boldsymbol{u_{ib}}$ is the target velocity for the solid obstacle and $\Omega_f$ is the region of the domain accessible to the fluid. $\alpha_p$ is a penalisation time-step parameter, which depends on the time step $\Delta t$ employed during discretisation, and is commonly defined for $\alpha_p \in [0, 1]$, where lower values increase accuracy but are more prone to numerical instabilities \citep{Kadoch2012_jcp}. Here, all case studies are numerically investigated using customised solvers developed within the open-source framework \citet{OpenFOAM}, which employs a finite-volume discretisation of equations (\ref{eqn:mass_eq})--(\ref{eqn:momentum_eq}). Time advancement is performed using the PISO scheme \citep{Issa1986_jcp} and a second-order backward scheme. Second-order centred schemes handle spatial discretisation. The customised solver has been specifically designed for the study of cylinder-based geometries with IBMs and has been previously validated and employed in related works \citep{Er2025}.

\subsection{Data Assimilation: Ensemble Kalman Filter for state estimation and parameter optimisation}
\label{sec:DataAssimilation}

The Ensemble Kalman Filter (EnKF) \citep{Evensen1994_jgr} is a Monte Carlo extension of the classical Kalman Filter (KF) \citep{Kalman1960_jbs}. It propagates over time both the physical system state $\boldsymbol{u}$ (commonly the velocity field discretised over the computational domain in CFD applications) and its associated uncertainty, represented by the error covariance matrix $\boldsymbol{P} = \mathbb{E}[(\boldsymbol{u} - \boldsymbol{u}^t)(\boldsymbol{u} - \boldsymbol{u}^t)^T]$, where $\boldsymbol{u}^t$ denotes the (unknown) true state. To avoid explicitly defining the high-dimensional matrix $\boldsymbol{P} \in \mathbb{R}^{n_{DF}\times n_{DF}}$, with $n_{DF}$ denoting the number of degrees of freedom (e.g., twice the number of mesh elements in two-dimensional CFD configurations), the EnKF represents the state as an ensemble of realisations, referred to as ensemble members, each with a distinct state vector $\boldsymbol{u}_i$ ($i \in \left[1, N_e \right]$, where $N_e << n_{DF}$ is the number of ensemble members). The ensemble mean $\boldsymbol{\overline{u}}$ approximates the true state $\boldsymbol{u}^t$, while the deviations of each member from the mean are collected in the \emph{anomaly matrix} $\boldsymbol{X}$, defined as:

\begin{eqnarray}
    \boldsymbol{\overline{u}} &=& \frac{\sum_i^{N_e} \boldsymbol{u}_{i}}{N_e} \\
    \left[\boldsymbol{X} \right]_i &=& \frac{\boldsymbol{u}_{i}-\boldsymbol{\overline{u}}}{\sqrt{N_e-1}} \label{eqn:anomalies_state}
\end{eqnarray}
If the ensemble is significantly large to represent the system statistics and its members are statistically independent, the error covariance matrix $\boldsymbol{P}$ can then be efficiently approximated as:

\begin{equation}
\boldsymbol{P}_k \approx \boldsymbol{X} \left(\boldsymbol{X} \right)^T,
\end{equation}
where $\boldsymbol{X} \in \mathbb{R}^{n_{DF} \times N_e}$.
%To enhance performance, the \emph{extended-state} EnKF approach \citep{Evensen2009_ieee} can be employed. In this method, certain model parameters $\theta$ are also propagated alongside the system state to reduce model bias. Hence, the system state $\boldsymbol{u}^\prime = [\boldsymbol{u} \,\theta]^T$.
The EnKF follows a two-step cycle. The first step, referred to as forecast, and represented by ($\boldsymbol{\cdot}^f$), advances each ensemble member from time $k-1$ to $k$ through a numerical model $\mathcal{M}(\theta)$, which denotes the discrete representation of the underlying physical model governed by the parameters $\theta$:

\begin{equation}
    \boldsymbol{u}^f_{i,k} = \mathcal{M}(\theta_{i,k:k-1}) \, \boldsymbol{u}_{i,k-1}
    \label{eqn:forecast}
\end{equation}
If observational data is available at time $k$, an analysis step follows, represented by ($\boldsymbol{\cdot}^a$), where the ensemble is updated to optimally combine the model forecast and the measurements, minimising the posterior error covariance matrix $\boldsymbol{P}_k^a$. An \emph{extended-state} formulation \citep{Evensen2009_ieee} can also be employed, in which selected model parameters $\theta$ are estimated jointly with the state, leading to a reduction in the model bias and yielding an augmented state vector $\boldsymbol{u}' = [\boldsymbol{u}, \theta]^T$.

To maintain dimensional consistency in the EnKF formulation, an ensemble of $N_e$ observation vectors is required. These are assembled into the observation matrix $\boldsymbol{Y}_k \in \mathbb{R}^{n_o \times N_e}$, obtained by perturbing the actual observation vector $\boldsymbol{y}_k \in \mathbb{R}^{n_o \times 1}$. Each perturbation is modelled as Gaussian noise $\boldsymbol{e}_i$ with zero mean and covariance $\boldsymbol{R}_k \in \mathbb{R}^{n_o \times n_o}$, which represents the observation uncertainty. The matrix $\boldsymbol{R}_k$ is typically assumed diagonal, implying that the measurement errors are Gaussian, unbiased, and mutually independent. In the limit of an infinitely large ensemble, $\lim_{N_e \to +\infty} \mathbb{E}[(\boldsymbol{e} - \mathbb{E}[\boldsymbol{e}]) (\boldsymbol{e} - \mathbb{E}[\boldsymbol{e}])^T] = \boldsymbol{R}$, which becomes time-independent. Thus, each ensemble realisation of the observation matrix can be expressed as $\left[\boldsymbol{Y}_k \right]_i=\boldsymbol{y}_k+\boldsymbol{e}_i$, where $\boldsymbol{e}_i \sim \mathcal{N}(0, \boldsymbol{R})$. 

\begin{figure}
    \centering
    \includegraphics[width=0.7\linewidth]{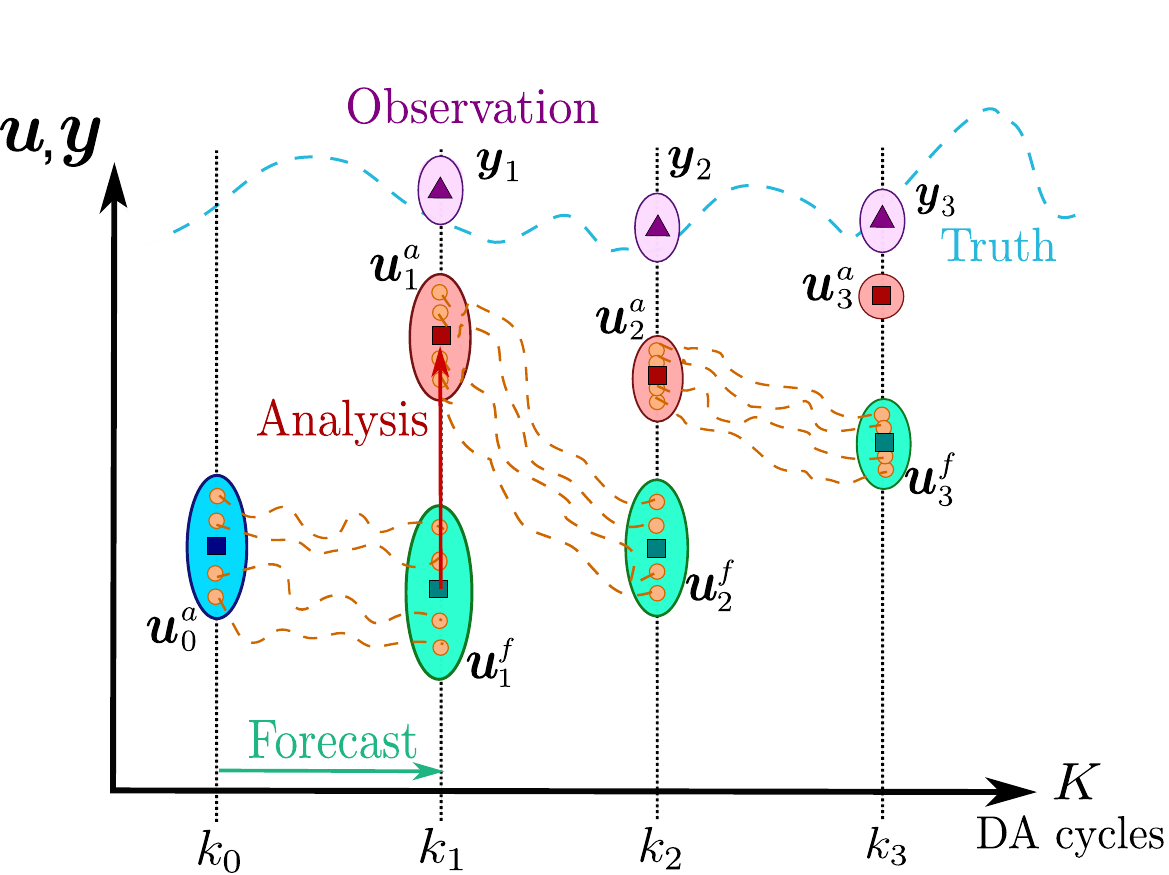}
    \caption{Ensemble Kalman Filter algorithm, highlighting the forecast and analysis steps}
    \label{fig:EnKF_algorithm}
\end{figure}

The EnKF also requires a sampling operator $\mathcal{H}(\boldsymbol{u}_k^f)$, where $\mathcal{H}$ serves as an interpolation operator that maps the forecasted system state from the model space onto the observation space. Analogous to the anomaly matrix with the forecasted system state $\boldsymbol{X}^f$ following (\ref{eqn:anomalies_state}), this projected ensemble is represented by the matrix $\boldsymbol{S}_k^f$, whose columns correspond to normalised anomalies of the sampled states. These are computed from the ensemble mean $\overline{\mathcal{H}(\boldsymbol{u}_k^f)}$ as:

\begin{eqnarray} 
    \overline{\mathcal{H}(\boldsymbol{u}_{k}^f)} &=& \frac{\sum_i^{N_e} \left[\mathcal{H}(\boldsymbol{u}_{k}^f) \right]_i}{N_e} \\
    \left[\boldsymbol{S}_k^f \right]_i &=& \frac{\left[\mathcal{H}(\boldsymbol{u}_{k}^f) \right]_i - \overline{\mathcal{H}(\boldsymbol{u}^f_k)}}{\sqrt{N_e-1}}
    \label{eqn:anomalies_stateSampling}
\end{eqnarray}
The Kalman gain matrix $\boldsymbol{K}_k$ providing the most optimal weight between the state and the observations can be expressed as in (\ref{eqn:KalmanGain}) \citep{Carrassi2018_WIREs}:

\begin{equation}
    \boldsymbol{K}_k = \boldsymbol{X}_k^f \left(\boldsymbol{S}_k^f \right)^T \left[\boldsymbol{S}_k^f \left(\boldsymbol{S}_k^f \right)^T + \boldsymbol{R} \right]^{-1}
    \label{eqn:KalmanGain}
\end{equation}
All the elements described above are required to calculate the updated system's state $\boldsymbol{u}_k^a$ and model parameters $\theta_k^a$ as in (\ref{eqn:updatedState}), which are subsequently used to advance the numerical model $\mathcal{M}(\theta_{i, k+1:k})$ during the next iteration at time step $k+1$. A detailed step-by-step implementation of the EnKF is provided in Alg.~\ref{alg:EnKF}, while Fig.~\ref{fig:EnKF_algorithm} shows a schematic of the two-step procedure.

\begin{equation}
    \boldsymbol{u}_k^a = \boldsymbol{u}_k^f + \boldsymbol{K}_k \left( \boldsymbol{Y}_k - \mathcal{H}(\boldsymbol{u}_k^f)\right)
    \label{eqn:updatedState}
\end{equation}

\begin{algorithm}
    \caption{Scheme of the EnKF used in the present study.}
    \label{alg:EnKF}
    \textbf{Input:} $\mathcal{M}(\theta_0)$, $\boldsymbol{R}$, $\boldsymbol{y}_k$, and a prior/initial state system $\boldsymbol{u}_{i,0}$, where usually $\boldsymbol{u}_{i,0} \sim \mathcal{N}(\boldsymbol{\overline{u}}_0, \sigma_0^2)$. Initial model parameters are also perturbed: $\theta_{i,1:0}\sim\mathcal{N}(\overline{\theta_{1:0}},\eta_0^2)$. \\
    \For{$k = 1, 2,..., K$}{
        \For{$i = 1, 2, ..., N_e$}{
    \nl Advancement in time of the state vectors:\\
    \qquad $\boldsymbol{u}_{i,k}^f = \mathcal{M}(\theta_{i,k:k-1})\,\boldsymbol{u}_{i,k-1}$ \\
    \nl Generation of an observation matrix from the confidence level given to the observation data:\\
    \qquad$ [\boldsymbol{Y}_k]_i = \boldsymbol{y}_k + \boldsymbol{e}_i$, with $\boldsymbol{e}_i \sim \mathcal{N}(0,\boldsymbol{R})$\\
    \nl Estimation of the ensemble means (system state and projection matrix):\\
    \qquad$\boldsymbol{\overline{u}}_k^f = \frac{1}{N_e}\sum_{i = 1}^{N_e}\boldsymbol{u}_{i,k}^f$,\,
    $\overline{\mathcal{H}(\boldsymbol{u}^f_k}) = \frac{1}{N_e}\sum_{i = 1}^{N_e} \left[\mathcal{H}(\boldsymbol{u}_{k}^f) \right]_i$ \\
    \nl Computation of the anomaly matrices (system state and projection matrix):\\
    \qquad$\boldsymbol{X}_k = \frac{\boldsymbol{u}_{i,k}-\boldsymbol{\overline{u}}_k}{\sqrt{N_e-1}}$,\,
    $\boldsymbol{S}_k = \frac{\left[\mathcal{H}(\boldsymbol{u}_{k}^f)\right]_i - \overline{\mathcal{H}(\boldsymbol{u}^f_k)}}{\sqrt{N_e-1}}$ \\
    \nl Calculation of the Kalman gain:\\
    \qquad$\boldsymbol{K}_k = \boldsymbol{X}_k^f(\boldsymbol{S}_k)^T \left[\boldsymbol{S}_k(\boldsymbol{S}_k)^T + \boldsymbol{R}\right]^{-1}$\\
    \nl Update of the state vectors and model parameters $\theta_{i,k+1:k}$:\\
    \qquad$\boldsymbol{u}_{i, k}^a = \boldsymbol{u}_{i,k}^f + \boldsymbol{K}_k \left(\left[\boldsymbol{Y}_{k} \right]_i- \left[\mathcal{H}(\boldsymbol{u}_{k}^f) \right]_i \right)$
    }
    }
\end{algorithm}

In previous studies \citep{Valero2025_jcp}, it was observed that state estimation and the local parameter optimisation of $\alpha_p=\alpha_p(\boldsymbol{x})$ via EnKF could enhance the near-wall resolution of wall-bounded turbulent flows with an under-refined mesh, achieving accuracy comparable to that of high-fidelity simulations. It was also observed that these parameters can be learned by a machine learning algorithm, yielding improved computational efficiency compared to the data assimilation approach \citep{Valero2025_cf}, which can be particularly costly in ensemble-based or adjoint formulations. However, the analyses were conducted under specific conditions, namely:

\begin{itemize}
    \justifying
    \item The flow statistics were steady with time, such that the same physical behaviour was represented throughout the assimilation period (observation window). Furthermore, spatial inhomogeneity was solely restricted to the wall-normal direction, resulting in all grid layers equidistant from the immersed walls experiencing identical flow physics.
    \item No-slip boundary conditions were imposed at the walls, ensuring that the wall velocity $\boldsymbol{u_{ib}}$ remained zero at all times. This eliminated the influence of time-dependent wall motion, which would otherwise arise in cases involving moving bodies.
\end{itemize}

In the current work, we aim to extend this framework to an oscillatory regime, where the flow exhibits a rich variety of dynamical features that vary both with the phase of the oscillatory cycle and locally within the computational domain, and where the solid body undergoes time-dependent motion. This introduces an additional layer of complexity beyond what has been previously explored.

\subsection{Sliding Windows and Their Role with Data Assimilation and Machine Learning}

%\textcolor{red}{My intention is to develop a section similar to 2.2 in C\&F in which we explain the importance of the observation windows, and how the presence/absence of data and the different physics across the oscillating period (there is a rich variety of dynamical features within a single cycle and the information content of the observations depends strongly on the phase in which they are collected) makes it appropriate to divide the full observation window into a sequence of smaller, consecutive windows. This strategy enables the combined Data Assimilation-Machine Learning framework to operate more efficiently by focusing on localised temporal dynamics while still capturing the global structure of the oscillatory cycle.}\\

Both Data Assimilation and hybrid Machine Learning are data-driven techniques that blend a physical model and observations of the same physical phenomenon to improve predictive accuracy. However, they employ different strategies to achieve it. For the former, sparse observations in time and space are combined with the governing equations to accurately predict the system state with the highest level of precision. In contrast, for the latter, some black-box models trained with a large, densely sampled set of data typically replace or correct some of the terms of the governing equations that do not offer globally accurate predictive capabilities (e.g., Reynolds stress tensor or subgrid-scale model) \citep{Kutz2017_jfm, Beck2019_jcp}, or that are computationally expensive to solve (e.g., Poisson equation) \citep{Weymouth2022_cf}. Despite these differences, both approaches must contend with the fact that the underlying flow physics can vary significantly over time, meaning that the information content of the observations and the appropriate modelling strategy may change throughout the simulation period, motivating the use of smaller windows that better capture the local flow dynamics.

In view of these circumstances, within the data assimilation field, interest in dividing the full \textit{observation window} into a sequence of smaller, consecutive (and, optionally, overlapping) sub-windows arises for several reasons. First, in the case of variational approaches, each sub-window---referred to in this context as \textit{assimilation windows}---can focus on a narrower temporal frame, making it easier to assume ``quasi-stationarity'' of the flow within that window (or at least reduced variation) and simplify the modelling/assimilation task. \citet{Chandramouli2020_jcp, Wang2025_jfm} argued that, in turbulent flows, the assimilation window used in 4DVar formulations must be shorter than the Lyapunov timescale to ensure that the physical model remains close to the truth over each window and that the adjoint gradients remain well-conditioned. \citet{Albarakati2024_jcp} claimed that, when projecting the model and the data onto a reduced-dimensional subspace through Properly Orthogonal Decomposition, employing a sliding subset of snapshots (i.e., \textit{sliding windows}) enables the assimilation technique to more accurately capture the non-linear evolution, non-Gaussian uncertainty and the high dimensionality of the state by extracting the dominant spatiotemporal modes governing the observed data. 

Additionally, although in sequential statistical methods updates are performed at each observation time independently, analogous ideas arise in how these methods manage the temporal frequency of assimilation---also referred to as the \textit{assimilation window} even if it spans only a single observation time. \citet{Meldi2018_ftc} studied the effect of the frequency of time sampling of observation on the accuracy of the prediction by the assimilation operator, which depends on the characteristic average advection time. Also, the concept of sliding windows can be highly relevant to non-sequential statistical approaches such as the Ensemble Kalman Smoother \citep{Evensen2000_mwr, Bocquet2014_qjrms}, which provides a natural way to choose the smoothing lag (past dependence) and to incorporate limited future information without performing a full fixed-interval reanalysis.

\begin{figure}
    \centering
    \includegraphics[scale=0.45]{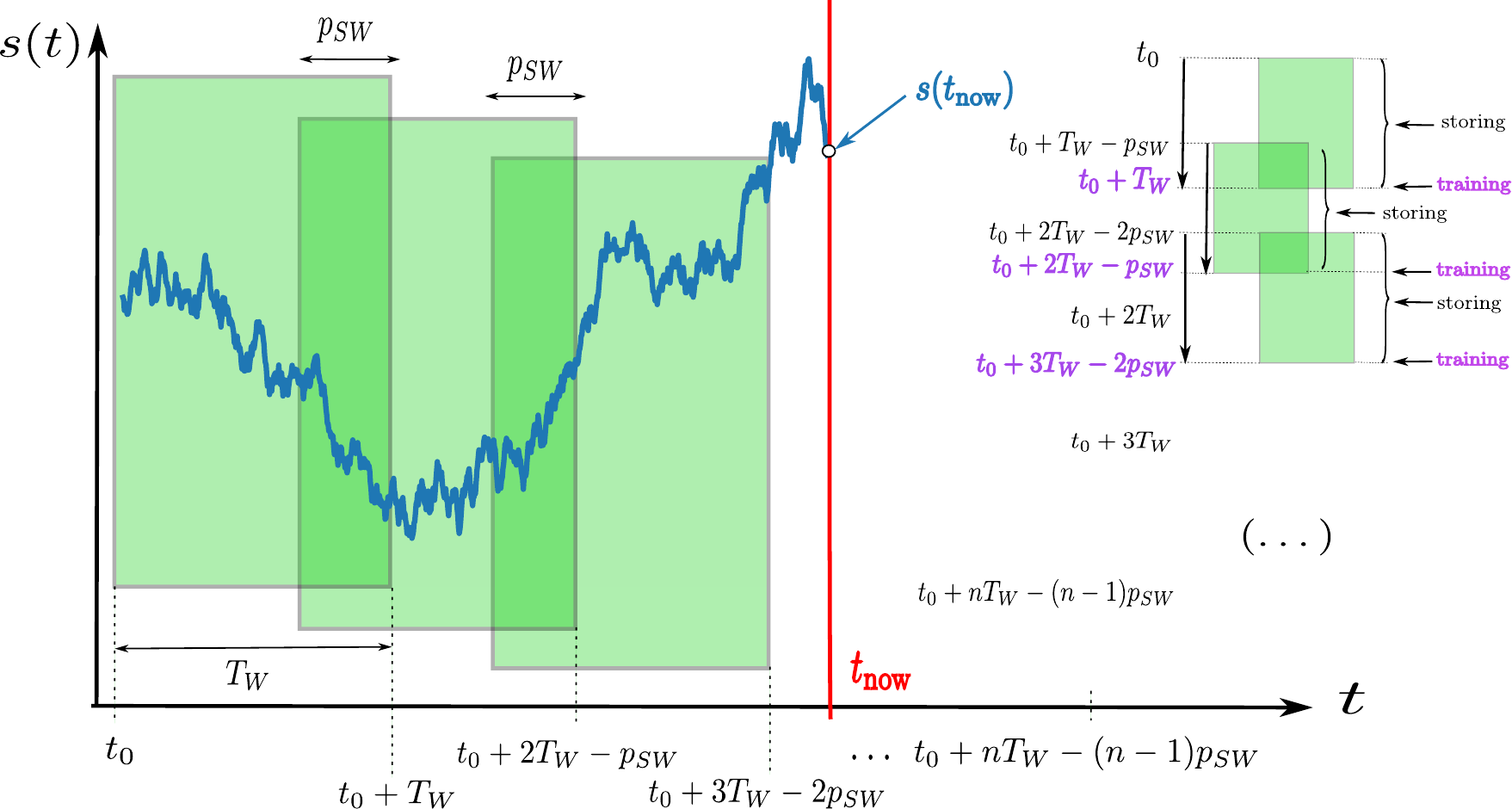}
    \caption{Sliding-window approach for a generic signal $s(t)$ from an initial time $t_0$ up to the current time $t_\textrm{now}$. Each green rectangle denotes a window of size $T_W$, while $p_{SW}$ indicates the uniform sliding period between consecutive windows}
    \label{fig:SWA_generic}
\end{figure}

Yet it is in the field of machine learning that the concept of sliding windows has recently gained attention, particularly for time-series forecasting, where we need to process streaming data and possibly confront concept drift \citep{Webb2016_dmkd}. In fact, thanks to the window moving forward in time, one can track time-evolving dynamics, enhance the resolution of time-instantaneous information, and, unlike a single global model, compute the learning process across several black-box models. Each model can then specialise in representing distinct physical behaviours or corrections tailored to its local temporal region, thereby reducing the overall modelling bias. Furthermore, different from recurrent architectures, by training each model on the locally coherent dynamics of a particular time snapshot, the approach produces simpler, more accurate surrogates that specialise in prominent flow features and avoid the tendency of recurrent models to accumulate prediction error over long sequences \citep{Karpathy2015}. Nonetheless, this approach is not without challenges: it might introduce discontinuities at window boundaries, require careful balancing of data across windows, and depend critically on the appropriate selection of window size $T_W$, which is strongly influenced by data availability and the relevant temporal scales of the governing physics. Even so, studies have shown that introducing overlapping sliding periods $p_{SW}$ between consecutive windows can help mitigate prediction noise \citep{Gama2014_acm}, making this method suitable for the current investigation. The sliding-window framework is illustrated in Fig.~\ref{fig:SWA_generic}.

%\textcolor{red}{(CONTINUE WITH THE CONCEPT OF SLIDING WINDOWS FOR MACHINE LEARNING AND COMPARISON WITH LSTM. MAYBE A FIGURE AT THE END OF THIS SECTION WOULD BE NICE).}

\section{Test case: oscillating cylinder in a quiescent fluid}
\label{sec:test_case}

\begin{figure}
    \centering
    \begin{tabular}{cc}
    \includegraphics[width=0.4\linewidth]{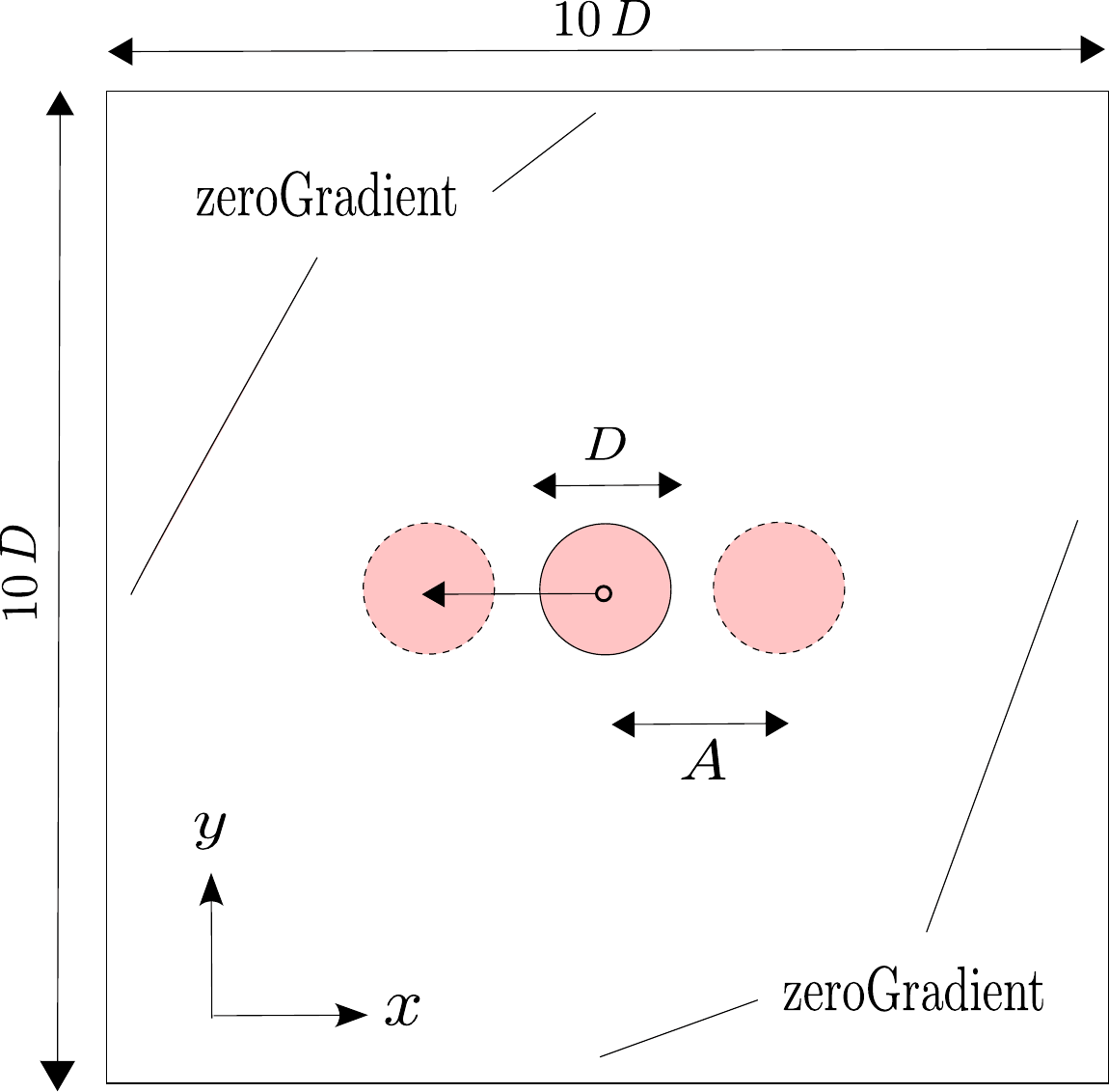} &
    \includegraphics[width=0.4\linewidth]{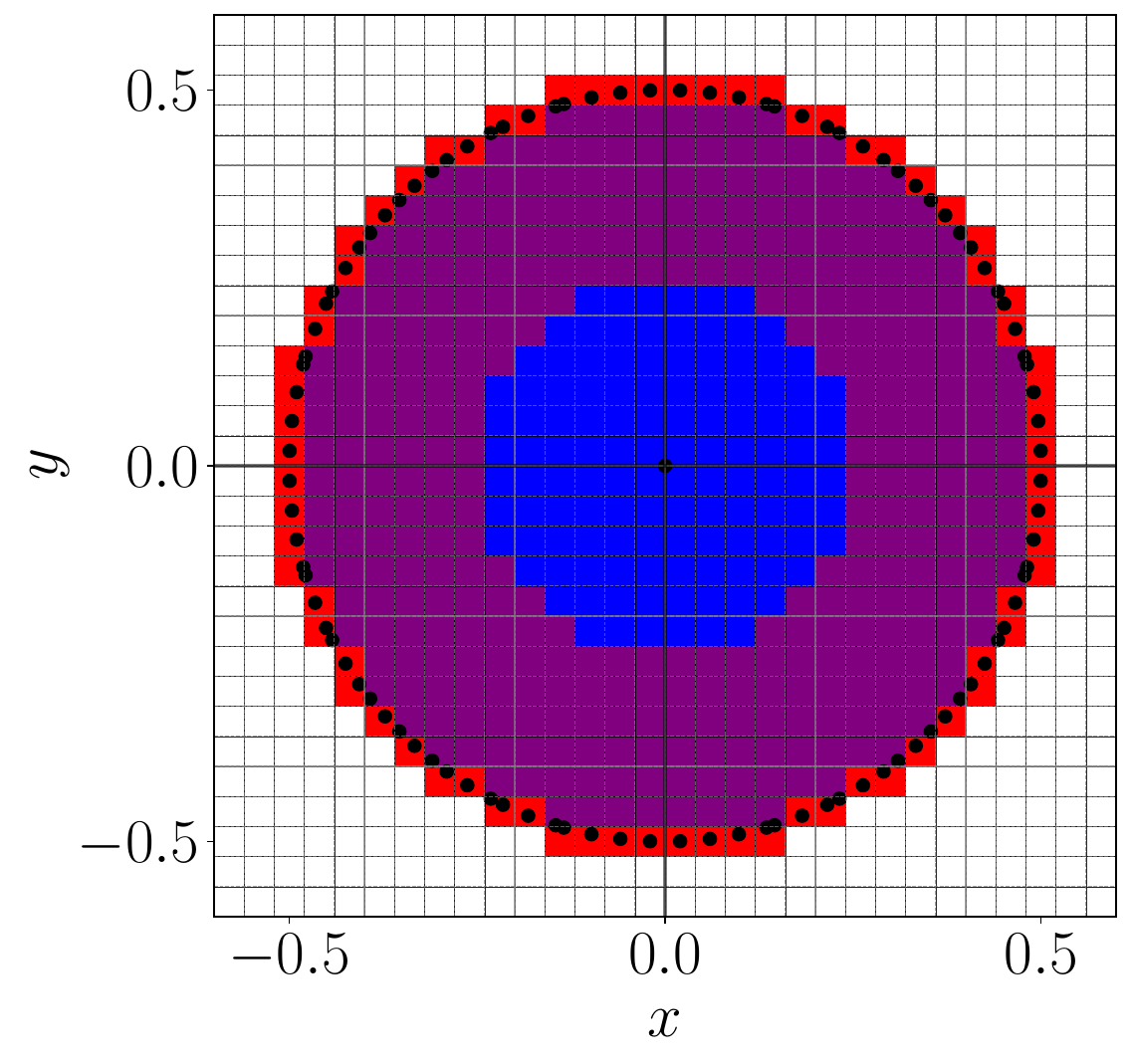} \\
    \textit{(a)} & \textit{(b)}
    \end{tabular}
    \caption{Illustration of the test case, where \textit{(a)} denotes the computational domain employed for the penalty IBM simulation, and \textit{(b)} represents the initial configuration of the cylinder, highlighting the discretised immersed boundary and penalised cells (the coloured ones) where $\boldsymbol{f}_P$ is applied}
    \label{fig:cylinder_scheme}
\end{figure}

The test case investigated consists of a rigid body moving horizontally in a fluid at rest, exhibiting oscillatory behaviour. Specifically, it considers harmonic motion in the $x-$direction of a two-dimensional cylinder with diameter $D$ that is initially placed at ($x_0, y_0$), corresponding to the centre of a square domain of size $10D \times 10D$. A representation of the domain can be found in Fig. \ref{fig:cylinder_scheme}\textit{(a)}. The main interest of this configuration is that the flow around an oscillating cylinder can show very different vortex structures and mechanisms depending on the governing non-dimensional numbers \citep{Tatsuno1990_jfm}. The two primary dimensionless numbers are the Reynolds number $Re_D = U_\textrm{max} D/ \nu$ and the Keulegan-Carpenter number $KC = U_\textrm{max} / f D$. The former relates the ratio between the convective and viscous forces, while the latter describes the relative importance of the drag forces over the inertial ones for a bluff body in an oscillatory flow. Here, $f$ is the frequency of the oscillation and $U_\textrm{max}$ corresponds to its maximum velocity. The cylinder's centre $\boldsymbol{x_{ib}}$ is estimated with the following expression:

\begin{equation}
    \boldsymbol{x_{ib}} = \begin{pmatrix}
        x_{ib} \\
        y_{ib}
    \end{pmatrix} = \begin{pmatrix}
        x_0 - A \, \textrm{sin}(2\pi f t) \\
        y_0 
    \end{pmatrix}
    \label{eqn:motionCylinder}
\end{equation}

Hence, the velocity of the cylinder $\boldsymbol{u_{ib}}$ is just the derivative with respect to time of (\ref{eqn:motionCylinder}), leading to:

\begin{equation}
    \boldsymbol{u_{ib}} = \begin{pmatrix}
        u_{x_{ib}} \\
        u_{y_{ib}}
    \end{pmatrix} =
    \begin{pmatrix}
        -2\pi f A\, \textrm{cos} (2\pi f t) \\
        0
    \end{pmatrix}
    \label{eqn:velocityCylinder}
\end{equation}
where $A$ is the amplitude of the oscillation. From (\ref{eqn:velocityCylinder}), we obtain $U_\textrm{max} = 2\pi f A$. In the following, we investigate a configuration with $Re = 100$ and $KC = 5$, which corresponds to a laminar flow regime characterised by symmetric flow conditions with respect to the centre plane in the $y$-direction, which is hereafter referred to as $y = 0$, and periodic vortex structures with each oscillation. To numerically reproduce this configuration, the computational domain boundaries are treated with \emph{zeroGradient} conditions, i.e., $\partial()/\partial \boldsymbol{n} = 0$ for both velocity $\boldsymbol{u}$ and pressure $p$, where $\boldsymbol{n}$ denotes the outward normal. Alternative boundary conditions, such as periodic ones, were tested and yielded no significant deviations in the results. Specifically for this configuration, \citet{Dutsch1998_jfm} carried out some experimental measurements with a stationary tank with fluid at rest, in which the cylinder was sinusoidally actuated by a crankshaft gear drive and a Laser Doppler Anemometer (LDA) probe was mounted on the top of the tank, allowing for velocity measurements. Also, the moving body makes it an excellent benchmark for validating new Immersed Boundary Methods (IBMs). \citet{Tsetoglou2024_nmf} validated the so-called \emph{Volume-of-Solid Implicit Volume Penalty method} (VOS-IVP) against this test case, ensuring that the system of equations is fully mass conservative while improving computational efficiency. \citet{Cai2016} examined this problem using the \emph{Moving Immersed Boundary Method} (M-IBM), which shows high computational efficiency with solids undergoing large displacements.

\subsection{Penalisation IBM for the oscillating cylinder}
\label{sec:penIBM_cylinder}

In this section, we perform a parametric study of the penalisation IBM developed by \citet{Angot1999}, in which the cylinder is accounted for by including a penalty forcing term $\boldsymbol{f}_P$ defined as in (\ref{eqn:forcePenDarcy}). 
%for the incompressible Navier--Stokes equations (\ref{eqn:mass_incompressible})--(\ref{eqn:momentumY_incompressible}). The solver follows a PISO loop where $\boldsymbol{f}_P$ is solved implicitly in the momentum equations.
Fig.~\ref{fig:cylinder_scheme}\textit{(b)} illustrates the discretisation of the cylinder at the initial time $t=0$. The coloured cells are penalised by the inclusion of the term $\boldsymbol{f}_P$ in the dynamic equations. The red mesh elements denote an additional layer in the fluid region $\Omega_f$, which might improve stability, although it prevents mass conservation in this layer. We discuss in \S\ref{sec:DA_experiment} that this additional cell layer will help us reduce the discontinuities between the solid and fluid regions during the Data Assimilation experiment.

In this parametric study, we analyse the influence of the penalisation time-step parameter $\alpha_p$ introduced in \S\ref{sec:numericalSolver}, and the uniform grid resolution $\Delta \boldsymbol{x} = (\Delta x, \Delta y)$, with $\Delta x = \Delta y$. Our goal is twofold: (i) to obtain a configuration that closely matches literature data, allowing the generation of sensors to use as observations in a Data Assimilation framework, and (ii) to test the model's sensitivity to these parameters, which will provide a basis for the assimilation process. We analyse two quantities to determine them. First, we study the velocity profiles at the oscillation phase $\Delta\phi = 2\pi f \tau = \pi$, i.e., when $u_{ib_x} = U_\mathrm{max}$, where $\tau$ is the time shift from the beginning of the oscillation. Second, we evaluate the resistive force $F_x$ over one period $T=1/f$. To estimate the resistive force, as explained in \citet{Tsetoglou2024_nmf}, several aspects need to be considered. It is estimated from the \emph{interaction solid/fluid} force $\boldsymbol{f}_{\textrm{interaction}}$, defined as a forcing term responsible for stopping the external fluid from penetrating the solid region. If the cylinder did not move, it would coincide with the penalty force $\boldsymbol{f}_P$ introduced into the Navier--Stokes equations. However, one needs to consider that in the case of a moving body, $\boldsymbol{f}_P$ also acts to move the internal fluid enclosed in the solid region $\Omega_b$ with a velocity $\boldsymbol{u_{ib}}$. This additional contribution, referred to as $\boldsymbol{f}_\mathrm{internal} \approx \partial \boldsymbol{u} / \partial t$, represents the momentum of the moving internal fluid. This means that $\boldsymbol{f}_P$ can be separated into two contributions:

\begin{equation}
    \begin{aligned}
        \boldsymbol{f}_P &= \boldsymbol{f}_{\textrm{internal}} + \boldsymbol{f}_{\textrm{interaction}} \\
        \boldsymbol{f}_{\textrm{interaction}} &= -\frac{1}{\alpha_p \Delta t} \left(\boldsymbol{u}- \boldsymbol{u_{ib}} \right) - \frac{\partial \boldsymbol{u}}{\partial t}
    \end{aligned}
\end{equation}
Integrating $\boldsymbol{f}_\textrm{interaction}$ over the cylinder's volume or solid part $\Omega_b$, one obtains the total force $\boldsymbol{F} = (F_x, F_y)$, from which $F_x$ corresponds to the resistive force. This force is typically expressed in $N$ (Newtons) in the literature:

\begin{equation}
    \boldsymbol{F} = \rho \int_{\Omega_b} \boldsymbol{f}_\textrm{interaction} \,d\Omega_b = -\frac{\rho}{\alpha_p \Delta t} \int_{\Omega_b} \left(\boldsymbol{u}- \boldsymbol{u_{ib}} \right) \,d\Omega_b - \rho \int_{\Omega_b} \frac{\partial \boldsymbol{u}}{\partial t} \,d\Omega_b
    \label{eqn:force_integral}
\end{equation}
For a two-dimensional test case, when discretising the mesh, considering $\Delta \Omega_b = \Delta z\,\Delta\Sigma_b$, where $\Delta z$ is the spanwise length (assumed unitary), and $\Delta\Sigma_b$ is the surface of a mesh element inside the cylinder, (\ref{eqn:force_integral}) becomes:

\begin{equation}
    \boldsymbol{F} = \rho \Delta z \sum_{\Sigma_b} \boldsymbol{f}_\textrm{interaction} \,\Delta \Sigma_b = -\frac{\rho \Delta z}{\alpha_p \Delta t} \sum_{\Sigma_b} \left(\boldsymbol{u}- \boldsymbol{u_{ib}} \right) \,\Delta \Sigma_b - \rho \Delta z\sum_{\Sigma_b} \frac{\partial \boldsymbol{u}}{\partial t} \,d\Sigma_b
    \label{eqn:force_discretised}
\end{equation}

\begin{figure}[!h]
    \centering
    \begin{tabular}{ccc}
    \includegraphics[width=0.32\linewidth]{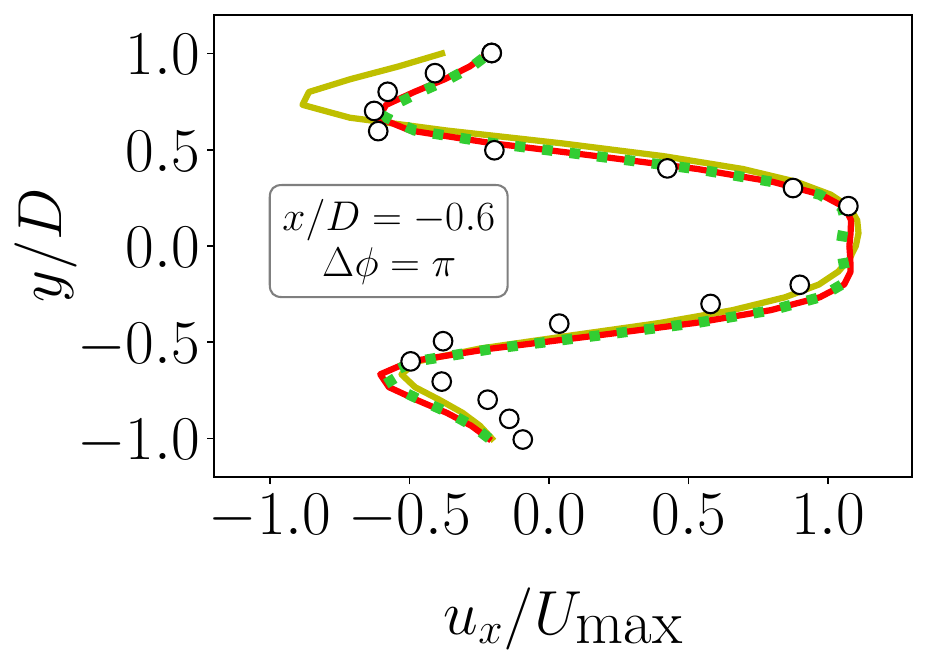} &
    \includegraphics[width=0.32\linewidth]{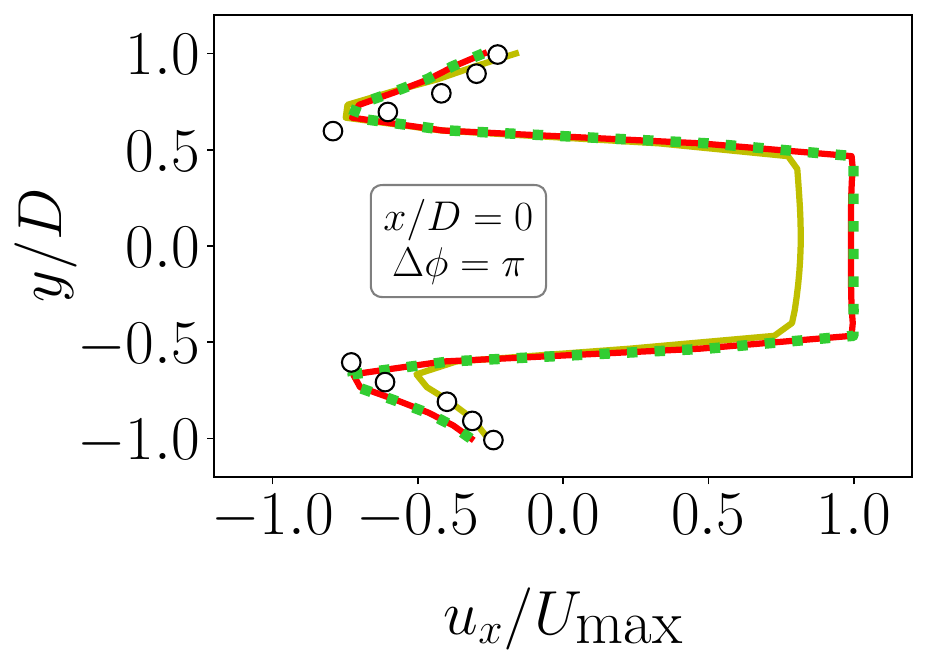} &
    \includegraphics[width=0.32\linewidth]{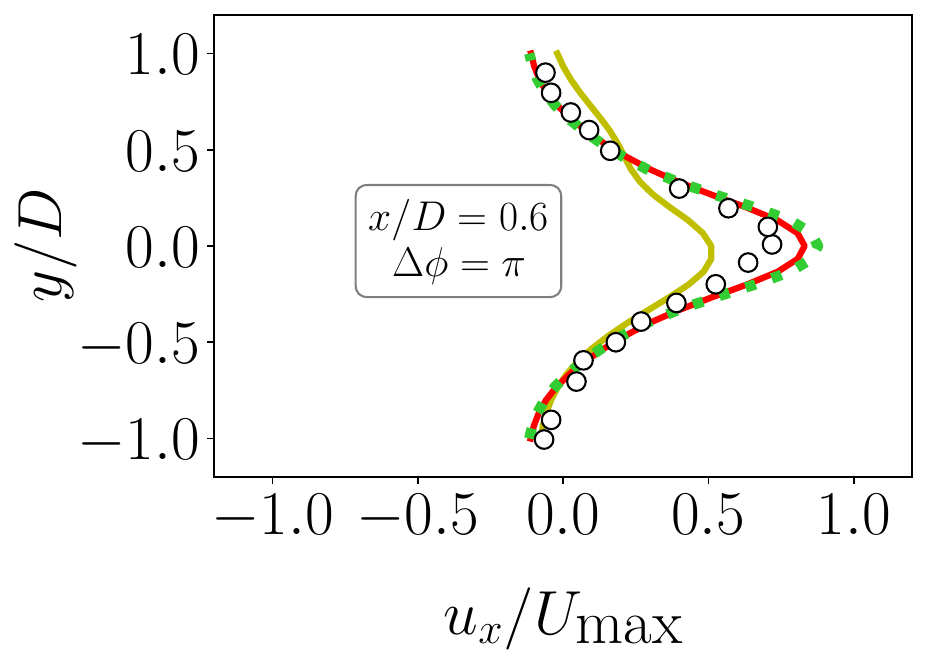} \\
    \textit{(a)} & \textit{(b)} & \textit{(c)} \\
    \includegraphics[width=0.32\linewidth]{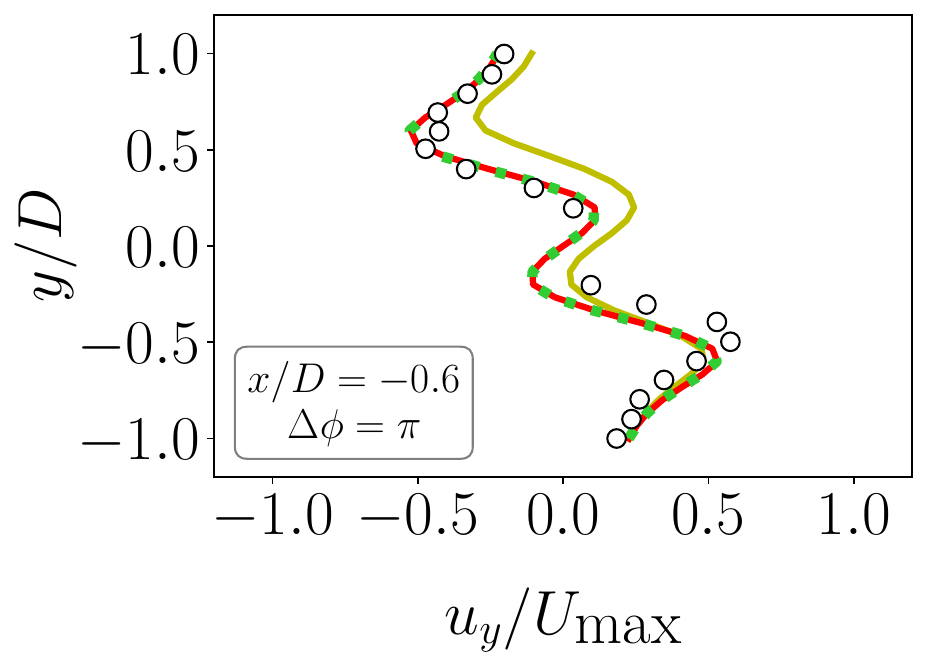} &
    \includegraphics[width=0.32\linewidth]{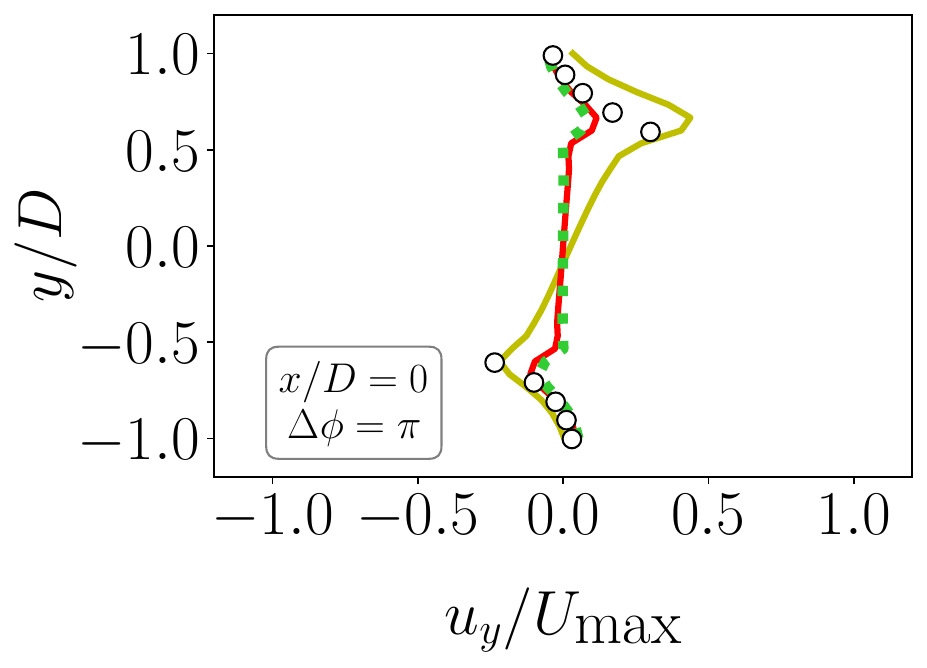} &
    \includegraphics[width=0.32\linewidth]{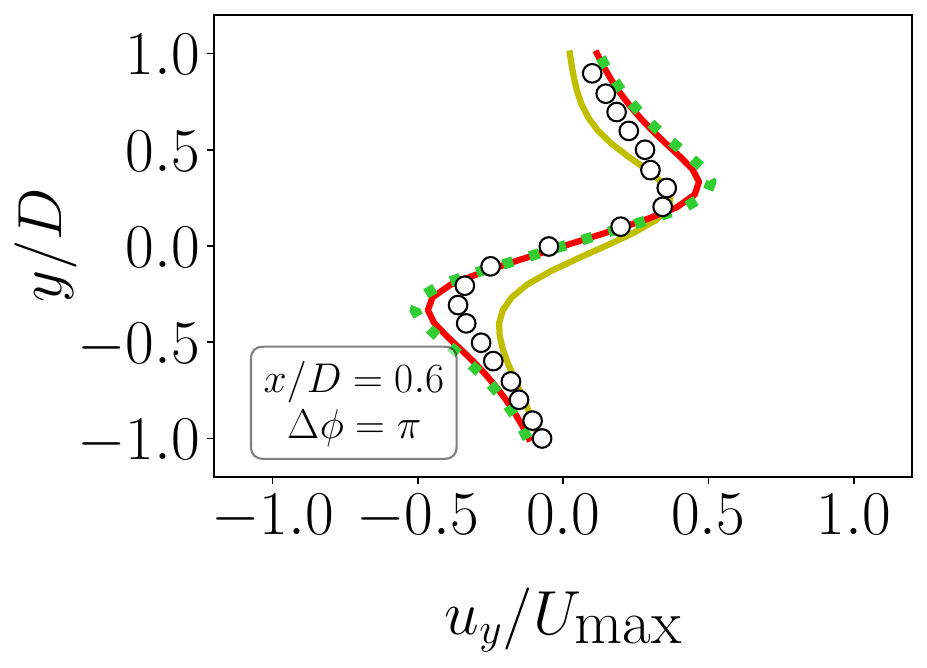} \\
    \textit{(d)} & \textit{(e)} & \textit{(f)}     
    \end{tabular}
    \caption{Velocity profiles when $\Delta\phi = \pi$ at three different planes $x/D \in [ -0.6, 0, 0.6]$ for (\protect\yellowline) $\alpha_p = 40$, (\protect\redline) $\alpha_p = 4$, and (\protect\greenlimelinedotted) $\alpha_p = 0.4$. The three simulations are performed with a resolution $\Delta \boldsymbol{x} = D/25$. (\protect\circlewhite) represents the data from the experiments conducted in \citet{Dutsch1998_jfm}}
    \label{fig:velocity_penalisation}
\end{figure}

For the analysis of the penalty time-step parameter $\alpha_p$, we study three configurations in which $\alpha_p \in [0.4, 4, 40]$. As one can observe, some of the studies are beyond the scope of $\alpha_p \in [0, 1]$ recommended in the literature; therefore, the method would be expected to become ineffective in driving the fluid velocity $\boldsymbol{u}$ towards the imposed velocity $\boldsymbol{u_{ib}}$ at the immersed boundary. 
% However, as we commented before, we need a \emph{prior} condition for which the Data Assimilation methodology can efficiently enhance the simulation. 
All simulations are performed with a coarse-grained resolution of $\Delta \boldsymbol{x} = D/25$ and a time step $\Delta t=0.0025\,t_A$, where $t_A = D/U_{\textrm{max}}$ is the advective time. Results are now presented and discussed. In Fig.~\ref{fig:velocity_penalisation}, one can see that there is not much difference between the profiles obtained with $\alpha_p = 0.4$ and $\alpha_p = 4$, showing in general good agreement with the literature, represented by the experiences performed by \citet{Dutsch1998_jfm}. On the other hand, $\alpha = 40$ shows greater discrepancies and solutions that are not physically consistent with the test case. One can observe, for instance, that the velocity profile $u_x$ is not symmetric with respect to $y = 0$ at the planes $x/D = -0.6$ and $x/D = 0$. Also, in $x/D = 0$, neither $u_x$ nor $u_y$ matches with the asymptotic value of $\boldsymbol{u_{ib}} = (U_\mathrm{max}, 0)$ inside the cylinder region $y \in [-D/2, D/2]$.

\begin{figure}[!h]
    \centering
    \begin{tabular}{cc}
        \includegraphics[width=0.48\linewidth]{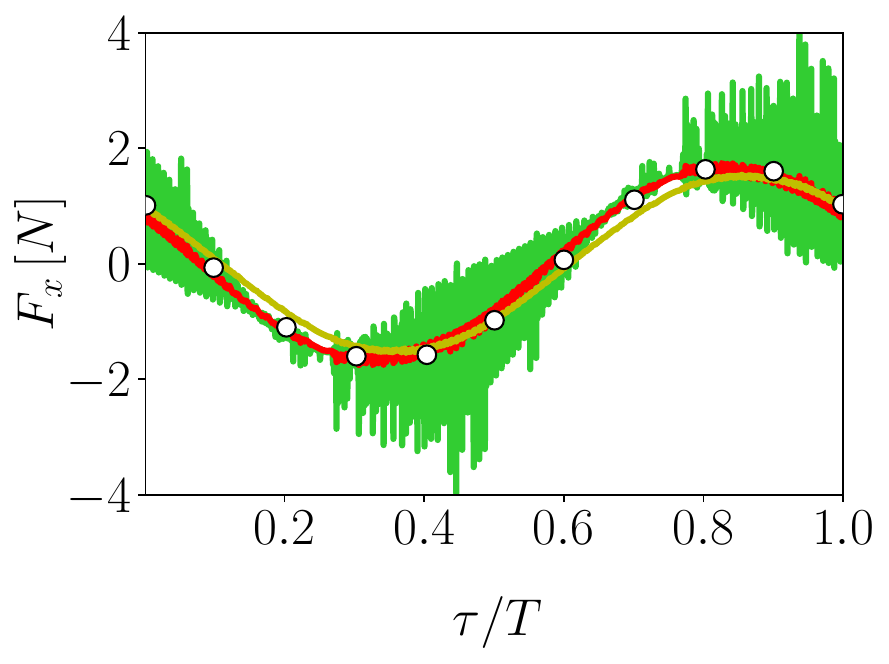} & 
        \includegraphics[width=0.52\linewidth]{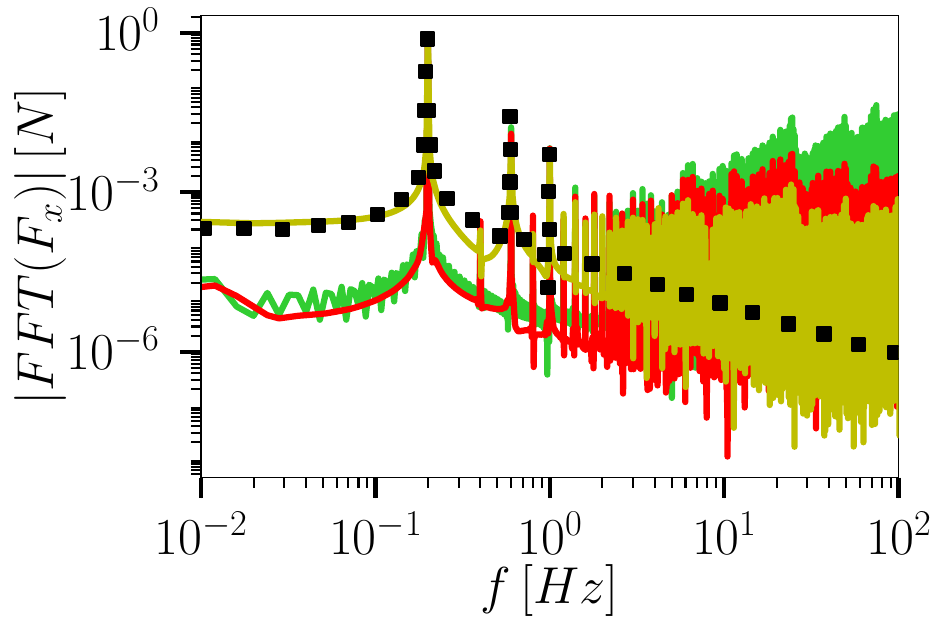} \\
         \textit{(a)} & \textit{(b)}
    \end{tabular}
    \caption{\textit{(a)} Resistive force $F_x$ over one period $T$, and \textit{(b)} frequency resistive force spectra. The simulations represented are (\protect\yellowline) $\alpha_p = 40$, (\protect\redline) $\alpha_p = 4$, and (\protect\greenlimeline) $\alpha_p = 0.4$. The three simulations are performed with a grid resolution $\Delta \boldsymbol{x} = D/25$. (\protect\circlewhite) represents the data from the experiments conducted in \citet{Dutsch1998_jfm} and ($\blacksquare$) is a body-fitted simulation run in \citet{Tsetoglou2024_nmf}}
    \label{fig:resistiveforce_penalisation}
\end{figure}

Results for the investigation of the resistive force $F_x$ are shown in Fig. \ref{fig:resistiveforce_penalisation}\textit{(a)}. Data is represented here in the reference frame of the force acting on the body by the fluid, which is the additive inverse of the forcing term defined in (\ref{eqn:force_integral}) and (\ref{eqn:force_discretised}). One could say that the simulation with $\alpha_p = 4$ best aligns with the experimental data since $\alpha_p = 0.4$ provokes meaningful discontinuities between the solid and the fluid regions that cause significant noise in the $F_x$ curve estimation. The results from the simulation with $\alpha_p = 40$ are very smooth, but they present discrepancies in some specific phases of the oscillation. In Fig.~\ref{fig:resistiveforce_penalisation}\textit{(b)}, we represent the spectra of the resistive force and compare it with a body-fitted simulation described in \citet{Tsetoglou2024_nmf}. One can see that all IBM simulations exhibit high-frequency (HF) noise, a well-known problem for IBM that has already been documented in the literature \citep{Mittal2005_arfm}. However, a concern arises when studying the low frequencies. All simulations correctly predict the peaks estimated with the body-fitted run, particularly the maximum peak located at $f = 0.2\,Hz$, but only the simulation with $\alpha_p = 40$ complies with the spectrum of the body-fitted run. The spectra are computed by using the 1D-Discrete Fourier Transform ($DFT$) \citep{Press2017_cambridge}, but they are sped up with the Fast Fourier Transform ($FFT$) algorithm \citep{Cooley1965_mc}:

\begin{equation}
    FFT(F_x)_k = \frac{1}{N} \sum_{n=0}^{N-1} e^{-2 \pi j \frac{n k}{N}} F_x (n\Delta t) \quad k=0, \ldots, N-1
\end{equation}
$N=100\,000$ is the total number of samples or window length, equivalent to $50T$. The frequencies are given by $f_k = k / (\Delta t N)$ for the $k-$th index.

\begin{figure}[!h]
    \centering
    \begin{tabular}{ccc}
    \includegraphics[width=0.32\linewidth]{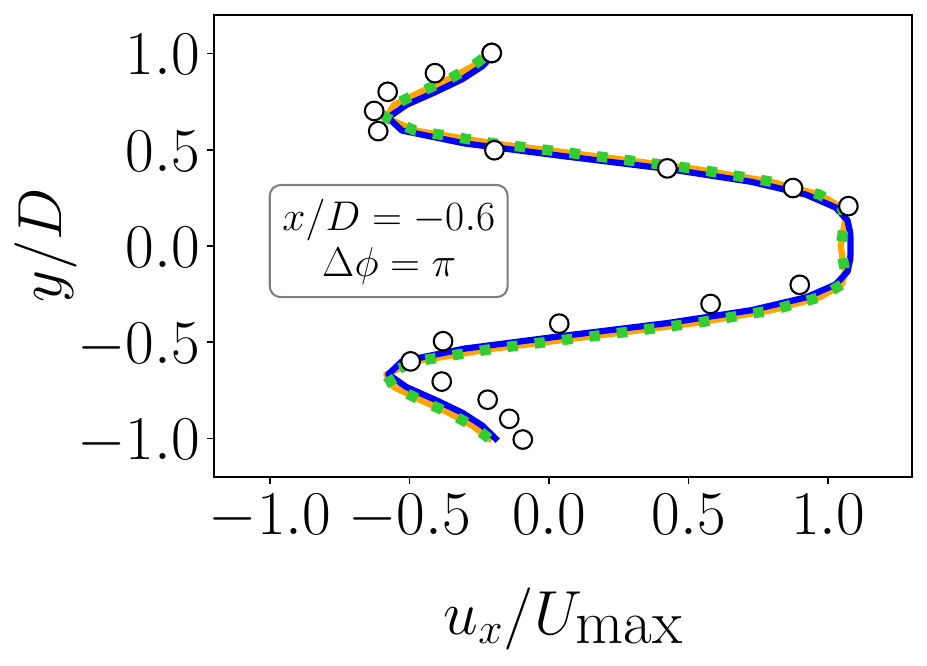} &
    \includegraphics[width=0.32\linewidth]{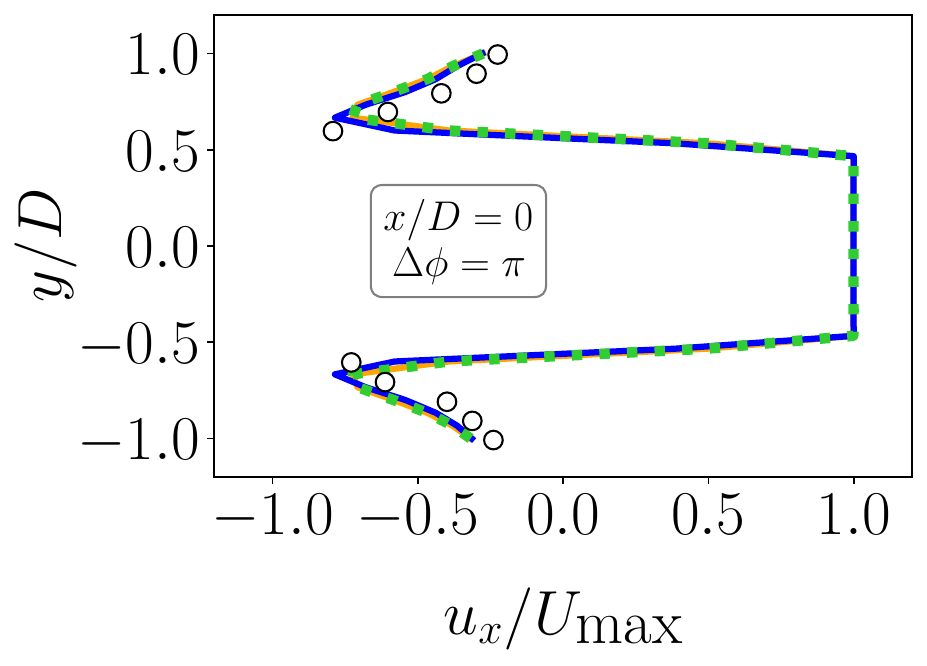} &
    \includegraphics[width=0.32\linewidth]{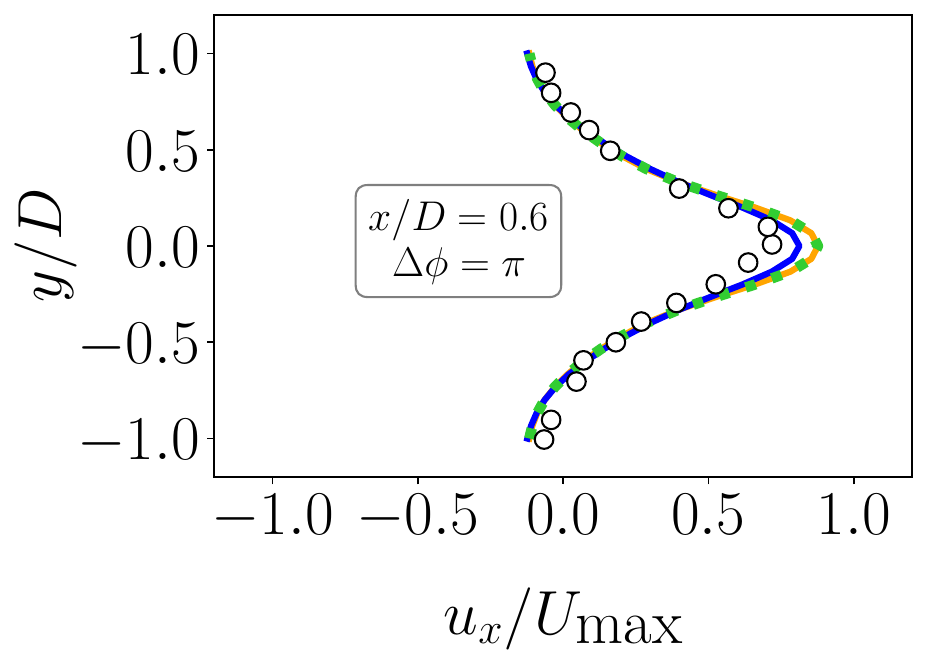} \\
    \textit{(a)} & \textit{(b)} & \textit{(c)} \\
    \includegraphics[width=0.32\linewidth]{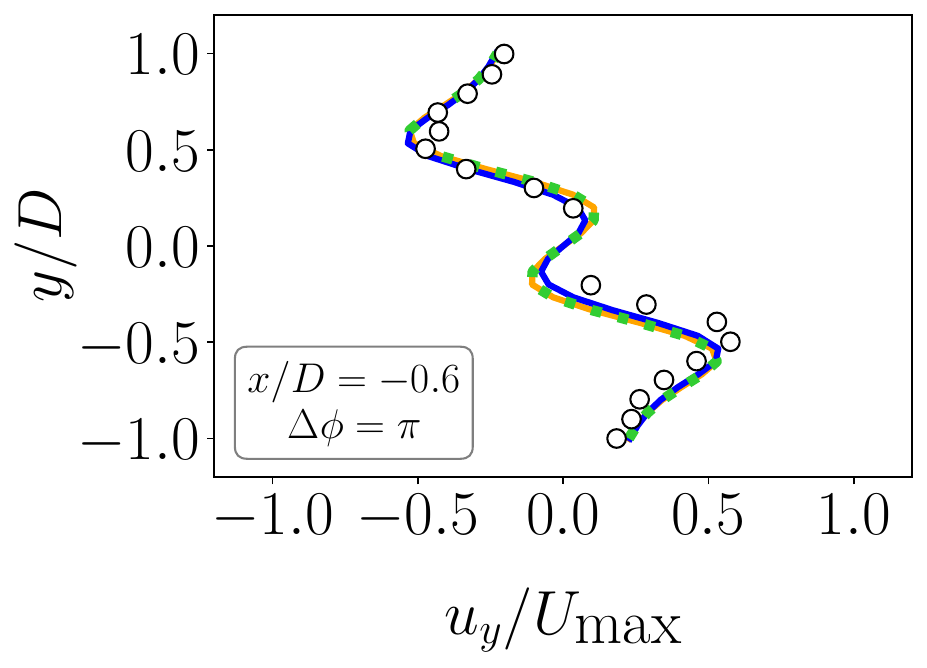} &
    \includegraphics[width=0.32\linewidth]{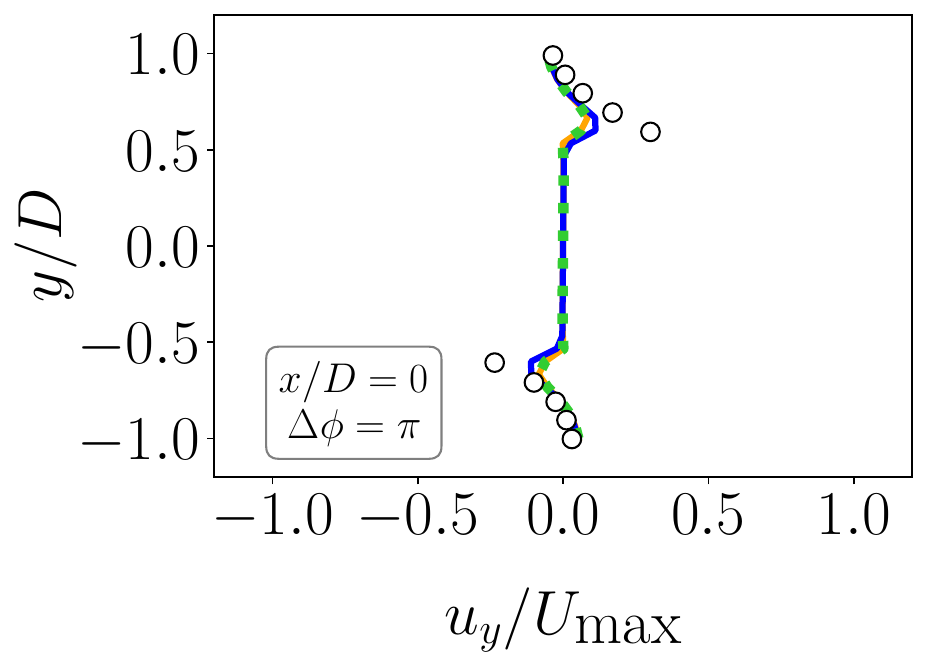} &
    \includegraphics[width=0.32\linewidth]{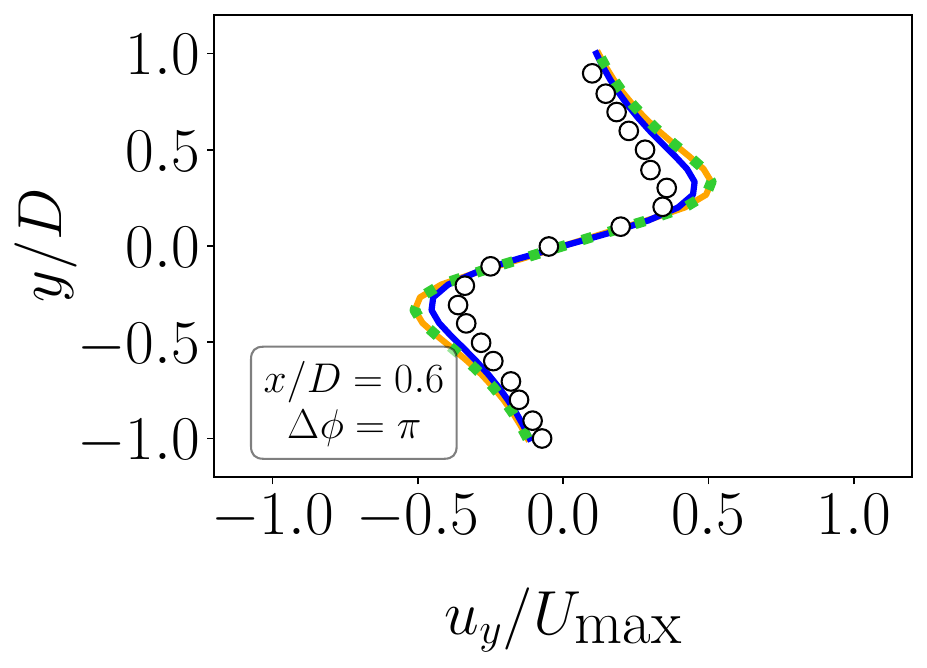} \\
    \textit{(d)} & \textit{(e)} & \textit{(f)}     
    \end{tabular}
    \caption{Velocity profiles when $\Delta \phi = \pi$ at three different planes $x/D \in [ -0.6, 0, 0.6]$ for (\protect\blueline) $\Delta \boldsymbol{x} = D/100$, (\protect\orangeline) $\Delta \boldsymbol{x} = D/40$, and (\protect\greenlimelinedotted) $\Delta \boldsymbol{x} = D/25$. The three simulations are performed with $\alpha_p = 0.4$. (\protect\circlewhite) represents the data from the experiments conducted in \citet{Dutsch1998_jfm}}
\label{fig:velocity_mesh}
\end{figure}

\begin{figure}[!h]
    \centering
    \begin{tabular}{cc}
        \includegraphics[width=0.48\linewidth]{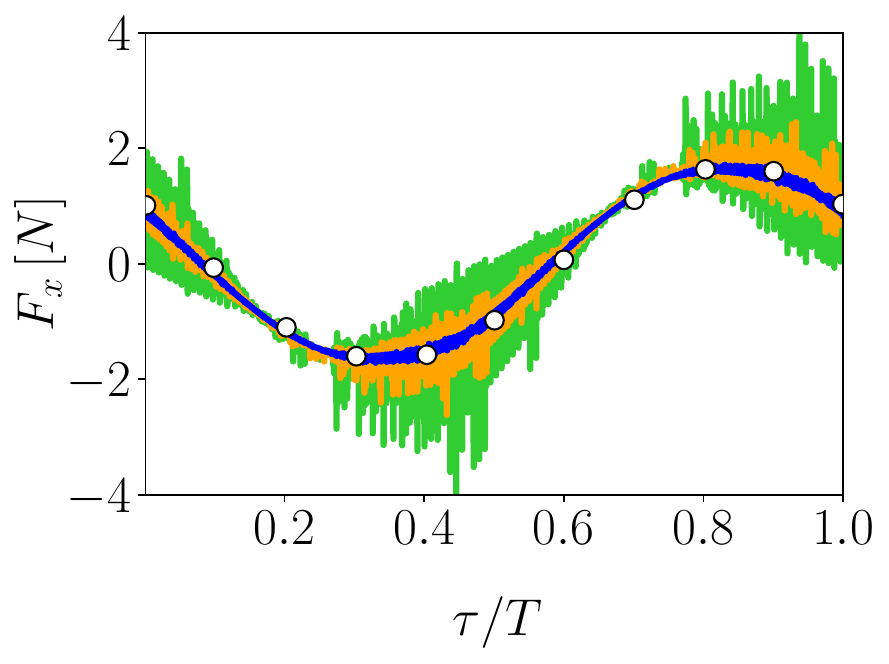} & 
        \includegraphics[width=0.52\linewidth]{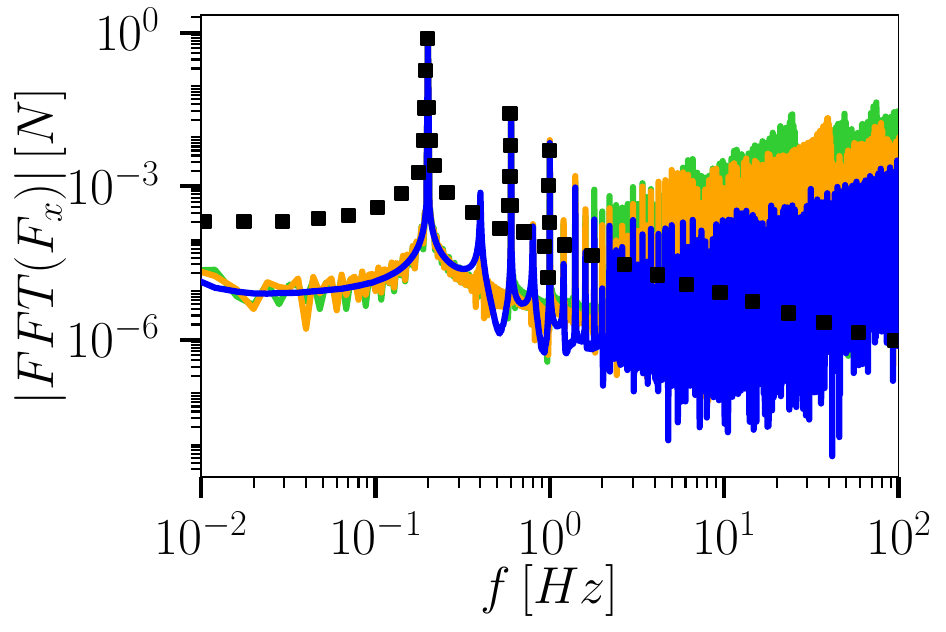} \\
         \textit{(a)} & \textit{(b)}
    \end{tabular}
    \caption{\textit{(a)} Resistive force $F_x$ over one period $T$, and \textit{(b)} frequency resistive force spectra. The simulations represented are (\protect\blueline) $\Delta \boldsymbol{x} = D/100$, (\protect\orangeline) $\Delta \boldsymbol{x} = D/40$, and (\protect\greenlimeline) $\Delta \boldsymbol{x} = D/25$. The three simulations are performed with $\alpha_p = 0.4$. (\protect\circlewhite) represents the data from the experiments conducted in \citet{Dutsch1998_jfm} and ($\blacksquare$) is a body-fitted simulation run in \citet{Tsetoglou2024_nmf}}
    \label{fig:resistiveforce_mesh}
\end{figure}

The sensitivity to the grid resolution $\Delta \boldsymbol{x}$ is studied next. Simulations are run for $\Delta \boldsymbol{x} \in [D/25, D/40, D/100]$, while keeping $\alpha_p = 0.4$ and $\Delta t = 0.0025\,t_A$. Concerning the velocity profiles shown in Fig. \ref{fig:velocity_mesh}, they exhibit no significant discrepancies among the different resolutions, with all correctly capturing the expected physics of the test case (penalisation of the velocity to its imposed value and symmetry about $y=0$). However, a slight improvement is observed for the fine-grained run with $\Delta \boldsymbol{x} = D/100$, particularly noticeable in Fig. \ref{fig:velocity_mesh}\textit{(c)} and \textit{(f)} for the plane $x/D=0.6$. The resistive force $F_x$ in Fig. \ref{fig:resistiveforce_mesh}\textit{(a)} shows that all curves are capable of following the experimental data, but the presence of noise is clearly related to the degradation of the mesh. About the spectra in Fig. \ref{fig:resistiveforce_mesh}\textit{(b)}, HF noise decreases in strength with the reduction of the grid spacing.

This parametric study indicates that, in general, lower values of $\alpha_p$ and finer mesh resolutions $\Delta x$ yield the best flow conditions in terms of velocity and force profiles. However, higher values of $\alpha_p$ provide greater stability, particularly by reducing noise in the force profiles, improving the representation of the low-frequency spectrum, and attenuating the high-frequency range. Therefore, in the Data Assimilation experiment described in \S\ref{sec:DA_experiment}, the objective is to optimise $\alpha_p$ to simultaneously enhance flow conditions and accurately capture the resistive force spectrum.

\section{Data Assimilation experiment}
\label{sec:DA_experiment}

\begin{figure}[!h]
    \centering
    \begin{tabular}{cc}
    \includegraphics[width=0.44\linewidth]{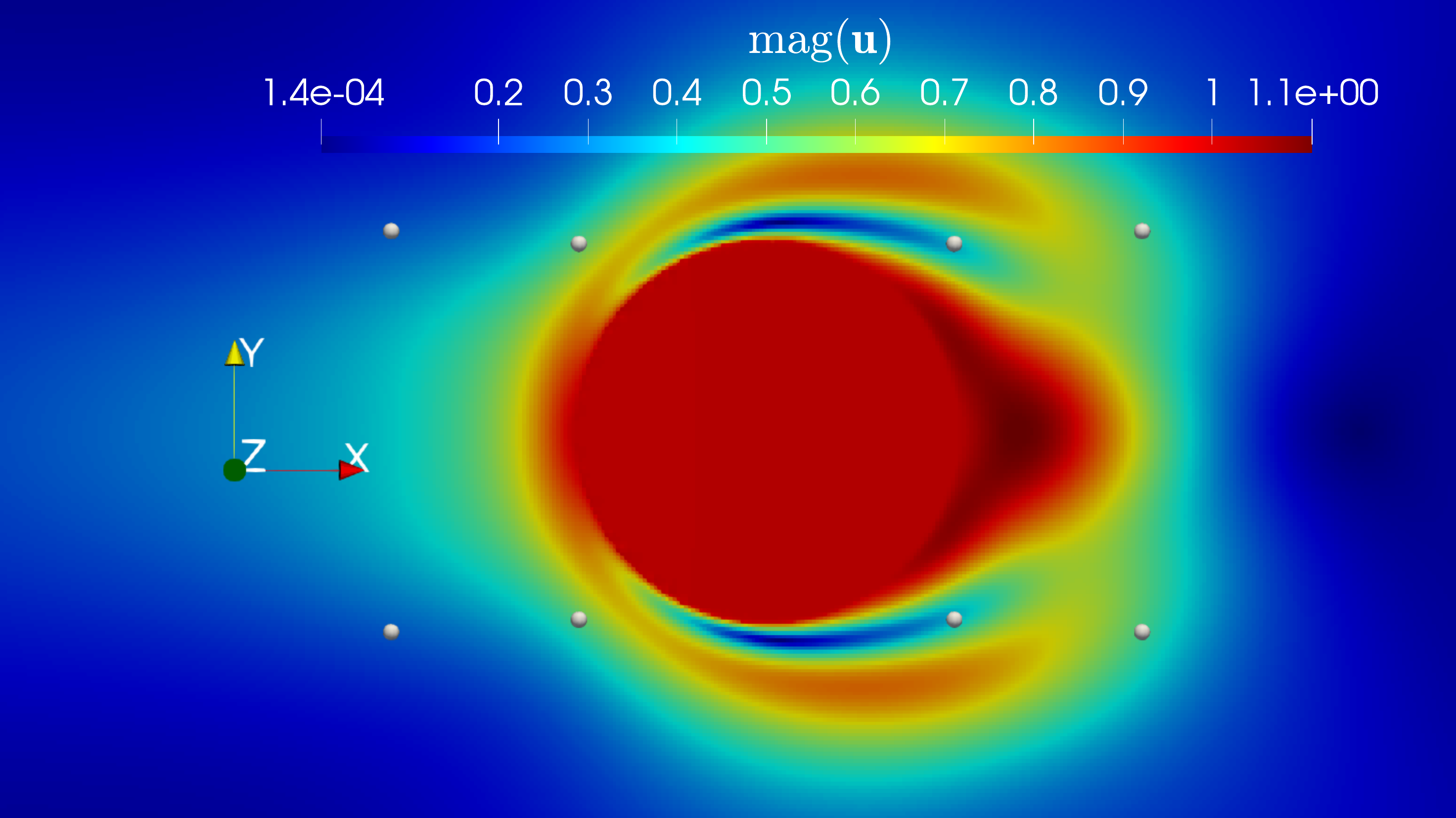} &
    \includegraphics[width=0.56\linewidth]{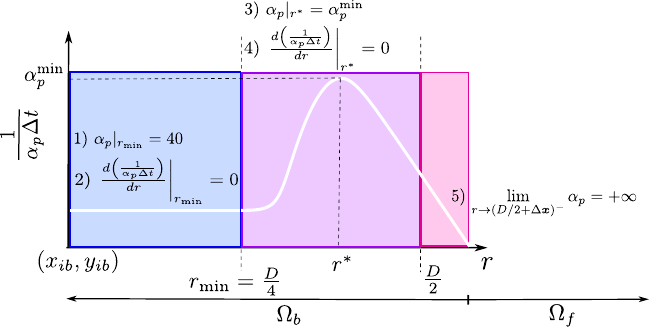} \\
    \textit{(a)} & \textit{(b)} 
    \end{tabular}
    \caption{\textit{(a)} Location of the high-fidelity observation with respect to the position of the cylinder when its centre $(x_{ib}, y_{ib}) = (0,0)$, and \textit{(b)} definition of all the constraints for the penalisation time-step parameter $\alpha_p$ as a function of the radial position}
    \label{fig:cylinder_dataAss}
\end{figure}

In this section, we consider two simulations from the database described in \S\ref{sec:penIBM_cylinder}. First, the simulation with $\alpha_p = 0.4$ and $\Delta \boldsymbol{x} = D/100$ serves as a reference, from which $16$ high-fidelity observations are extracted from $8$ local sensors that measure both the streamwise $u_x$ and crosswise $u_y$ velocities. As one can see in Fig.~\ref{fig:cylinder_dataAss}\textit{(a)}, the $8$ sensors are strategically selected to help ensure the symmetry condition around $y=0$. This simulation is referred to as HF-IBM (\emph{high-fidelity} IBM). The second simulation, in which $\alpha_p = 40$ and $\Delta \boldsymbol{x} = D/25$, represents the \emph{prior} condition for the state estimation of our assimilation experiment. This simulation was observed to exhibit problems in ensuring proper conditions for the velocity and force profiles; therefore, data assimilation is used here to improve its performance. This simulation is referred to as LF-IBM (\emph{low-fidelity} IBM). An Ensemble Kalman Filter (EnKF) with $N_e = 40$ ensemble members is employed to carry out the assimilation, which uses the CONES library \citep{Villanueva2024_cof} adapted for OpenFOAM solvers and available on \url{https://gitlab.ensam.eu/pe431/cones-dev}. CONES provides a framework for performing online, ensemble-based Data Assimilation (DA) at runtime, enabling continuous field and model parameter corrections without interrupting or restarting the simulation. The step-by-step procedure specific to the present study is described in \S\ref{sec:DA_PISO}. A total number of $K = 5\,000$ analysis phases over a total time window of $50\,T$ are performed, with constantly spaced assimilation windows of $20\Delta t$. The uncertainty prescribed for the observations (i.e., their standard deviation) is set to $0.05\,U_{\mathrm{max}}$, which is a plausible value for realistic applications.

Regarding the ensemble's variability, since the flow is laminar, perturbing only some of the model's parameters would not be sufficient. Thus, the perturbation is also applied to the physical field (i.e., velocity field $\boldsymbol{u}$). To achieve this, $40$ instantaneous velocity snapshots from the LF-IBM run over a single period $T$ serve as initial conditions for each ensemble member. As a result, each realisation of the ensemble starts with a different phase in the cycle/initial position of the cylinder ($x_0, y_0$). The purpose of the DA experiment is to reduce the ensemble's variability, in order to identify a suitable \textit{state estimation} that complies with the observations in a probabilistic sense. This is obtained here via the optimisation of a set of free parameters $\theta$. Determining these parameters allows the ensemble of realisations to synchronise with the high-fidelity observation. In this context, the cylinder's initial position in the streamwise direction, $x_0$, is treated as a free parameter to be inferred by the EnKF. Consequently, the cylinder’s position becomes a dynamic variable influenced by the DA algorithm, which must be accounted for within the PISO algorithm (see Alg.~\ref{alg:IBM_DA_cylinder}). Additionally, the value of $\alpha_p$ is locally optimised. This variable is expressed via an expansion in the radial direction $r=\sqrt{(x-x_{ib})^2+(y-y_{ib})^2}$ which, unlike the azimuthal coordinate $\Theta = \arctan{(y-y_{ib})/(x-x_{ib})}$, does not depend on the oscillation phase. This phase-independence allows us to assume $\alpha_p \neq \alpha_p (t)$, thereby simplifying the assimilation procedure within the EnKF framework. This function presents the following shape:

\begin{equation}
    \boldsymbol{f}_P (r) \propto -\frac{1}{\alpha_p \Delta t} = -(a_0+a_1r+a_2r^2+a_3r^3+a_4 r^4)
    \label{eqn:radial_expansion}
\end{equation}

The choice for the expansion describing $1/(\alpha_p \Delta t)$ provides a more flexible parametric space. The coefficients $a_j, \, j \in(0,...,4)$ in (\ref{eqn:radial_expansion}) will converge to time-independent values through the assimilation procedure, and are computed through an algebraic operation based on the optimised parameters $\theta^a_k$ estimated via the EnKF. While these model parameters remain constant over time, the state estimation procedure dynamically corrects the flow field, thereby capturing the system's inherent unsteady behaviour through the assimilation of instantaneous observations.  These coefficients capture information about the critical points of the radial expansion $\left(\frac{\partial (1/(\alpha_p \Delta t))}{\partial r} = 0 \right)$. A fourth-degree polynomial is employed to enforce five specific constraints. The detailed formulation of both the parameters and the conditions they satisfy is presented next. First, the LF-IBM simulation achieved the desired resistive spectrum at low frequencies. Then, we define an inner region of the cylinder $r < r_\textrm{min}$ (corresponding to the blue region in Fig.~\ref{fig:cylinder_scheme}\textit{(b)}) where $\alpha_p$ is kept constant and equal to its LF-IBM value, $\alpha_p=40$. This condition ensures that the low-frequency noise in the resistive force $F_x$ remains consistent with the body-fitted simulation spectrum. Later, when $r \geq r_\textrm{min}$, we allow for a smooth transition of $\alpha_p$ to let the value evolve toward a minimum $\alpha_p^\mathrm{min}$, expected to approach $\alpha_p=0.4$, so the simulations match the HF-IBM observations. Last, we want that $\lim\limits_{r \to (D/2+\Delta \boldsymbol{x})^-} \alpha_p = +\infty$ to reduce the discontinuity between the fluid and the solid regions. One can see that the limit considers the size of the additional cell layer represented in red in Fig.~\ref{fig:cylinder_scheme}\textit{(b)}. A sketch of all the constraints with the function $1/(\alpha_p \Delta t)$ is represented in Fig. \ref{fig:cylinder_dataAss}\textit{(b)}, with the distinction between the zone where $\alpha_p$ does not change (in blue), the zone affected by the polynomial within the cylinder (in purple), and the additional external cell layer (in red). Since OpenFOAM solves the Navier--Stokes equations in the Cartesian coordinate framework, we optimise, separately, the radius $r^\ast$ and $\alpha_p^\textrm{min}$ for each component of the forcing term $\boldsymbol{f}_P = (f_{P_x}, f_{P_y})$, and we employ their values to recalculate the coefficients of the polynomial expansion written in (\ref{eqn:radial_expansion}) at each analysis phase $k$ up to convergence. The application of this strategy for the IBM forcing required the optimisation of five model parameters, which are grouped in the vector $\theta = \{ x_0, r^\ast_x, \alpha_{p_x}^\textrm{min}, r^\ast_y, \alpha_{p_y}^\textrm{min} \}$.  The matrix formulation used to compute the coefficients $a_j$ at each assimilation phase is presented in (\ref{eqn:matrix_parameters}), where the sub-index $i = (x,y)$ denotes the corresponding spatial direction.

\begin{gather}
    \begin{pmatrix}
        1 & r_{\textrm{min}} & r_{\textrm{min}}^2 & r_{\textrm{min}}^3 & r_{\textrm{min}}^4 \\
        0 & 1 & 2r_{\textrm{min}} & 3r_{\textrm{min}}^2 & 4r_{\textrm{min}}^3 \\
        1 & r^\ast_i & r^{\ast^2}_i & r^{\ast^3}_i & r^{\ast^4}_i \\
        0 & 1 & 2r^\ast_i & 3r^{\ast^2}_i & 4r^{\ast^3}_i \\
        1 & \frac{D}{2} + \Delta x_i & \left(\frac{D}{2} + \Delta x_i \right)^2 &
        \left(\frac{D}{2} + \Delta x_i \right)^3 & \left(\frac{D}{2} + \Delta x_i \right)^4
    \end{pmatrix}
    \begin{pmatrix}
        a_0 \\ a_1 \\ a_2 \\ a_3 \\ a_4
    \end{pmatrix} =
    \begin{pmatrix}
        \frac{1}{40\Delta t} \\ 0 \\ \frac{1}{{\alpha_{p_i}^{\textrm{min}}} \Delta t} \\ 0 \\ 0
    \end{pmatrix}
    \label{eqn:matrix_parameters}
\end{gather}

The \emph{prior} for $r^\ast$ and $\alpha_p^\textrm{min}$ in the two directions correspond to normal distributions such as $r^\ast = \mathcal{N} \left(\frac{3D}{8}, \left(0.05\frac{3D}{8} \right)^2 \right)$ and $\alpha_p^\textrm{min} = \mathcal{N}(40, 4)$. For the former, the average lies between $r_\mathrm{min}=D/4$ and the cylinder's boundary at $D/2$, with a standard deviation of $5\%$ among ensemble members. For the latter, the average is set to the value imposed at $r < r_\textrm{min}$, with a variability of again $5\%$.

\subsection{Results}
\label{sec:ResultsDA_cylinder}

\begin{figure}[!h]
    \centering
    \begin{tabular}{cc}
    \includegraphics[width=0.5\linewidth]{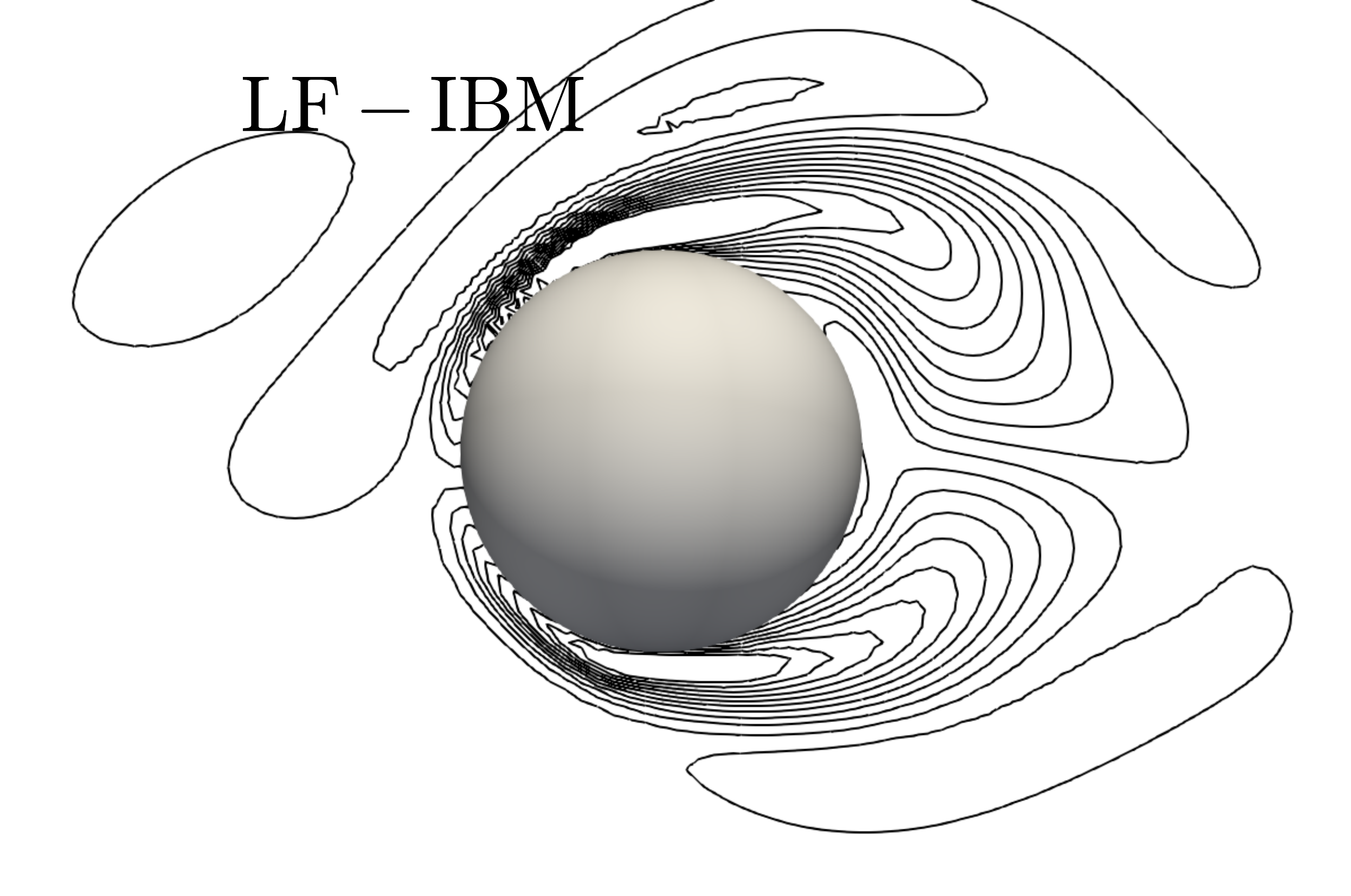} &
    \includegraphics[width=0.5\linewidth]{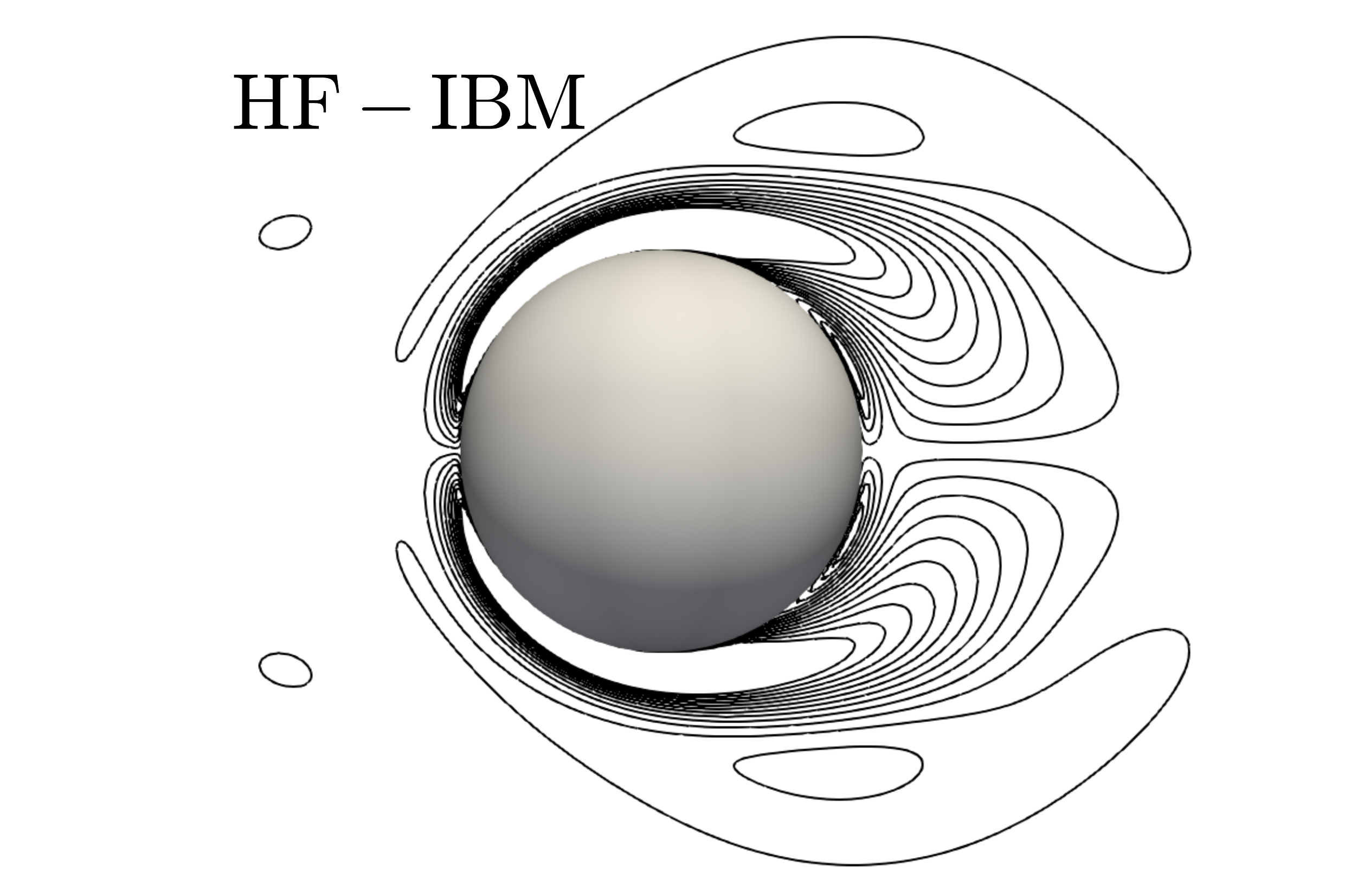} \\
    \textit{(a)} & \textit{(b)} \\
    \includegraphics[width=0.5\linewidth]{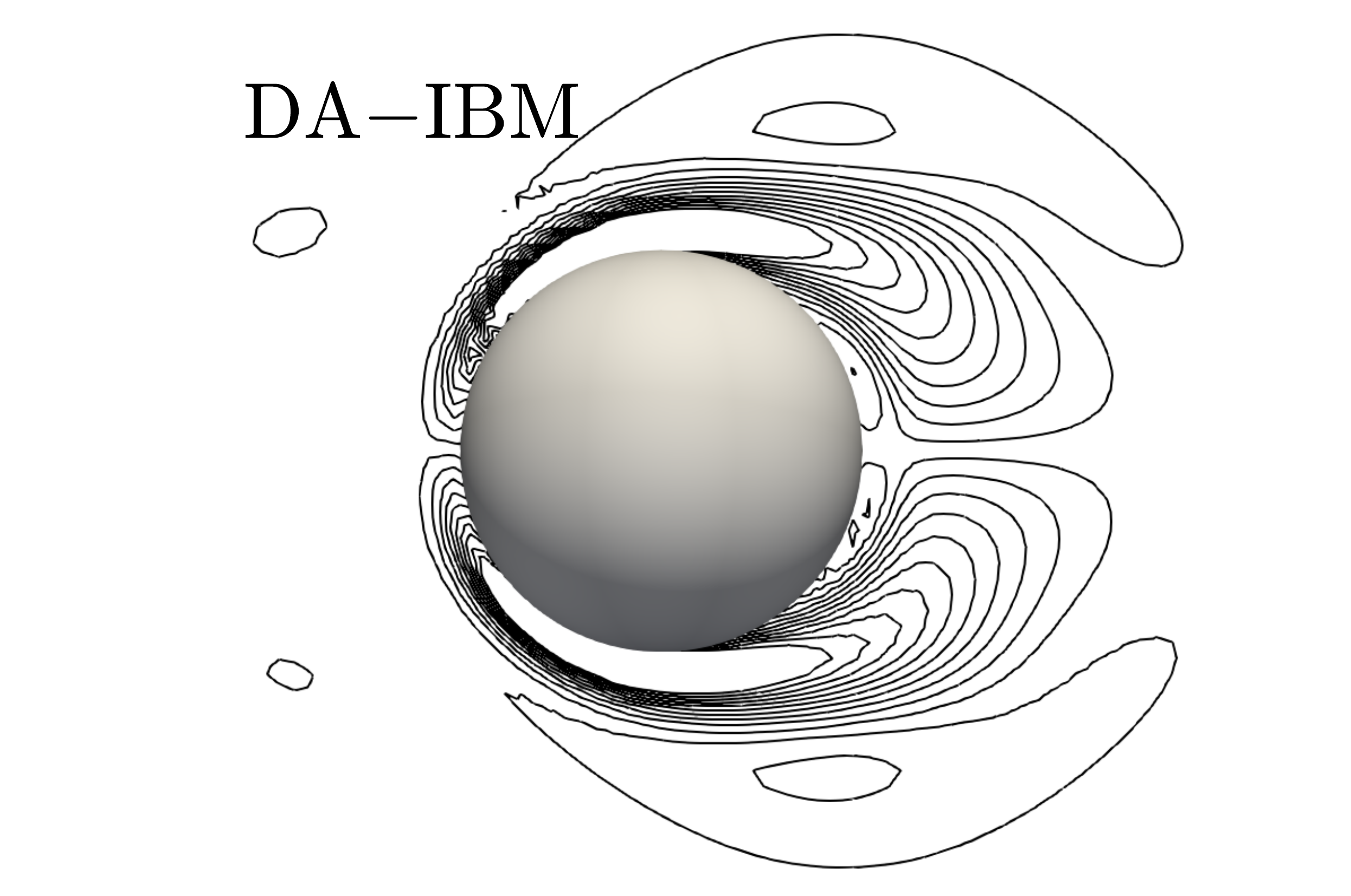} &
    \includegraphics[width=0.5\linewidth, height=4.8cm]{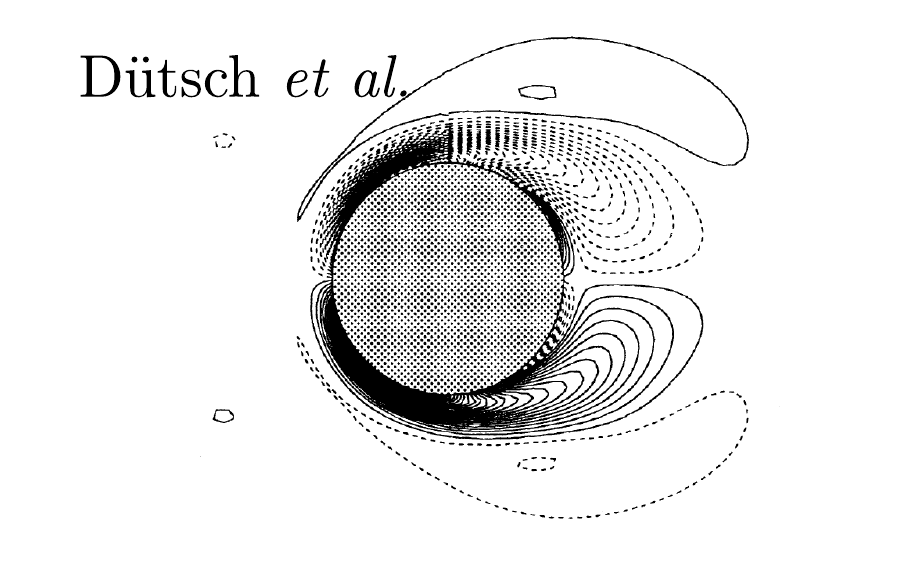} \\
    \textit{(c)} & \textit{(d)}    
    \end{tabular}
    \caption{Vorticity contours $\omega_z$ when $\Delta \phi = 0$ for \textit{(a)} the \emph{prior} LF-IBM, \textit{(b)} the reference simulation from which we obtain the high-fidelity data HF-IBM, \textit{(c)} the experiment with Data Assimilation DA-IBM, and \textit{(d)} the experiment performed by \citet{Dutsch1998_jfm}}
    \label{fig:vorticity_contours}
\end{figure}

The analysis of the performance of the DA experiment includes an investigation of several flow quantities. First, we examine the vorticity $\boldsymbol{\omega} = \boldsymbol{\nabla} \times \boldsymbol{u}$ when the phase difference $\Delta\phi = 0$, i.e., $u_{ib_x} = -U_\mathrm{max}$. In Fig.~\ref{fig:vorticity_contours}, one can see that the DA-IBM run outperforms the \emph{prior} LF-IBM simulation, closely matching the high-fidelity HF-IBM results. Minor perturbations, mostly near the leading edge, are observed, which may be associated with the coarser mesh resolution. However, this does not prevent the results obtained with DA from recovering the symmetry condition at $y = 0$, which was lost in the LF-IBM run.

\begin{figure}[!h]
    \centering
    \begin{tabular}{ccc}
    \includegraphics[width=0.32\linewidth]{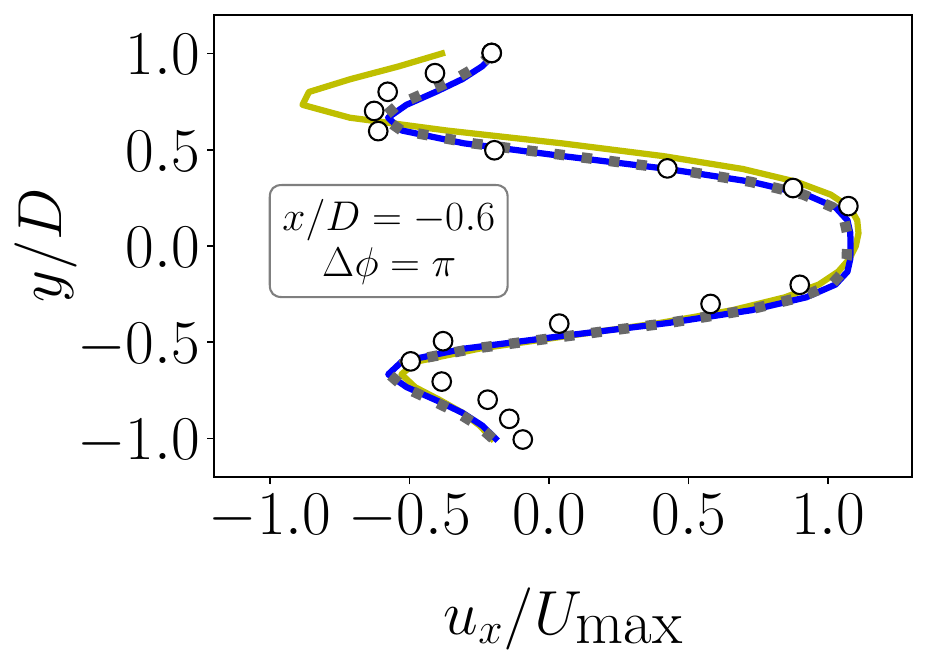} &
    \includegraphics[width=0.32\linewidth]{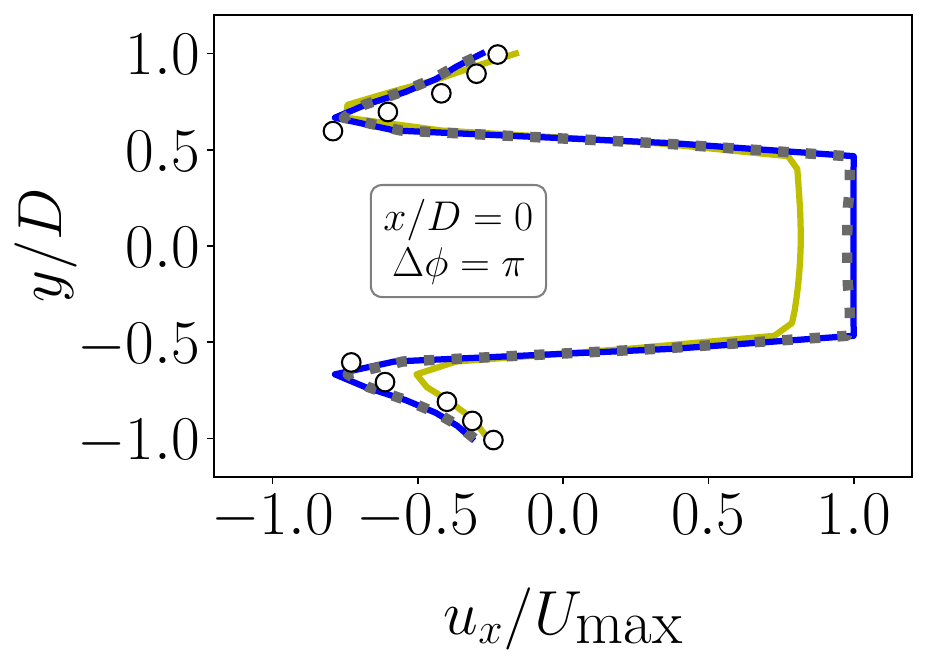} &
    \includegraphics[width=0.32\linewidth]{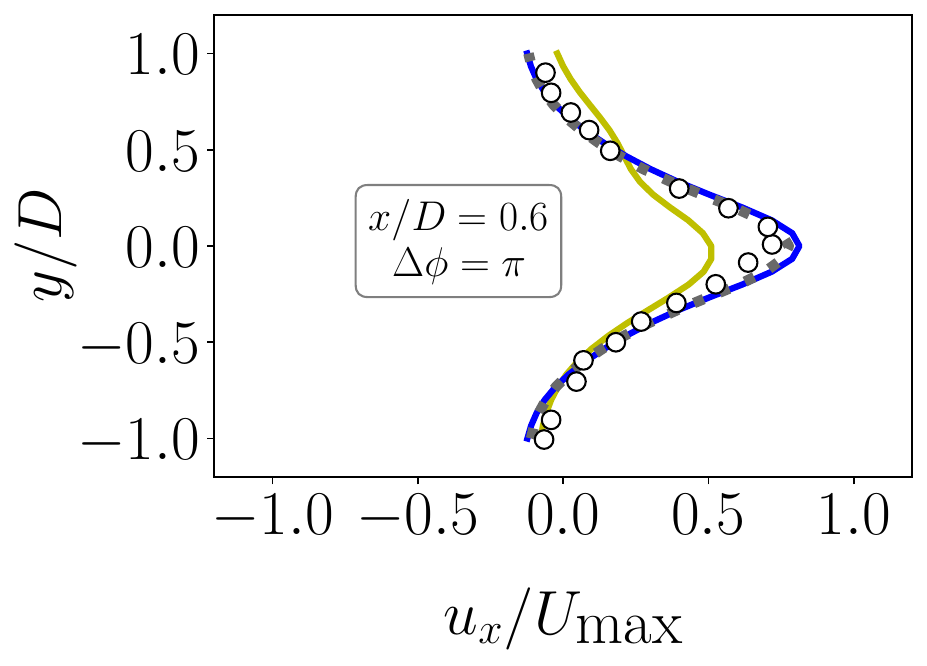} \\
    \textit{(a)} & \textit{(b)} & \textit{(c)} \\
    \includegraphics[width=0.32\linewidth]{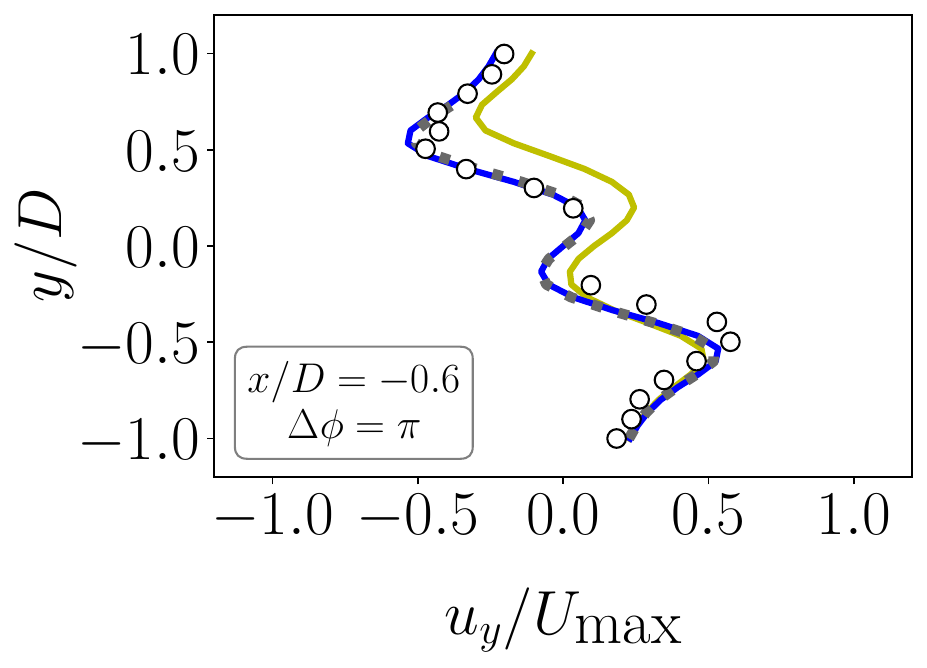} &
    \includegraphics[width=0.32\linewidth]{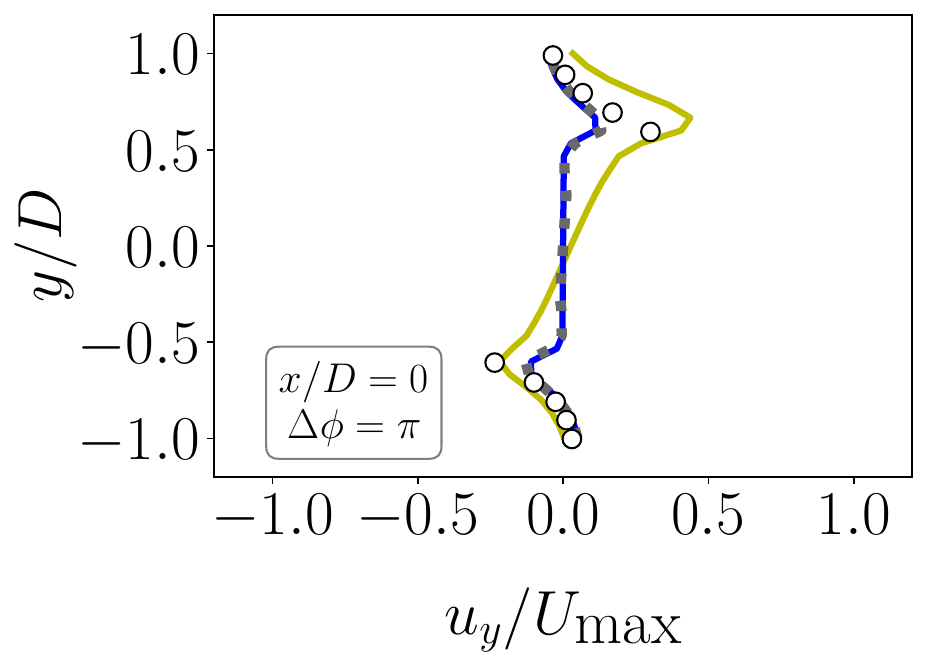} &
    \includegraphics[width=0.32\linewidth]{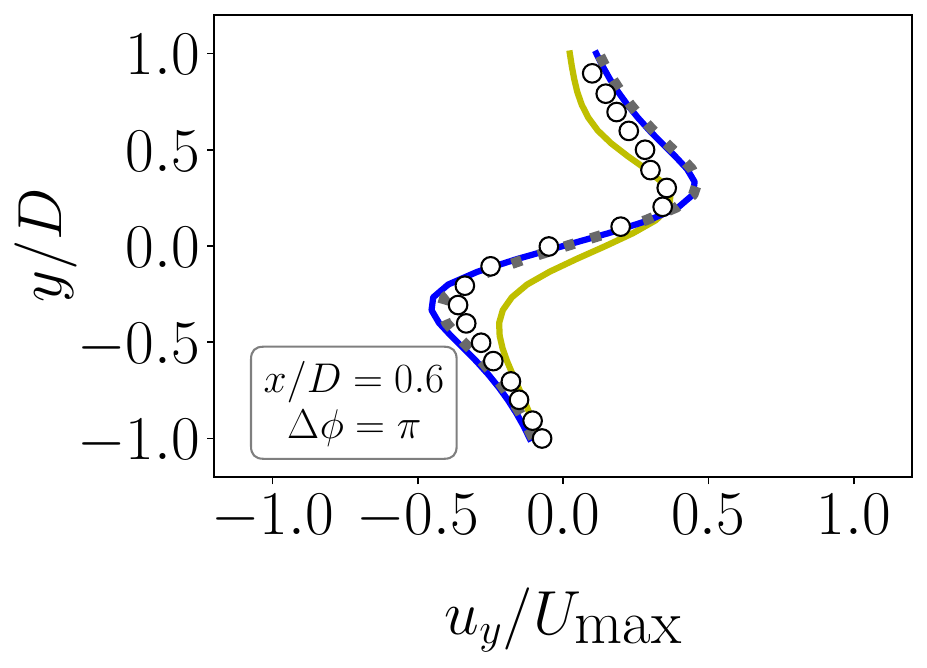} \\
    \textit{(d)} & \textit{(e)} & \textit{(f)}     
    \end{tabular}
    \caption{Velocity profiles when $\Delta \phi = \pi$ at three different planes $x/D \in [ -0.6, 0, 0.6]$ for (\protect\greylinedottedstrong) the DA-IBM experiment, (\protect\blueline) the HF-IBM run, and (\protect\yellowline) the LF-IBM simulation. (\protect\circlewhite) represents the data from the experiments conducted in \citet{Dutsch1998_jfm}}
    \label{fig:velocity_DA}
\end{figure}

\begin{figure}[!h]
    \centering
    \begin{tabular}{ccc}
    \includegraphics[width=0.32\linewidth]{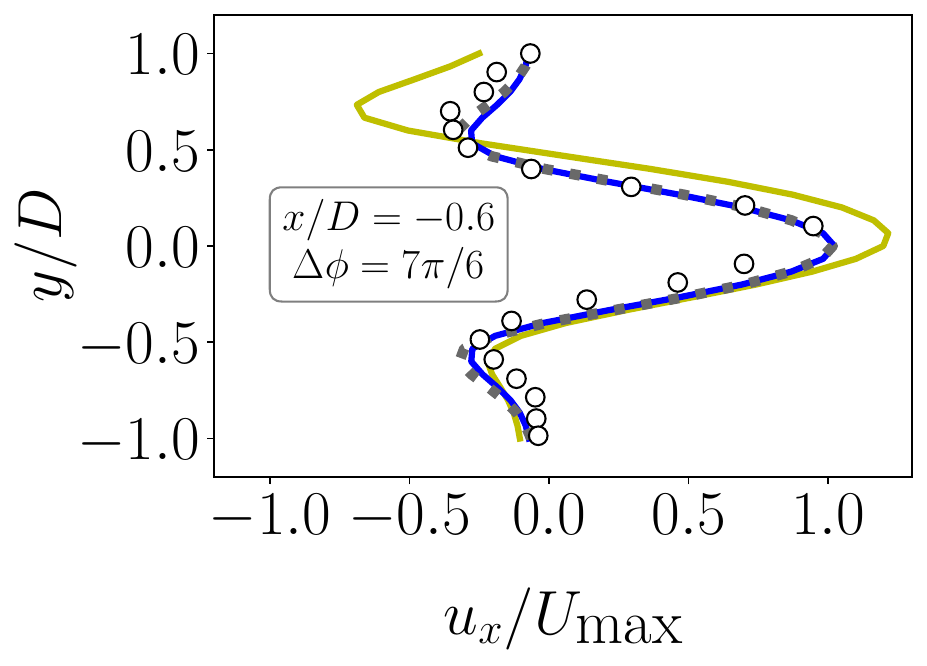} &
    \includegraphics[width=0.32\linewidth]{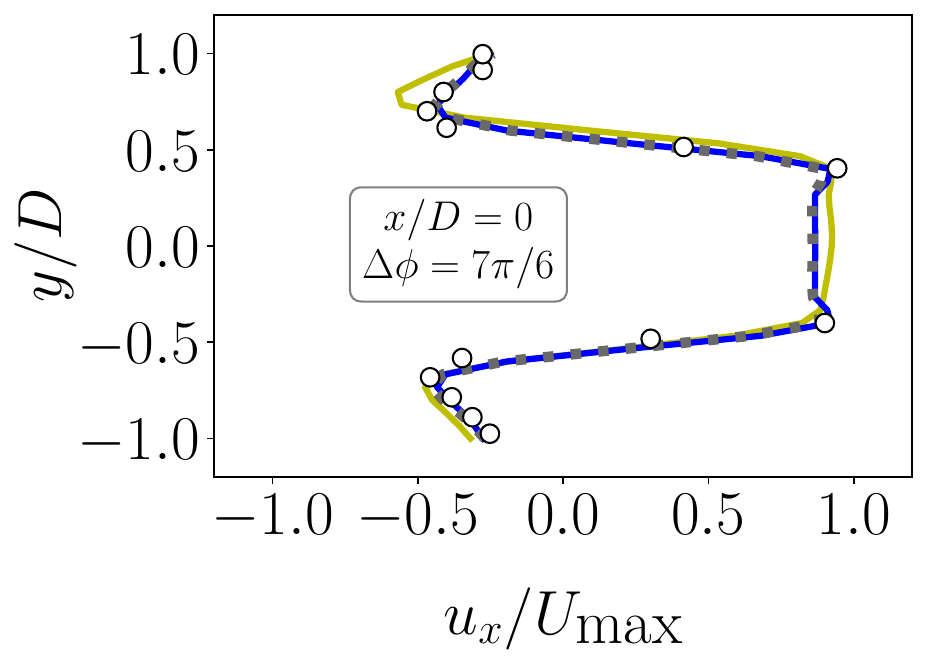} &
    \includegraphics[width=0.32\linewidth]{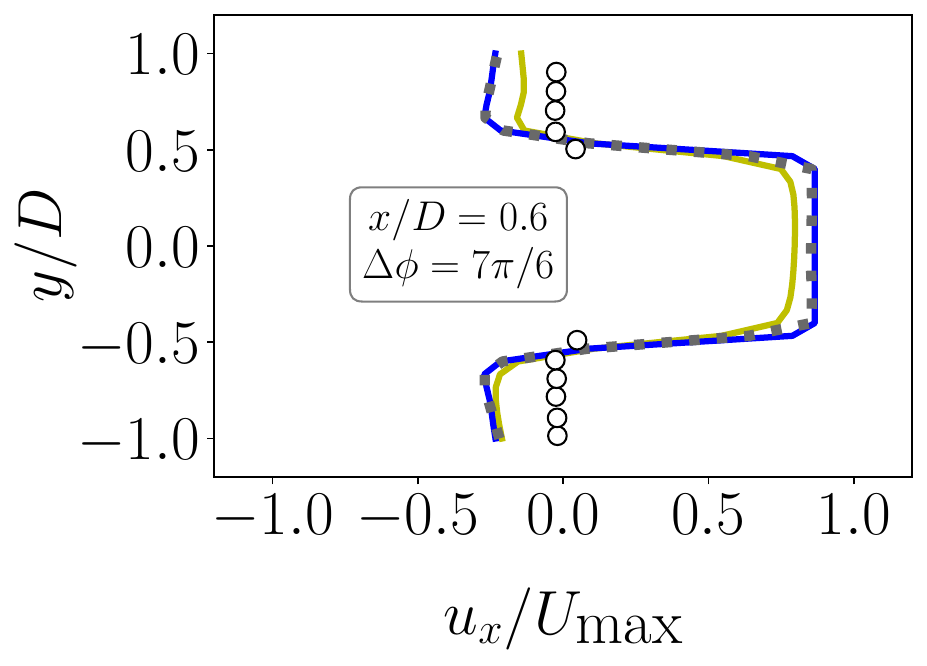} \\
    \textit{(a)} & \textit{(b)} & \textit{(c)} \\
    \includegraphics[width=0.32\linewidth]{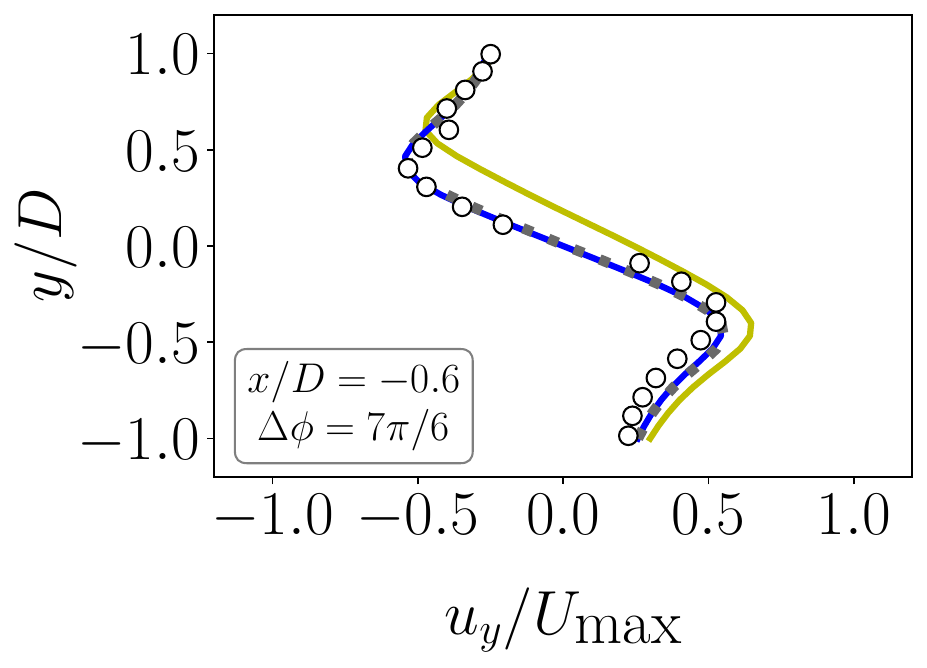} &
    \includegraphics[width=0.32\linewidth]{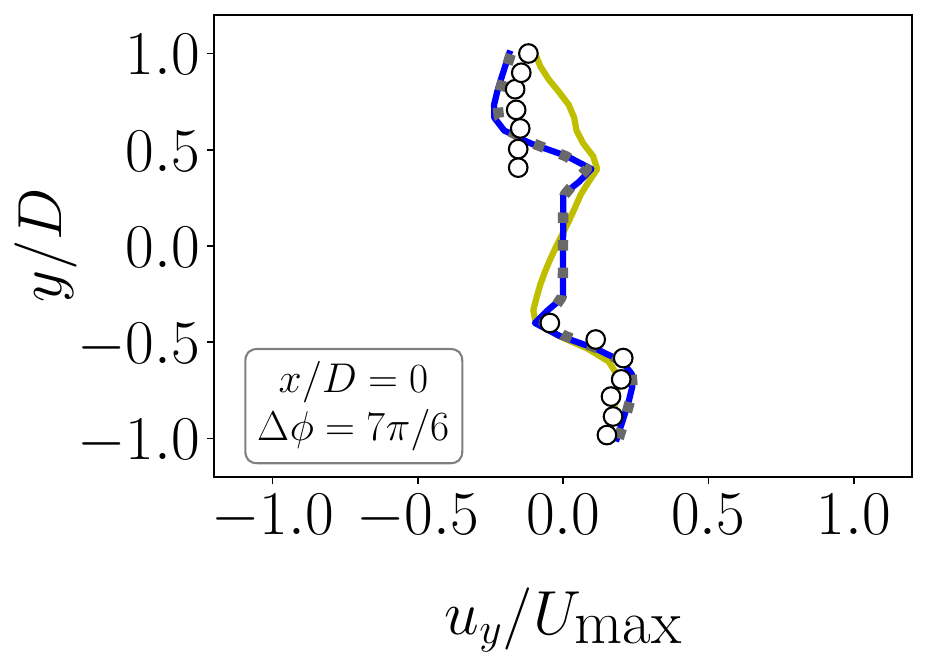} &
    \includegraphics[width=0.32\linewidth]{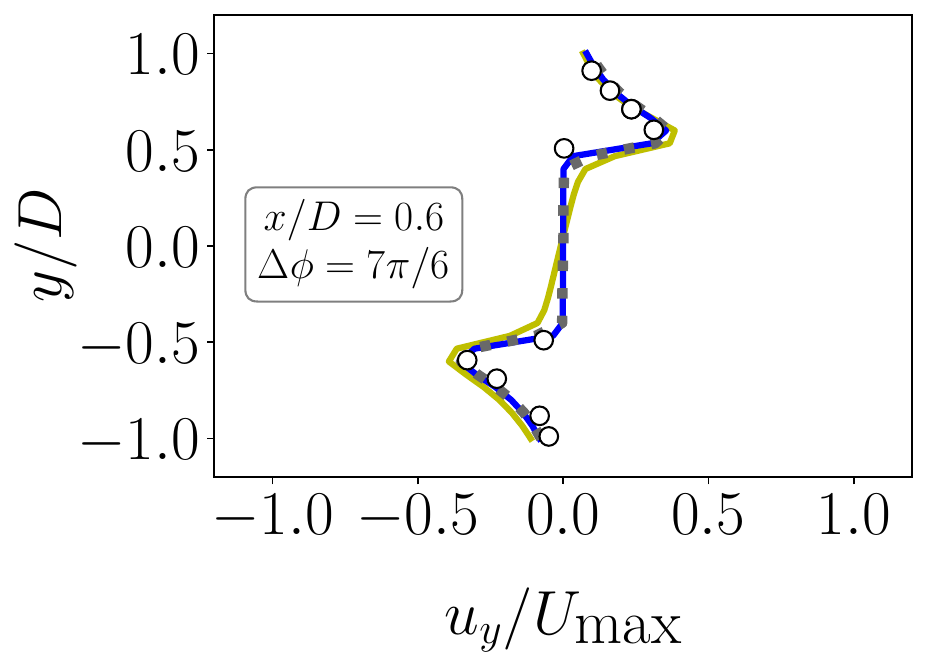} \\
    \textit{(d)} & \textit{(e)} & \textit{(f)}     
    \end{tabular}
    \caption{Velocity profiles when $\Delta \phi = 7\pi/6$ at three different planes $x/D \in [ -0.6, 0, 0.6]$ for (\protect\greylinedottedstrong) the DA-IBM experiment, (\protect\blueline) the HF-IBM run, and (\protect\yellowline) the LF-IBM simulation. (\protect\circlewhite) represents the data from the experiments conducted in \citet{Dutsch1998_jfm}}
    \label{fig:velocity_DA_phase210}
\end{figure}

Second, we represent again the velocity profiles at three different planes when the phase difference $\Delta \phi = \pi$, as illustrated in Fig.~\ref{fig:velocity_DA}. Again, the DA-IBM run is symmetric around $y = 0$, and properly constrains the velocity within the solid region. In particular, the condition at $x/D = 0$, in which $(u_x, u_y) = (U_\mathrm{max}, 0)$ is satisfied within the cylinder $y \in [-D/2, D/2]$. Overall, the velocity profiles from the DA-IBM run closely match those from the HF-IBM simulation. Additionally, we represent the profiles at $\Delta \phi = 7\pi/6$ in Fig.~\ref{fig:velocity_DA_phase210}, which corresponds to a cylinder position at $x_{ib}/D = 0.4$ and velocity $u_{x_{ib}}/U_\textrm{max} = 0.87$. In this case, one can see how DA profiles nearly overlap those from the HF-IBM run, demonstrating a clear improvement over the LF-IBM profiles. The only noticeable exception occurs in $u_x$ for the plane $x/D=0.6$, where the LF-IBM solution shows slightly better agreement above the cylinder compared to the experimental data; however, it still fails to preserve the symmetry with respect to $y=0$, so the velocity profiles represented in HF-IBM and DA-IBM runs are physically more consistent and reliable.

\begin{figure}[!h]
    \centering
    \begin{tabular}{cc}
        \includegraphics[width=0.48\linewidth]{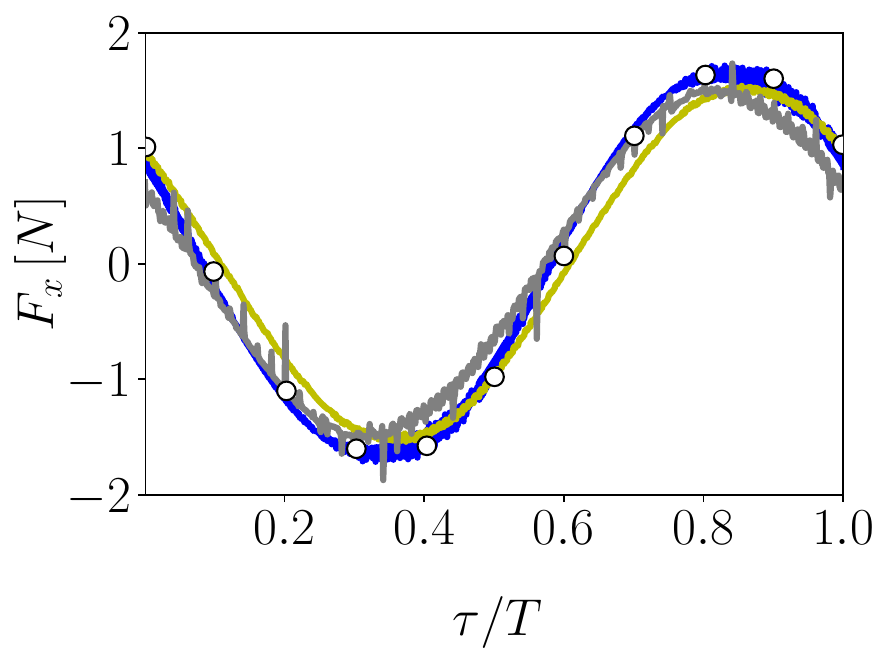} & 
        \includegraphics[width=0.52\linewidth]{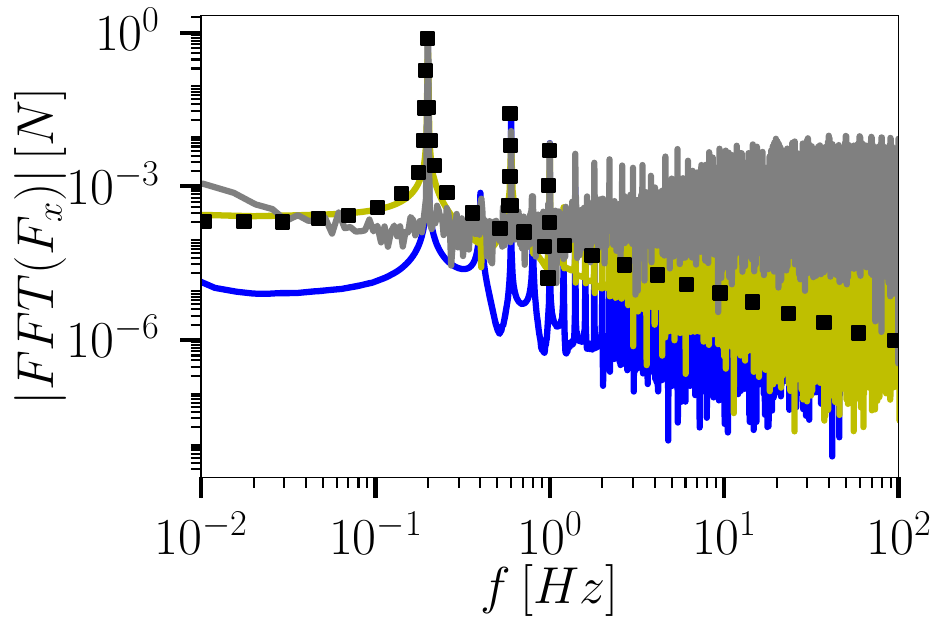} \\
         \textit{(a)} & \textit{(b)}
    \end{tabular}
    \caption{\textit{(a)} Resistive force $F_x$ over one period $T$, and \textit{(b)} frequency resistive force spectra. The simulations represented are (\protect\greylinesolidstrong) DA-IBM, (\protect\blueline) HF-IBM, and (\protect\yellowline) LF-IBM. (\protect\circlewhite) represents the data from the experiments conducted in \citet{Dutsch1998_jfm} and ($\blacksquare$) is a body-fitted simulation run in \citet{Tsetoglou2024_nmf}}
    \label{fig:resistiveforce_DA}
\end{figure}

Finally, we present again the interaction or resistive force $F_x$ over a single period $T$ in Fig. \ref{fig:resistiveforce_DA}. The DA-IBM profile matches the HF-IBM simulation with high precision, with just some slight discrepancies in the acceleration parts within the period ($d (\textrm{mag}(\boldsymbol{u_{ib}}))/dt > 0$), specifically when $\tau/T \in [0.25, 0.5]$ and $\tau/T \in [0.75, 1]$, corresponding to $\Delta \phi \in [\pi/2, \pi]$ and $\Delta \phi \in [-\pi/2, 0]$, respectively. In any case, these discrepancies are smaller than the differences in the deceleration phases ($d (\textrm{mag}(\boldsymbol{u_{ib}}))/dt < 0$) in the LF-IBM simulation for $\tau/T \in [0, 0.25]$ and $\tau/T \in [0.5, 0.75]$, i.e., $\Delta \phi \in [0, \pi/2]$ and $\Delta \phi \in [-\pi, -\pi/2]$. Regarding the spectral analysis, in the DA method combined with the radial expansion of $\alpha_p(r)$, low frequencies are better captured since $\alpha_p$ is left to the stable value $\alpha_p=40$ within the central part of the cylinder ($r < r_\textrm{min}$), and the HF noise observed matches the equivalent noise observed in Fig.~\ref{fig:resistiveforce_penalisation}\textit{(b)} for the run with $\alpha_p = 0.4$ and grid refinement of $\Delta \boldsymbol{x} = D/25$. In fact, as shown in Tab. \ref{tab:parameters_cylinder_DA}, the averaged value from all ensemble members $(\overline{\cdot})$ for the minimum value $\alpha_P^\textrm{min}$ obtained with DA in both directions $x$ and $y$ is very close to the $\alpha_p = 0.4$ value from the HF-IBM run, with this maximum occurring near the initial condition described in \S\ref{sec:DA_experiment} at $r/D = 3/8$.

% \begin{table}[!h]
%     \centering
%     \begin{tabular}{ c | c c c c c c}
%          & $F^\prime_x \, \mathrm[{N}]$ & $\varepsilon \,(F^\prime_x) \, \mathrm{[\%]}$ & $f_{\mathrm{fund}} \,\mathrm{[Hz]}$ & $\alpha_p$ & $L_x \times L_y$ & $\Delta \boldsymbol{x}$ \\
%          & & & & & & \\
%          LF-IBM & $1.06$ & $7.02$ & $0.2$ & $40$ & $10D\times10D$ & $[D/25, D/25]$\\
%          HF-IBM & $1.19$ & $4.39$ & $0.2$ & $0.4$ & $10D\times10D$ & $[D/100, D/100]$\\
%          DA-IBM & $1.05$ & $7.89$ & $0.2$ & $\alpha_p(r)$ & $10D\times10D$ & $[D/25, D/25]$ \\
%          & & & & \\
%          VSI-IVP \cite{Tsetoglou2024_nmf} & $1.11$ & $2.63$ & $0.2$ & $1$ & $10D\times10D$ & $[D/100, D/100]$ \\
%          body-fitted \cite{Tsetoglou2024_nmf} & $1.14$ & $-$ & $0.2$ & $-$ & NaN & NaN
%     \end{tabular}
%     \caption{}
%     \label{tab:my_label}
% \end{table}

\begin{table}[!h]
    \centering
    \begin{tabular}{c | c c c c c}
         & $\overline{x_0}$ & $\overline{r_x^\ast}/D$ & $\overline{\alpha_{P_x}^\textrm{min}}$ & $\overline{r_y^\ast}/D$ & $\overline{\alpha_{P_y}^\textrm{min}}$  \\ \hline
      \emph{prior} & $0$ & $0.375$ & $40$ & $0.375$ & $40$ \\
       DA-IBM  & $-1.8639e^{-4}$ & $0.3915$ & $0.5310$ & $0.3909$ & $0.3716$
    \end{tabular}
    \caption{Comparison between the parameters employed as a \emph{prior} or initial condition in the DA experiment, and the values optimised after the DA experiment}
    \label{tab:parameters_cylinder_DA}
\end{table}

\section{Machine Learning experiment}
\label{sec:ML_experiment}

This section outlines the use of Data Assimilation (DA)-augmented profiles to construct black-box Machine Learning (ML) models that integrate a sliding-window strategy with the Random Forest Regression (RFR) algorithm (detailed in \S\ref{sec:RFR}) to learn the parameter optimisation process within the DA framework. Throughout this work, because the ML components operate on DA outputs, the models are trained and applied under the assumption that the relevant flow information (specifically the velocity field and the oscillation cycle) is already available from the preceding DA stage. The methodology is demonstrated for the oscillating-cylinder case using the penalty forcing term $\boldsymbol{f}_P$. As defined in (\ref{eqn:forcePenDarcy}), this term is treated as a local field whose behaviour is expected to vary with the oscillation phase. Accordingly, (i) separate models are developed for distinct phases of the oscillation cycle $\Delta \phi$, and (ii) suitable features must be identified to ensure an accurate characterisation of $\boldsymbol{f}_P$. This work is performed on the assumption that the features of the unsteady behaviour of the flow are known, as previously discussed. This permits training and using ML tools over well-defined time windows. This is clearly a strong simplification relative to real flows, where unsteady variations are unknown and must be accounted for online. However, this simplified analysis can highlight the potential for coupling DA and ML, and, in particular, the numerical and algorithmic difficulties that need to be addressed.   

\begin{figure}
    \centering
    \includegraphics[width=0.8\linewidth]{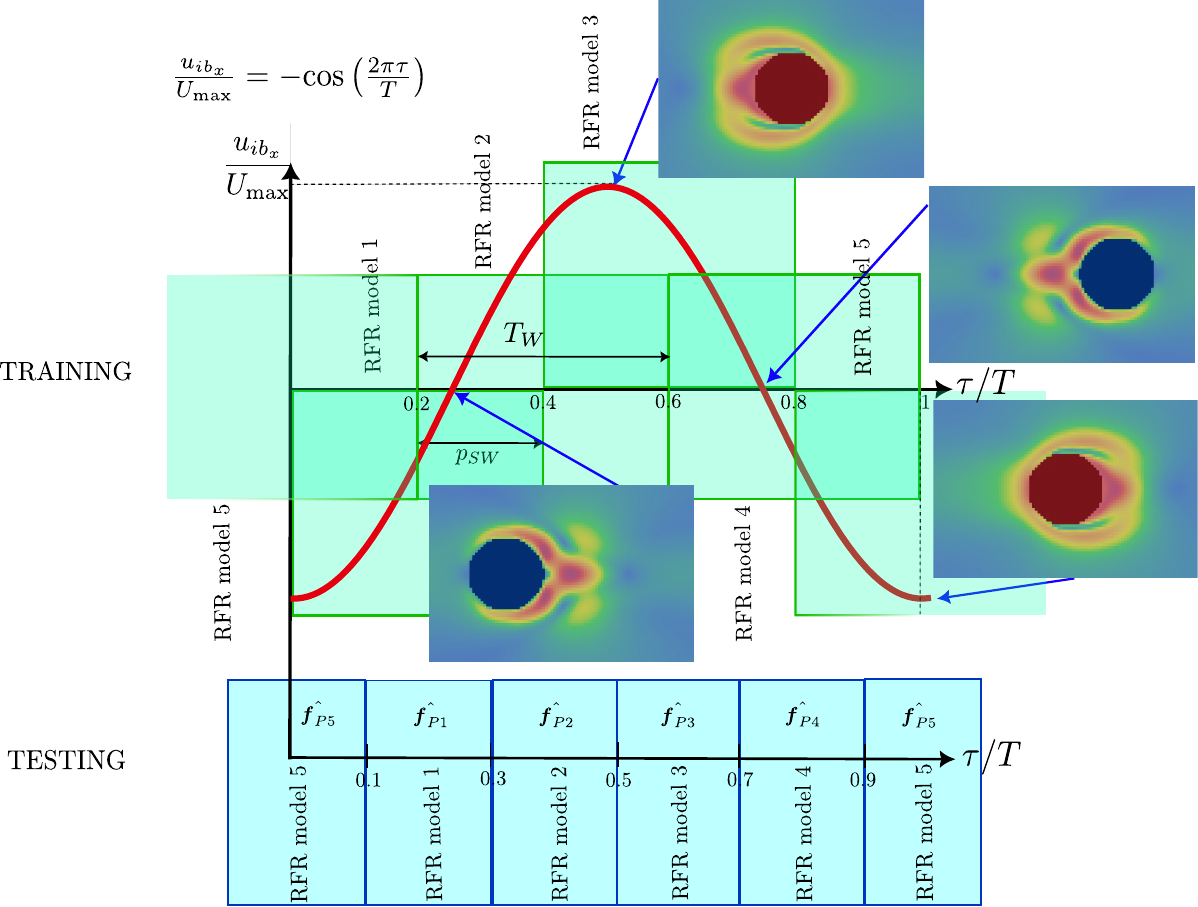}
    \caption{Sliding-window approach for the oscillating cylinder, where each green rectangle represents a different window during training, and blue rectangles indicate the utilisation sequence of the models during the prediction of the penalty forcing term $\boldsymbol{f}_P$}
    \label{fig:sw_cylinder}
\end{figure}

Concerning the first listed point (i.e, generation of separate models), we train five ML models online to capture different flow dynamics across the oscillation cycle (i.e., flow acceleration, deceleration, and direction changes) using streaming data during the assimilation procedure. Since the regime under consideration is laminar, the training data consists of the assimilated fields within the solid region $\Omega_b$ over a single oscillation cycle, comprising approximately $550$ mesh elements at each time step, extracted for the $N_e=40$ ensemble members. Additionally, to avoid abrupt interruptions in predictions at the boundaries between models, we employ a sliding window approach with a sliding period $p_{SW}=0.5 T_W$, where the window size is defined as $T_W = 2T/5$, ensuring a uniform length across the five models. Hence, the first model is trained on data from $\tau/T \in (0, 0.4]$, the second on $\tau/T \in (0.2, 0.6]$, the third on $\tau/T \in (0.4, 0.8]$, the fourth on $\tau/T \in (0.6, 1]$, and the last one on $\tau/T \in (0, 0.2]$ and $\tau/T \in (0.8, 1]$. By sampling data every $10\Delta t$ time steps (half immediately following the assimilation phase and half midway between two assimilation phases), each black-box model is trained with approximately $m_T = 1.76M$ samples, which is generally sufficient to ensure convergence of the RFR algorithm \citep{Valero2025_cf}. Of these samples, $80\%$ are explicitly used for training and $20\%$ are employed for validation. The remaining hyperparameters include the number of decision trees, set to $N=100$, and the minimum number of samples per leaf node, set to $m_{sl}=5$, ensuring robustness against overfitting. Each model is subsequently applied only within the central portion of its corresponding window, i.e., for $\tau/T \in (\tau_{\textrm{init}}/T+T_W/4, \tau_{\textrm{end}}/T - T_W/4]$, with $\tau_{\textrm{init}}/T$ and $\tau_{\textrm{end}}/T$ the initial and final times for each training window, respectively, and where predictive accuracy is expected to be highest. Integration with OpenFOAM is facilitated by the C++ library \textit{dlib} \citep{King2009_jmlr}, which provides strong modularity within OpenFOAM solvers. Fig.~\ref{fig:sw_cylinder} illustrates all these concepts.

Regarding the feature selection, we recall that in the DA optimisation procedure, the penalty forcing term $\boldsymbol{f}_P$ is a local field which, for a given mesh element, depends solely on the local flow velocity $\boldsymbol{u}$, the prescribed cylinder motion $\boldsymbol{u_{ib}}$, and the radial distance $r$ (through $\alpha_P = \alpha_P(r)$). At first glance, these quantities appear to provide a sufficient set of input features for the ML model. However, predictions generated by black-box ML models often contain residual noise. In a laminar regime, where the flow exhibits limited inherent physical dissipation, such errors are unlikely to be smoothed out by the flow dynamics and could instead accumulate over time. This issue is particularly critical in the present framework, as the PISO algorithm computes (via calls to the ML models) an explicit forcing term $\hat{\boldsymbol{f}_P}$ three times per time step: once during the predictor stage and twice during the correction stages after each velocity update (see Alg.~\ref{alg:IBM_ML_cylinder}). From this point onward, the hat symbol ($\hat{\cdot}$) indicates quantities predicted by the ML models. Repeated evaluations of $\hat{\boldsymbol{f}_P}$ can potentially amplify errors in the fields introduced by previous ML predictions, leading to spurious flow gradients, especially near the domain boundaries. 

To mitigate these effects, the boundary conditions are switched from \emph{zeroGradient} to periodic (cyclic) on opposite sides of the domain. This modification ensures that both the flow field and its fluxes are continuous across boundaries, effectively preventing the artificial accumulation of momentum that might occur with non-periodic (Neumann-type) boundaries. The cyclic formulation thus promotes smoother flow dissipation and yields a more physically consistent representation of the near-boundary dynamics, independent of the initial cylinder motion.

Additionally, inspired by the work by \citet{Villiers2025_ftc}, we enrich the input space with the cylinder acceleration $\boldsymbol{a_{ib}}$, which is the time derivative of $\boldsymbol{u_{ib}}$, and is incorporated to reinforce the phase difference $\Delta \phi$:

\begin{equation}
    \boldsymbol{a_{ib}} = \begin{pmatrix}
        a_{x_{ib}} \\
        a_{y_{ib}}
    \end{pmatrix} =
    \begin{pmatrix}
        4\pi^2 f^2 A\, \textrm{sin} (2\pi f t) \\
        0
    \end{pmatrix}
    \label{eqn:accelerationCylinder}
\end{equation}
Furthermore, the inclusion of the vorticity $\boldsymbol{\omega}$ accounts for the velocity gradients between neighbouring mesh elements. It is, however, counterproductive to introduce the time derivatives of the flow velocity, as the state estimation in the DA procedure introduces noise that increases exponentially with local accelerations, thereby degrading the performance of the black-box model. Since $\boldsymbol{f}_P = (f_{P_x}, f_{P_y})$ is a vector field, each component depends on the flow features along its respective direction. Consequently, two separate models are trained, one for each component of $\boldsymbol{f}_P$. The input features $\mathcal{X}$ and target dependent variables $\mathcal{Y}$ are normalised and defined as $\mathcal{X}_i=\{ r/(D/2), u_i/U_\mathrm{max}, u_{ib_x}/U_\mathrm{max}, a_{ib_x}/(U_\mathrm{max}f), \omega_z/f \}$ and $\mathcal{Y}_i=\boldsymbol{f}_{P_i}/(U_\mathrm{max}f)$, respectively, where the sub-index $i=(x,y)$ denotes the corresponding spatial direction.

Two metrics are employed to evaluate the quality of the black-box models on the validation dataset (comprising $0.2m_T$ samples): the coefficient of determination $\mathcal{R}^2$ and the normalised root mean square error $\textrm{NRMSE}$. Together, these metrics quantify both the variance explained by the model and the magnitude of the average prediction error. Their expressions are given by:

\begin{eqnarray}
    \mathcal{R}^2 &=& 1 - \frac{\sum_{j=1}^{0.2m_T} (\mathcal{Y}_j - \hat{\mathcal{Y}_j})^2} {\sum_{j=1}^{0.2m_T} (\mathcal{Y}_j - \overline{\mathcal{Y}})^2} \\
    \textrm{NRMSE} &=& \frac{1}{Q_3 - Q_1} \sqrt{\frac{\sum_{j=1}^{0.2m_T} (\mathcal{Y}_j - \hat{\mathcal{Y}}_j)^2}{0.2m_T}}
\end{eqnarray}
where $j \in [1, 0.2m_T]$ indexes the individual validation samples. For normalisation, the metrics employ the mean of the reference data $\overline{\mathcal{Y}}$ and the interquartile range $Q_3 - Q_1$, corresponding to the $25$\textsuperscript{th} and $75$\textsuperscript{th} percentiles of the true values $\mathcal{Y}$, which is advantageous due to its robustness against outliers. The results, summarised in Tab.~\ref{tab:metrics_RFR_slidingWindow}, show consistently high $\mathcal{R}^2$ values and low $\textrm{NRMSE}$ levels, confirming the excellent predictive accuracy and generalisation capability of all trained RFR models.

\begin{table}[!h]
    \centering
    \begin{tabular}{c | c c c c}
         & $\textrm{NRMSE}_x$ & $\mathcal{R}^2_x$ & $\textrm{NRMSE}_y$ & $\mathcal{R}^2_y$  \\ \hline
      RFR model 1 & $0.053$ & $0.997$ & $0.014$ & $1$ \\
      RFR model 2  & $0.048$ & $0.997$ & $0.012$ & $1$ \\
      RFR model 3 & $0.051$ & $0.998$ & $0.016$ & $1$ \\
      RFR model 4  & $0.053$ & $0.996$ & $0.015$ & $1$ \\
      RFR model 5 & $0.045$ & $0.999$ & $0.015$ & $1$
    \end{tabular}
    \caption{Metrics computed for the two components ($x,y$) of the forcing term $\boldsymbol{f}_P$ for each of the five RFR models employed}
    \label{tab:metrics_RFR_slidingWindow}
\end{table}

\subsection{Results}
\label{sec:ResultsML_cylinder}

\begin{figure}[!h]
    \centering
    \begin{tabular}{ccc}
    \includegraphics[width=0.32\linewidth]{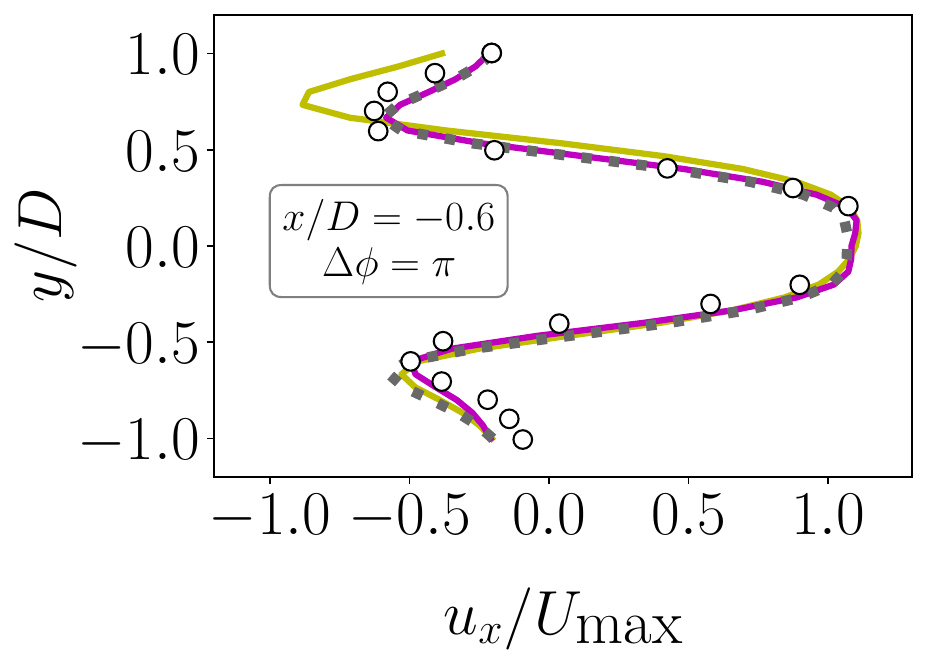} &
    \includegraphics[width=0.32\linewidth]{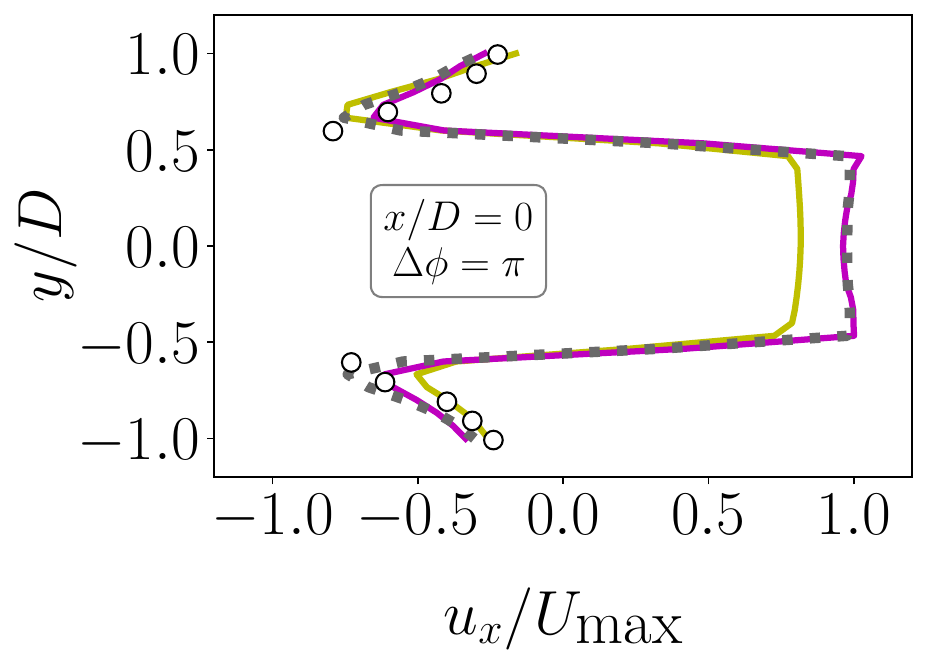} &
    \includegraphics[width=0.32\linewidth]{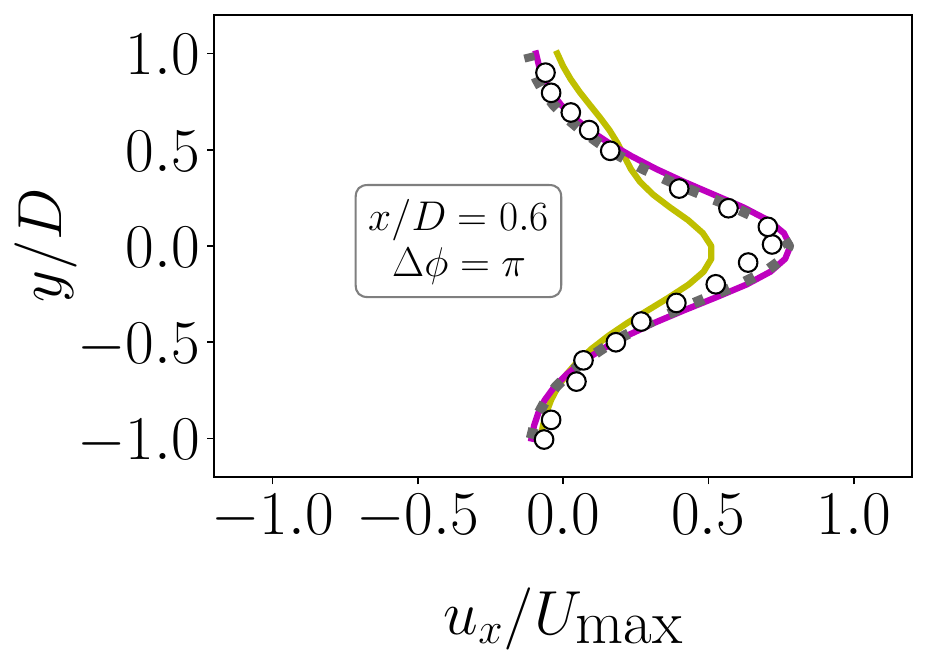} \\
    \textit{(a)} & \textit{(b)} & \textit{(c)} \\
    \includegraphics[width=0.32\linewidth]{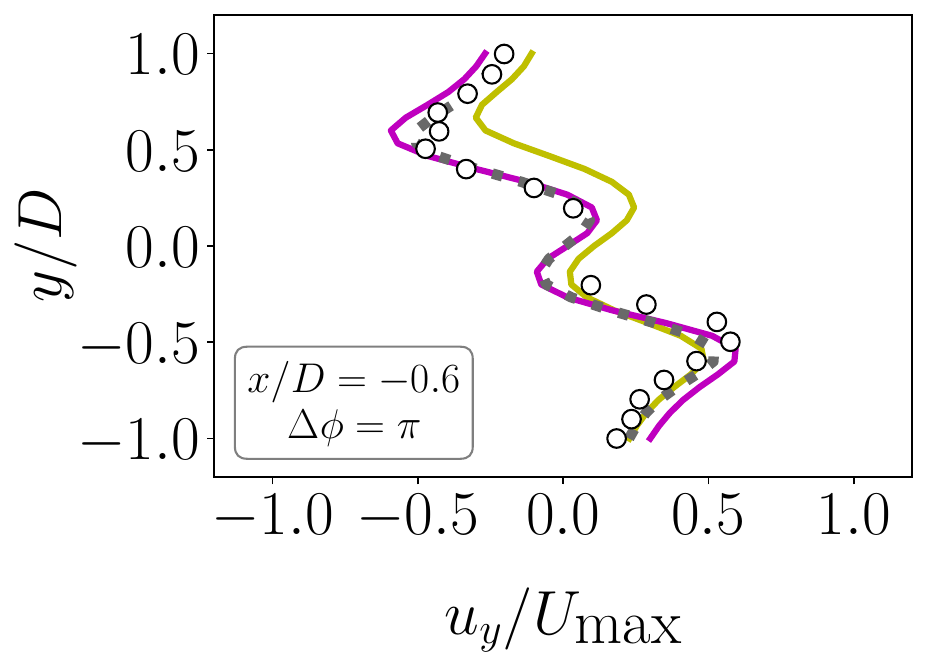} &
    \includegraphics[width=0.32\linewidth]{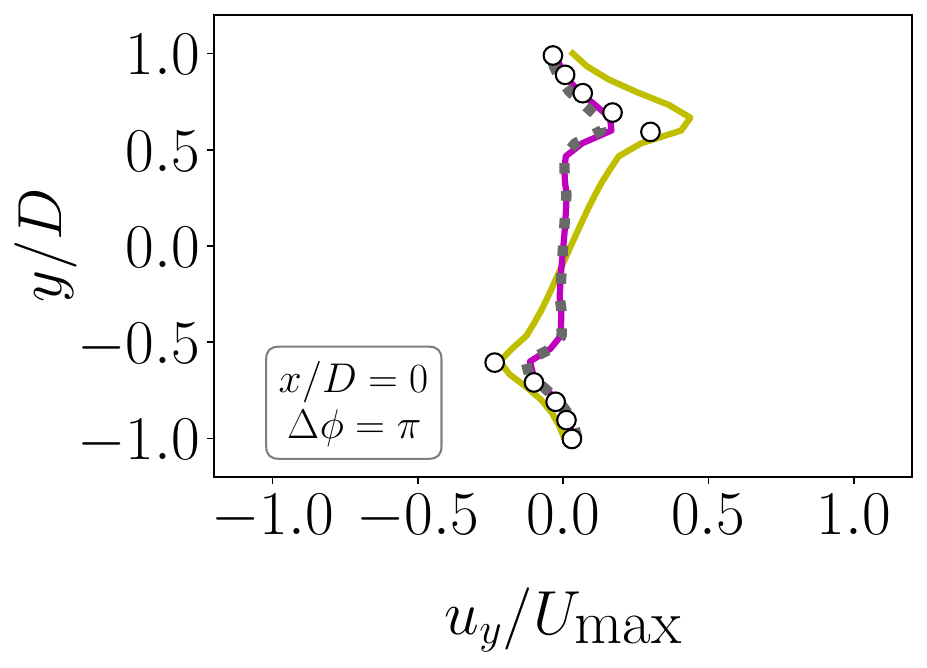} &
    \includegraphics[width=0.32\linewidth]{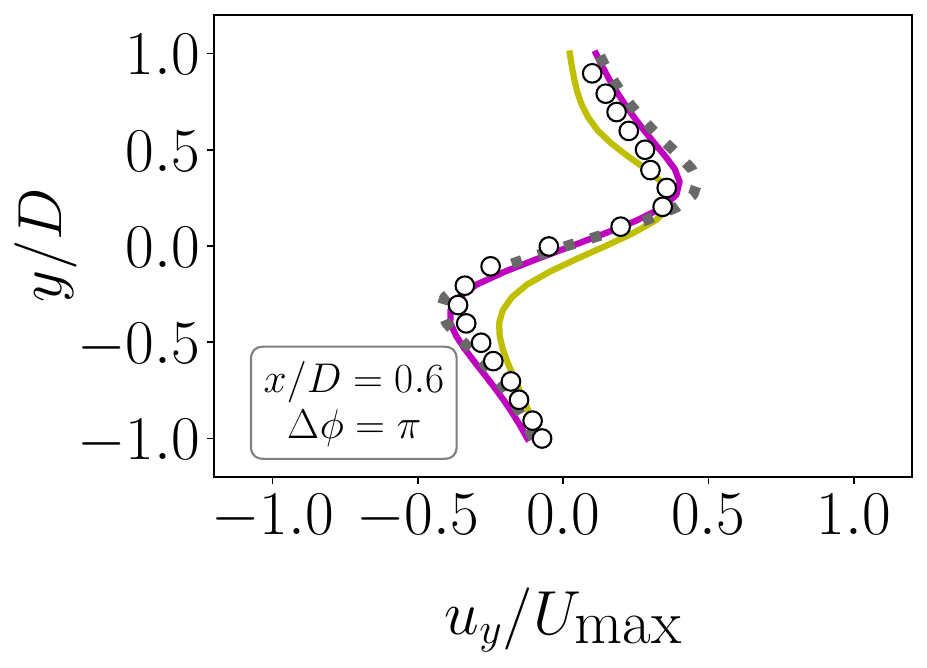} \\
    \textit{(d)} & \textit{(e)} & \textit{(f)}     
    \end{tabular}
    \caption{Velocity profiles when $\Delta \phi = \pi$ at three different planes $x/D \in [ -0.6, 0, 0.6]$ for (\protect\magentaline) the ML-IBM run, (\protect\greylinedottedstrong) the DA-IBM experiment, and (\protect\yellowline) the LF-IBM simulation. (\protect\circlewhite) represents the data from the experiments conducted in \citet{Dutsch1998_jfm}}
    \label{fig:velocity_ML_phi180}
\end{figure}

\begin{figure}[!h]
    \centering
    \begin{tabular}{ccc}
    \includegraphics[width=0.32\linewidth]{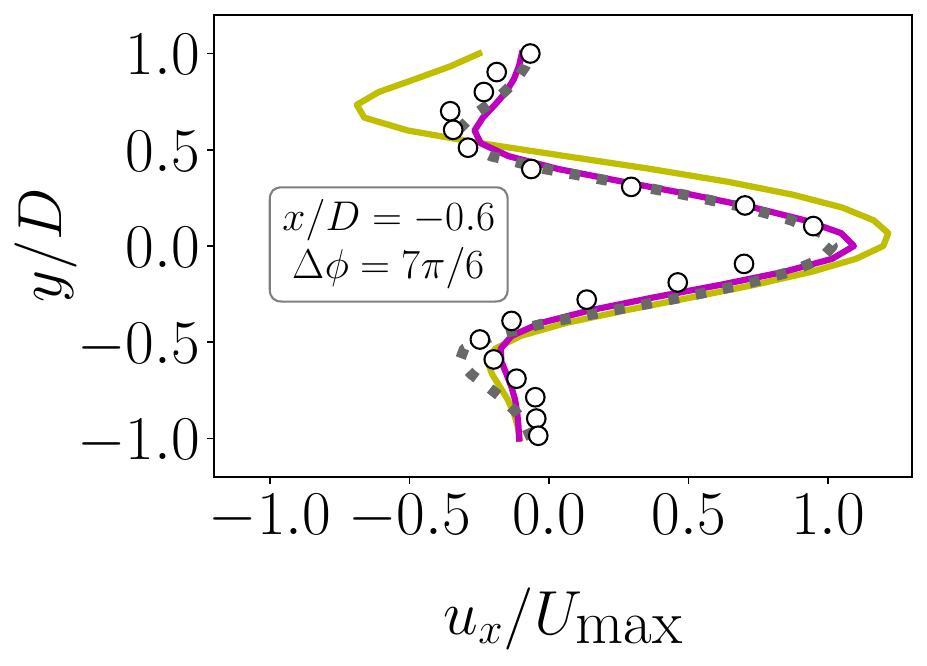} &
    \includegraphics[width=0.32\linewidth]{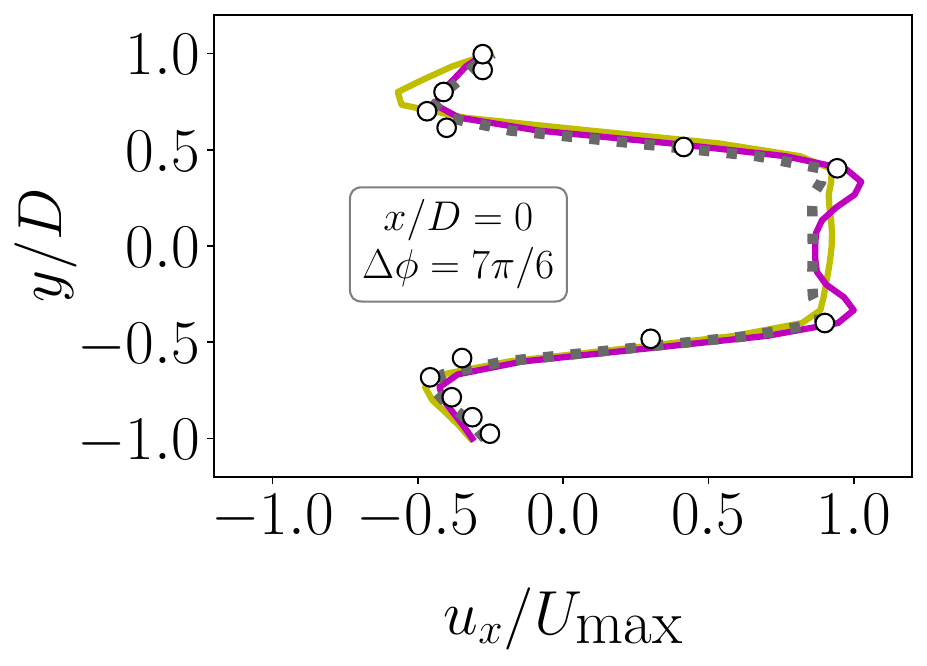} &
    \includegraphics[width=0.32\linewidth]{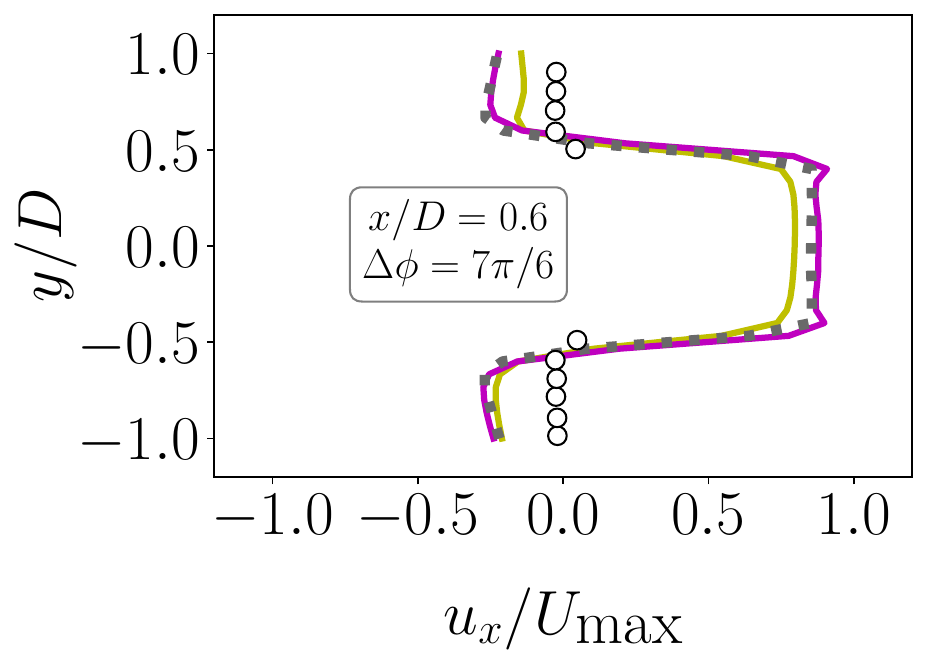} \\
    \textit{(a)} & \textit{(b)} & \textit{(c)} \\
    \includegraphics[width=0.32\linewidth]{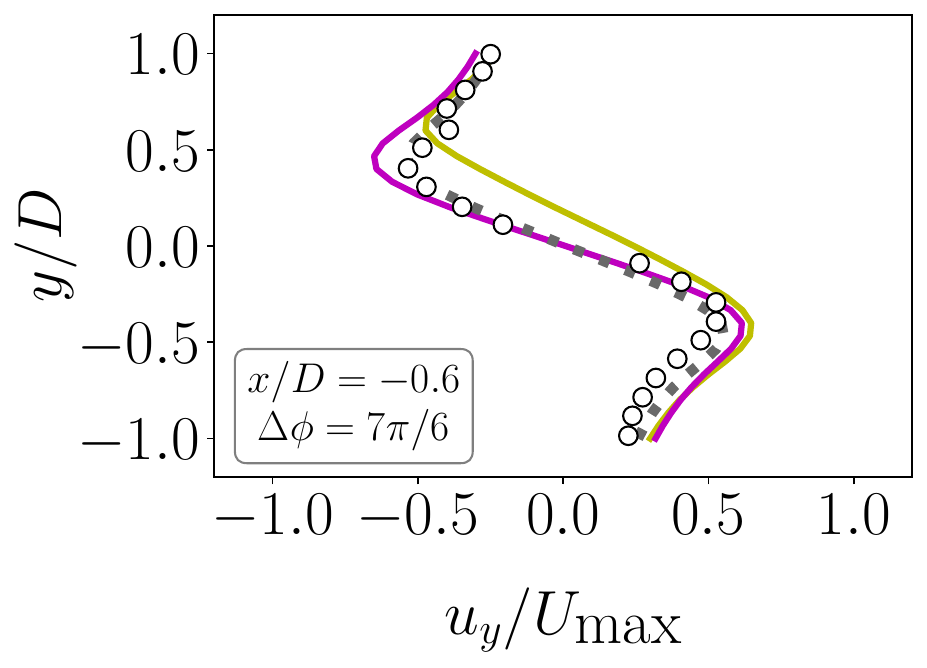} &
    \includegraphics[width=0.32\linewidth]{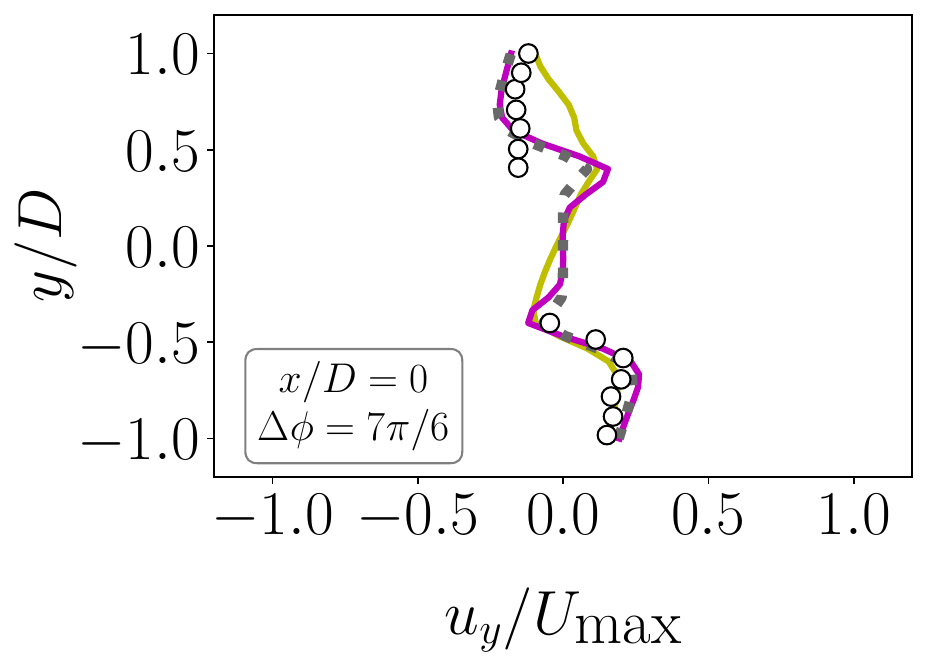} &
    \includegraphics[width=0.32\linewidth]{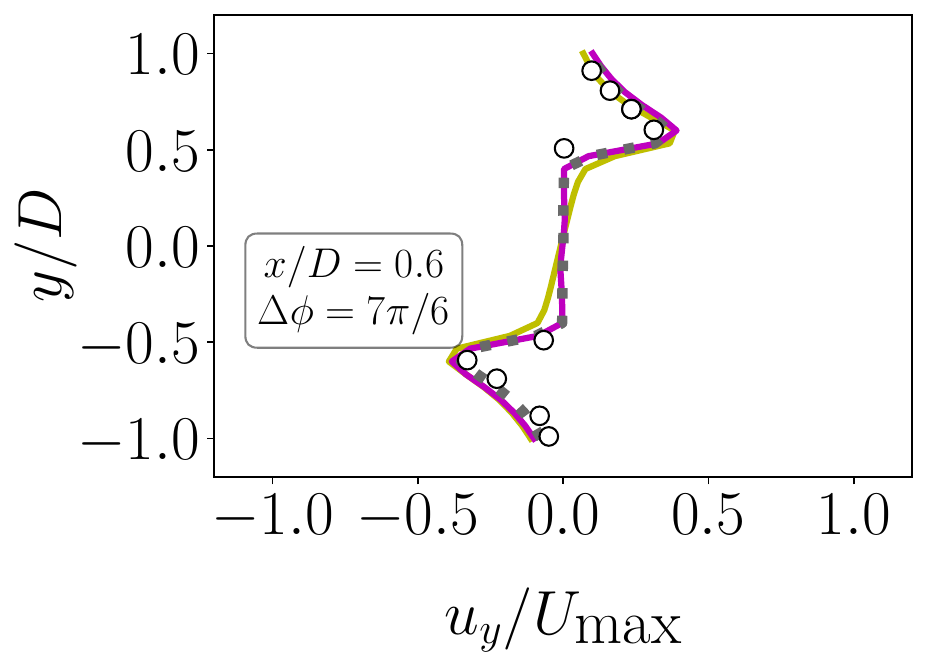} \\
    \textit{(d)} & \textit{(e)} & \textit{(f)}     
    \end{tabular}
    \caption{Velocity profiles when $\Delta \phi = 7\pi/6$ at three different planes $x/D \in [ -0.6, 0, 0.6]$ for (\protect\magentaline) the ML-IBM run, (\protect\greylinedottedstrong) the DA-IBM experiment, and (\protect\yellowline) the LF-IBM simulation. (\protect\circlewhite) represents the data from the experiments conducted in \citet{Dutsch1998_jfm}}
    \label{fig:velocity_ML_phi210}
\end{figure}

In this section, we assess the performance of the sliding-window strategy under conditions identical to those of the training dataset. This simulation is referred to as ML-IBM. We begin by examining the velocity profiles at three planes for the phase differences $\Delta \phi = \pi$ and $\Delta \phi=7\pi/6$, as previously analysed in \S\ref{sec:ResultsDA_cylinder}. Fig.~\ref{fig:velocity_ML_phi180} presents the profiles for $\Delta \phi = \pi$. The ML-IBM results nearly coincide with those obtained using the DA-IBM method and show a substantial improvement over the LF-IBM solution. Minor discrepancies with respect to DA-IBM appear in regions with pronounced velocity gradients, but these differences are negligible relative to the overall agreement. Fig.~\ref{fig:velocity_ML_phi210} illustrates the case for $\Delta \phi = 7\pi/6$, where some limitations of the ML approach become evident. Although the ML-IBM results generally follow the DA-IBM trends and continue to outperform LF-IBM, the method struggles to accurately reproduce the wake structure behind the cylinder. This deviation is most apparent in Fig.~\ref{fig:velocity_ML_phi210}\textit{(b)}, where the velocity profile exhibits a parabolic shape (with reduced velocity near $y/D = 0$) not captured by either the IBM or DA formulations. Nevertheless, this remains the only noticeable discrepancy, and overall, the ML-IBM results represent a marked improvement compared with the LF-IBM baseline.

\begin{figure}[!h]
    \centering
    \begin{tabular}{cc}
        \includegraphics[width=0.48\linewidth]{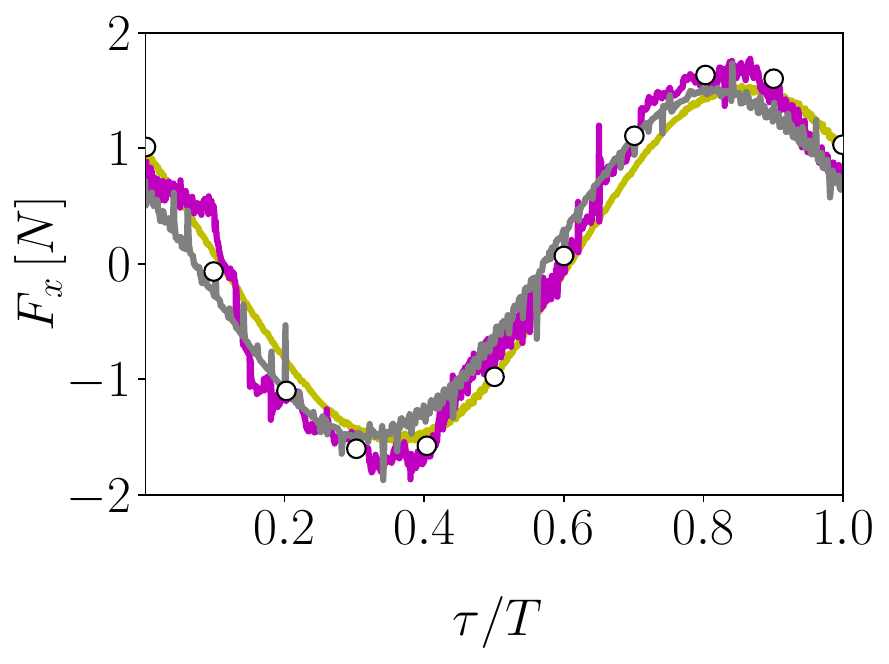} & 
        \includegraphics[width=0.52\linewidth]{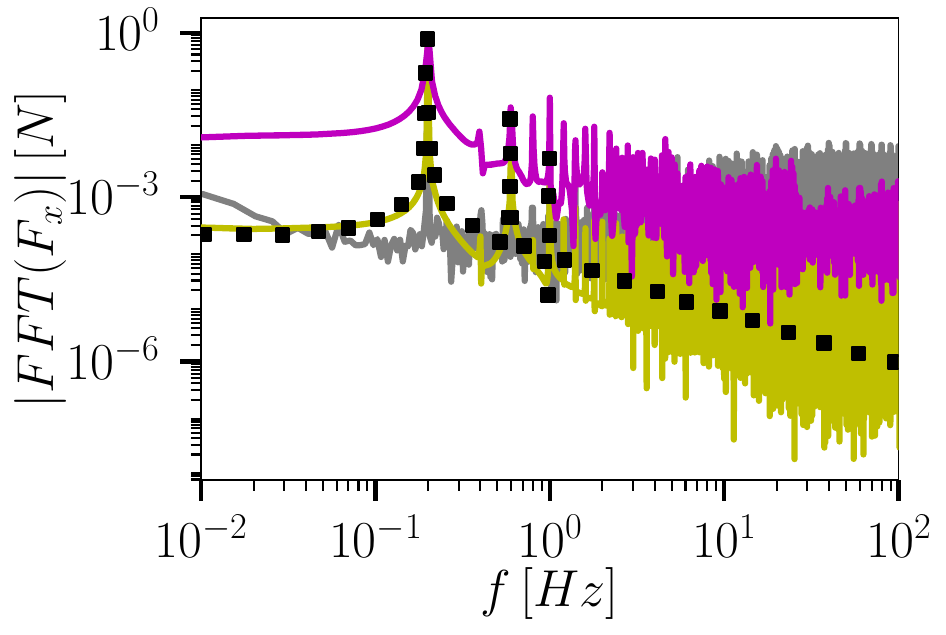} \\
         \textit{(a)} & \textit{(b)}
    \end{tabular}
    \caption{\textit{(a)} Resistive force $F_x$ over one period $T$, and \textit{(b)} frequency resistive force spectra. The simulations represented are (\protect\magentaline) ML-IBM, (\protect\greylinesolidstrong) DA-IBM, and (\protect\yellowline) LF-IBM. (\protect\circlewhite) represents the data from the experiments conducted in \citet{Dutsch1998_jfm} and ($\blacksquare$) is a body-fitted simulation run in \citet{Tsetoglou2024_nmf}}
    \label{fig:resistiveforce_ML}
\end{figure}

Fig.~\ref{fig:resistiveforce_ML} presents the analysis of the resistive force $F_x$ obtained with the ML models and illustrates the main areas of improvement for the current approach. The temporal profile over a single period in Fig.~\ref{fig:resistiveforce_ML}\textit{(a)} is noticeably noisier than that obtained with the DA method, with some phases of the oscillation cycle exhibiting sharper, less physical variations. Even so, the level of noise remains substantially lower than in the baseline IBM simulation performed at the same grid resolution ($\Delta \boldsymbol{x}=D/25$) and penalisation parameter $\alpha_p=0.4$, shown previously in Fig.~\ref{fig:resistiveforce_penalisation}\textit{(a)}. Furthermore, the spectral analysis in Fig.~\ref{fig:resistiveforce_ML}\textit{(b)} reveals that the ML prediction contains less high-frequency noise than the DA result. However, the low-frequency spectrum is remarkably overestimated, although the dominant vortex-shedding frequency is still correctly identified at $f=0.2\,Hz$. These discrepancies may stem not only from limitations of the ML models in reconstructing the flow during specific oscillation phases, but also from the absence of the \emph{state-estimation} corrections that the DA procedure applies to the flow field, which are not replicated here. Future investigations will therefore examine the inclusion of additional, phase-dependent ML models designed to provide local corrections to the field, emulating the role of state estimation within the DA framework.

\section{Conclusion}
\label{sec:Conclusions}

The present study introduces a dual data-driven framework that combines Data Assimilation (DA) and Machine Learning (ML) to optimise the performance of a penalisation Immersed Boundary Method (IBM). Using an Ensemble Kalman Filter (EnKF) and Random Forest Regression (RFR), the approach is applied to an oscillating cylinder undergoing harmonic motion in a quiescent fluid, a configuration in which flow statistics vary periodically in time. The results show that DA-ML techniques is more accurate than the baseline IBM, improving both local velocity profiles at different oscillation phases $\Delta \phi$ and integral quantities such as the resistive force $F_x$, even though the distance between the sensors and the body surface is not uniform throughout the cycle.

Building on these results, ongoing research seeks to extend the framework to more complex statistically unsteady flows, including cases with oscillatory features where neither the phase nor the flow field is known a priori. In such settings, training phase-specific ML models naturally motivate a temporal Mixture of Experts (MoE) strategy, in which specialised models handle different dynamics regimes and a separate gating mechanism assigns their relative contributions. Similar multi-expert approaches have proven effective in other contexts, such as the RANS-correction framework of \citet{deZordoBanliat2024_jcp} and the flow-class-dependent modelling strategy of \citet{Cherroud2025_jcp}.

The oscillating-cylinder configuration studied here lends itself directly to this MoE perspective, as its symmetry and strict periodicity already define clear phase-dependent regimes. The ML models developed in this work can thus be interpreted as initial ``experts,'' with a future gating model enabling their adaptive activation during prediction. Such an extension would offer the flexibility needed to handle more intricate or weakly periodic behaviour (such as asymmetry, amplitude modulation, or drift) and would provide a path toward generalising the methodology across a broader range of unsteady flows. In particular, exploring regimes characterised by different Reynolds numbers $Re$ and Keulegan-Carpenter numbers $KC$, where diverse vortex patterns emerge \citep{Tatsuno1990_jfm}, could further expand the applicability of ML-assisted IBMs to a wide spectrum of oscillatory configurations.

\begin{appendix}
\section{Appendix. Random Forest Regression and coupling with the PISO algorithm}
\label{sec:RFR}

Random Forest Regression (RFR) \citep{Breiman2001_ml} is an ensemble learning algorithm that builds a number of $N$ decision trees during the training phase. The final output is estimated as the average of the decision tree predictions, thereby enhancing accuracy and robustness, and making it a reliable model for regression tasks. By combining bootstrapping and random feature selection to account for each decision tree's predictions, it reduces overfitting risk and enhances model stability. Each decision tree $n \in [1,...,N]$ is constructed using a subset of the entire training dataset and selects splitting criteria or nodes to progress in tree depth. Commonly, the splitting criterion aims to minimise the variance of the target variable within resulting subsets, i.e., the split point is chosen to minimise the Mean Square Error (MSE) of the child nodes (MSE\textsubscript{child}) and to maximise the reduction in MSE ($\Delta\textrm{MSE}$) from the parent to the child nodes, which is computed as the difference between the MSE in the parent node and the weighted sum of the MSEs in the resulting child nodes. These quantities are calculated using the following expressions:
\begin{eqnarray}
 \textrm{MSE} &=& \frac{1}{m} \sum_{j = 1}^m \left(\mathcal{Y}_j - \hat{\mathcal{Y}_j} \right)^2 \\
 \Delta\textrm{MSE} &=& \mathrm{MSE_{parent}} - \overbrace{\left(\frac{m_{\mathrm{left}}}{m} \,\mathrm{MSE_{left}} + \frac{m_{\mathrm{right}}}{m} \,\mathrm{MSE_{right}} \right)}^{\text{MSE\textsubscript{child}}} 
\end{eqnarray}
where $m$ represents the number of samples of the parent node, $\mathcal{Y}_j$ and $\hat{\mathcal{Y}_j}$ point out to the actual and predicted target value for the sample $j \in [1,m]$, and $m_{\mathrm{left}}$ and $m_{\mathrm{right}}$ are associated with the number of samples in the left and right child nodes. The MSEs for the parent, left child, and right child nodes are $\mathrm{MSE_{parent}}$, $\mathrm{MSE_{left}}$ and $\mathrm{MSE_{right}}$, respectively. 

At each parent node, the dataset is divided into two subsets based on a specific feature $\mathcal{X}_l$ and a split point, yielding two child nodes. The left child node usually represents data meeting the split criteria  (e.g., feature value $\mathcal{X}_{j,l} \leq$ split point), while the right child node does not (e.g., feature value $\mathcal{X}_{j,l} >$ split point). This splitting-nodes procedure continues recursively for each child node until predefined stopping criteria are met, which, in our case, is the minimum number of samples required to form a leaf node $m_{sl}$.

Different from the penalty Immersed Boundary Method (IBM) methodology, which defines an implicit forcing term $\boldsymbol{f}_P=\boldsymbol{f}_P(\boldsymbol{u})$ in the momentum equation in (\ref{eqn:momentum_eq}), the RFR algorithm requires the computation of an explicit output term, leading to several estimations of $\hat{\boldsymbol{f}_P}$ within the PISO loop, as described in Alg. \ref{alg:IBM_ML_cylinder}.

\begin{algorithm}[!h]
    \caption{Scheme of penalisation ML-IBM for the oscillating cylinder.}
    \label{alg:IBM_ML_cylinder}
    Let us consider the initial time $t_0$, the time step $\Delta t$, and $j \in [1, 2]$ as a single iteration within the PISO loop: \\
    \For{$t = t_0 + \Delta t, t_0 +2\,\Delta t,..., t_{\mathrm{end}}$}
    {
    \nl Update of the position ${\boldsymbol{x_{ib}}}_t$ (which allows for the calculation of the radial distance $r_t$), the velocity ${\boldsymbol{u_{ib}}}_t$ and acceleration ${\boldsymbol{a_{ib}}}_t$ of the cylinder, and computation of the new mesh elements being $\Omega_b$. \\
    \nl Resolution of the momentum equation, from where $\boldsymbol{u}_{t,0}$ is obtained: \\
    \qquad  $\boldsymbol{A} \,\boldsymbol{u}_{t,0} - \boldsymbol{b} - \hat{\boldsymbol{f}}_{P_{t-\Delta t, 2}} = -\nabla p_{t-\Delta t, 2}$\\
    \nl Estimation of the dimensional forcing term $\hat{\boldsymbol{f}}_{P_{t,0}}$ from $\hat{\boldsymbol{f}}_{P_{t,0}}/(U_\textrm{max}f)$ with the predicted velocity $\boldsymbol{u}_{t,0}$ (employed to calculate $\boldsymbol{\omega}_{t,0}$) by means of RFR. By calling the operator $\mathcal{F}_{\textrm{p.o.}}^n$ to define the \emph{parameter optimisation} (p.o.) prediction of a single decision tree $n$: \\
    \qquad $\frac{\hat{\boldsymbol{f}}_{P_{t,0}}}{U_\textrm{max}f}= \frac{1}{N}\sum_{n = 1}^N \mathcal{F}_{\textrm{p.o.}}^n \left(\frac{r_t}{D/2}, \frac{\boldsymbol{u}_{t,0}}{U_\textrm{max}}, \frac{u_{ib_{x,t}}}{U_\textrm{max}}, \frac{a_{ib_{x,t}}}{U_\textrm{max}f}, \frac{\omega_{z_{t,0}}}{f}\right)$ \\
    \For{$j = 1, 2$}{
    \nl Computation of the pressure $p_{t, j}$ through the Poisson equation. $A$ is a scalar field calculated from $\boldsymbol{A}$, whereas $\boldsymbol{T}^\prime(\boldsymbol{u}_{t,j-1})$ is a tensor field containing the discretised form of all the terms on the left side of the momentum equation. To account for the update of the forcing term $\hat{\boldsymbol{f}}_{P_{t,j-1}}$, an additional term is included. \\
    \qquad $\boldsymbol{\nabla} \cdot \frac{1}{A} \nabla{p_{t, j}} = \boldsymbol{\nabla} \cdot \left(\frac{\boldsymbol{T}^\prime(\boldsymbol{u}_{t,j-1})}{A}\right) + \boldsymbol{\nabla} \cdot  \left(\frac{\hat{\boldsymbol{f}}_{P_{t,j-1}}}{A} \right)$ \\
    \nl Update of the velocity field $\boldsymbol{u}_{t, j}$ to satisfy the zero-divergence condition:\\
    \qquad $\boldsymbol{u}_{t,j} = \frac{\boldsymbol{T}^\prime(\boldsymbol{u}_{t,j-1})}{A} - \frac{1}{A} \nabla p_{t,j} + \frac{1}{A} \hat{\boldsymbol{f}}_{P_{t,j-1}}$ \\
    \nl Estimation of the forcing term $\hat{\boldsymbol{f}}_{P_{t,j}}$ with the new velocity field $\boldsymbol{u}_{t,j}$ (and $\boldsymbol{\omega}_{t,j}$) via RFR: \\
    \qquad $\frac{\hat{\boldsymbol{f}}_{P_{t,j}}}{U_\textrm{max}f}= \frac{1}{N}\sum_{n = 1}^N \mathcal{F}_{\textrm{p.o.}}^n \left(\frac{r_t}{D/2}, \frac{\boldsymbol{u}_{t,j}}{U_\textrm{max}}, \frac{u_{ib_{x,t}}}{U_\textrm{max}}, \frac{a_{ib_{x,t}}}{U_\textrm{max}f}, \frac{\omega_{z_{t,j}}}{f}\right)$ \\
    }
    }
\end{algorithm}

\section{Appendix. PISO algorithm with the IBM-DA methodology}
\label{sec:DA_PISO}

The original PISO loop incorporates additional operations to apply the Data Assimilation (DA) corrections online, accounting for the cylinder's motion while retaining the original implicit treatment of the forcing term $\boldsymbol{f}_P$ from the penalty Immersed Boundary Method (IBM). The full procedure is detailed in Alg.~\ref{alg:IBM_DA_cylinder}.

\begin{algorithm}[!h]
    \caption{Scheme of penalisation DA-IBM for the oscillating cylinder.}
    \label{alg:IBM_DA_cylinder}
    Let us consider the initial time $t_0$, the time step $\Delta t$, and $j \in [1, 2]$ as a single iteration within the PISO loop: \\
    \For{$t = t_0 + \Delta t, t_0 +2\,\Delta t,..., t_{\mathrm{end}}$}
    {
    \nl Update of the position ${\boldsymbol{x_{ib}}}_t$ and velocity ${\boldsymbol{u_{ib}}}_t$ of the cylinder, and calculation of the new mesh elements being $\Omega_b$. \\
    \nl Resolution of the momentum equation, from where $\boldsymbol{u}_{t,0}$ is obtained: \\
    \qquad $\boldsymbol{A}\, \boldsymbol{u}_{t,0} - \boldsymbol{b} - \boldsymbol{f}_P(\boldsymbol{u}_{t,0}, {\boldsymbol{u_{ib}}}_t, \theta_{t-\Delta t}) = -\nabla p_{t - \Delta t,2} $\\
    \For{$j = 1, 2$}{
    \nl Estimation of the pressure $p_{t, j}$ through the Poisson equation. $A^\prime$ is a scalar field calculated from $\boldsymbol{A}$ and $\boldsymbol{f}_P$, whereas $\boldsymbol{T}^{\prime\prime}(\boldsymbol{u}_{t,j-1})$ is a tensor field containing the discretised form of all the terms on the left side of the previous equation: \\
    \qquad $\boldsymbol{\nabla} \cdot \cfrac{1}{A^\prime} \nabla{p_{t, j}} = \boldsymbol{\nabla} \cdot \left(\cfrac{\boldsymbol{T}^{\prime\prime}(\boldsymbol{u}_{t,j-1})}{A^\prime}\right)$ \\
    \nl Update of the velocity field $\boldsymbol{u}_{t, j}$ to satisfy the zero-divergence condition:\\
    \qquad $\boldsymbol{u}_{t,j} = \cfrac{\boldsymbol{T}^{\prime\prime}(\boldsymbol{u}_{t,j-1})}{A^\prime} - \cfrac{1}{A^\prime} \nabla p_{t,j}$
    }
    \nl For state estimation and parameter optimisation at time $t$, the velocity $\boldsymbol{u}_{t,2}$ corresponds to $\boldsymbol{u}_k^f$ for the EnKF at the $k^{th}$ analysis phase. By employing the DA algorithm, we obtain $\boldsymbol{u}_k^a$ and $\theta_k^a = \{ x_{0,k}^a, r_{x,k}^{\ast,a}, \alpha_{P_{x,k}}^{\textrm{min},a}, r_{y,k}^{\ast,a}, \alpha_{P_{y,k}}^{\textrm{min},a}\}$. \\
    \nl Update of the position ${\boldsymbol{x}_{ib}}_t^a$ of the cylinder with the analysis $x_0^a$, and re-calculation of the new mesh elements being $\Omega_b$. \\
    \nl Resolution of $\boldsymbol{f}_P^a$ by estimating the coefficients of (\ref{eqn:radial_expansion}) with the updated $\Omega_b$, the updated parameters $r_{x,k}^{\ast,a}, \alpha_{P_{x,k}}^{\textrm{min},a}, r_{y,k}^{\ast,a}, \alpha_{P_{y,k}}^{\textrm{min},a}$, and the updated state $\boldsymbol{u}_k^a$.
    }
\end{algorithm}
\end{appendix}

\begin{Backmatter}

%\paragraph{Acknowledgments}
%We are grateful for the technical assistance of A. Author.

\paragraph{Funding Statement}
This research was supported by the grant from ANR-JCJC-2021 IWP-IBM-DA.

\paragraph{Competing Interests}
None.

\paragraph{Data Availability Statement}
Data is available upon reasonable request. Code with the online Data Assimilation framework (CONES) can be found in: \verb+\url{https://gitlab.ensam.eu/pe431/cones-dev}+.

\paragraph{Ethical Standards}
The research meets all ethical guidelines, including adherence to the legal requirements of the study country.

\paragraph{Author Contributions}
Conceptualisation: M.V; M.M. Data Curation: M.V. Formal analysis: M.V; M.M. Funding acquisition: M.M. Investigation: M.V. Methodology: M.V; M.M. Project administration: M.V; M.M. Resources: M.V; M.M. Software: M.V. Supervision: M.M. Validation: M.V. Visualisation: M.V. Writing---original draft: M.V; M.M. Writing---review \& editing: M.V; M.M. All authors approved the final submitted draft.

% If using any of the following journal options:
%   wet, dap, dce, eds, prm, flw, jdm, psy, rsm
% then use the \printbibliography line instead of:
%\bibliography{example}
\printbibliography

\end{Backmatter}

\end{document}